\documentclass[a4paper]{aa}%
\usepackage{graphicx}
\usepackage{natbib}
\usepackage{longtable,lscape}
\usepackage{slashbox}

\usepackage{txfonts}  

\begin{document}

\titlerunning{Spectral monitoring of 3C 390.3}

\title{Spectral optical monitoring of 3C 390.3 in 1995-2007:
II. Variability of the spectral line parameters}

\author{L.\v C. Popovi\'c\inst{1,2} \and A.I. Shapovalova\inst{3}
\and D. Ili\'c\inst{2,4} \and A. Kova\v cevi\'c\inst{2,4} \and
W. Kollatschny\inst{5} \and A.N. Burenkov\inst{3} \and V.H. Chavushyan\inst{6}
\and N.G. Bochkarev\inst{7} \and  J. Le\'on-Tavares\inst{8}}

\institute{Astronomical Observatory, Volgina 7, 11160 Belgrade 74, Serbia
\and
Isaac Newton Institute of Chile, Yugoslavia Branch
\and
Special Astrophysical Observatory of the Russian AS,
Nizhnij Arkhyz, Karachaevo-Cherkesia 369167, Russia
\and
 Department of Astronomy, Faculty of Mathematics, University
of Belgrade, Studentski trg 16, 11000 Belgrade, Serbia
\and
Institut f\"ur Astrophysik,  Friedrich-Hund-Platz 1, G\"ottingen, Germany
\and
Instituto Nacional de Astrof\'{\i}sica, \'{O}ptica y
Electr\'onica, Apartado Postal 51, CP 72000, Puebla, Pue. M\'exico
\and
Sternberg Astronomical Institute, Moscow, Russia
\and
Mets\"ahovi Radio Observatory, Helsinki University of Technology TKK,
Mets\"ahovintie 114, FIN-02540 Kylm\"al\"a, Finland
}

\authorrunning{L.\v C. Popovi\'c et al.}

\offprints{L.\v C. Popovi\'c, \\ \email{lpopovic@aob.rs}\\ }

\date{Received  / Accepted }

\abstract
{A study of the variability of the broad emission-line parameters of
3C390.3, an active galaxy with the double-peaked emission-line
profiles, is {presented}. Here we give a detail analysis of
variation in the broad H$\alpha$ and H$\beta$ emission-line
profiles, the ratios, and the Balmer decrement of different line
segments.}
{With investigation of the variability of the broad line profiles we
explore the disk structure, that is assumed to emit the broad
double-peaked H$\beta$ and H$\alpha$ emission lines in the spectrum
of 3C390.3.}
{We divided the observed spectra in two periods (before and after
the outburst in 2002) and analyzed separately the variation in these
two periods. First we analyzed the spectral emission-line profiles
of the H$\alpha$ and H$\beta$ lines, measuring the peak positions.
Then, we divided lines into several segments, and we measured the
line-segment fluxes. The Balmer decrement variation for total
H$\alpha$ and H$\beta$ fluxes, as well as for the line segments has
been investigated and discussed. Additionally, we modeled the line
parameters variation using an accretion disk model and compare our
modeled line parameter variations with observed ones.}
{We compared the variability in the observed line parameters with
the disk model predictions and we found that the variation in line
profiles and in the line segments corresponds to the emission of a
disk-like BLR. But, also there is probably one additional emission
component that contributes to the H$\alpha$ and H$\beta$ line
center. We found that  the variation in the line profiles is caused
by the variation in the  parameters of the disk-like BLR, first of
all in the inner (outer) radius which can well explain the line
parameter variations in the Period I. The Balmer decrement across
the line profile has a bell-like shape, and it is affected not only
by physical processes in the disk, but also by different emitting
disk dimension of the H$\alpha$ and H$\beta$ line.}
{The geometry of the BLR of 3C390.3 seems to be very complex, and
inflows/outflows might be present, but it is evident that the broad
line region with disk-like geometry has dominant emission.}


\maketitle

\section{Introduction}

{The broad emission lines (BELs) are often observed in optical
and ultraviolet spectra of acitve galactic nuclei (AGN). The study
of the profiles and intensities of BELs can give us relevant
information about the geometry and physics of the broad line region
(BLR). The physics and geometry of the BLR are  uncertain and
investigation of BEL shape variability  in a long period is very
useful for determination of the BLR nature. The profiles of the BELs
in AGN can indicate the geometry of emitting plasma in the BLR
\citep[see e.g.][etc.]{sul00,pop04,ga09,za10}.

Particularly, very interesting objects are AGNs with unusual broad
emission-line profiles, where broad Balmer lines show double peaks
or double ''shoulders'', so called double-peaked emitters. The
double peaked line profiles may be caused by an accretion disk
emission. On the other hand,} the presence of an accretion-disk
emission in the BLR is expected, and double-peaked line profiles of
some AGN indicate {this} \citep{pe88,eh94,eh04,er09}. One of the
well known AGN with broad double-peaked emission lines in its
spectrum is the radio-loud active galaxy 3C 390.3. Although, the
double-peaked line profiles can be explained by different hypothesis
\citep[see e.g.][]{vz91}, as e.g. super-massive binary black holes
\citep{gas96}, outflowing bi-conical gas streams \citep{zh96}), it
seems that in this case a disk emission is present in the BLR
\citep[][hereinafter Paper I]{sh10}. There is a possibility that a
jet emission can affect the optical emission in 3C 390.3
\citep{tig10}, and some perturbations in disk could be present
\citep{jov10} and they can also affect the double-peaked line
profiles.

Long-term variability in the line/continuum flux as well as in the
line profiles is observed in objects with broad double peaked lines
\citep[see e.g.][]{di88,sh01,se02,se10,sh10,le10}. Long-term
variability of broad line profiles is intriguing because it is
usually unrelated to more rapid changes in the continuum flux, but
probably is related to physical changes in the accretion disk, as
e.g. brightness of some part of the disk, or changes in the disk
size and distance to the central black hole. The double-peaked broad
line profile variability can be exploited to test various models for
the accretion disk (as e.g. circular or elliptical). {Moreover, the
double-peaked broad line studies can provide important information
about the accretion disk, as e.g. inclination, dimension and
emissivity of the disk as well as  probes of dynamical phenomena
that may occur in the disk \citep[see in more details][and reference
therein]{er09}.}

In Paper I we have presented the results of the long-term
(1995--2007) spectral monitoring of \object{3C~390.3}. We have
analyzed the light curves of the broad H$\alpha$ and H$\beta$ line
fluxes and the continuum flux in the 13-year period.  We also found
that quasi-periodical oscillations (QPO) may be present in the
continuum and H$\beta$ light curves. We studied averaged profile of
the H$\alpha$ and H$\beta$ line in two periods (Period I from 1995
to 2002, and Period II from 2003 to 2007) and their characteristics
(as e.g. peaks separation and their intensity ratio, or FWHM). From
the cross-correlations (ICCF and ZCCF) between the continuum flux
and H$\beta$ and H$\alpha$ lines we found the lag of $\sim$95 days
for H$\beta$ and $\sim$120 days for H$\alpha$ (see Paper I for
details). We concluded that the broad emission region has disk-like
structure, but there could probably exists an additional component,
non-disk or also disk-like, with different parameters that
contributes to the line emission. {We found a differences in the
H$\alpha $ and H$\beta $ line profiles before and after the
beginning of the activity phase in 2002, consequently we divided our
spectra into two periods (before March 05, 2002 -- Period I and
after that -- Period II, see Paper I).}

In this paper  we study in more details the H$\alpha$ and H$\beta$
line profiles and ratios, taking into account the changes during the
monitoring period. The aim of this paper is to investigate the
changes in the the BLR structure of 3C 390.3 that cause the line
profile variations. To perform this investigation we  analyzed the
peak separation variations, variations in line segments and
variation in Balmer decrement. Using a relatively simple disk model,
we try to explain qualitatively the changes in disk structure that
can cause the line parameter variations.

The paper is organized as follow: in \S 2 we {describe} of our
observations; in \S 3 we {present} the analysis of the H$\alpha$
and H$\beta$ line profiles variability, the peak-velocity
variability and Balmer decrement; in \S 4 we study the line-segment
variations; in \S 5 the Balmer decrement variation is analyzed; in
\S 6 we  {discuss obtained results} , and finally in \S 7 we
outline our conclusions.

\section{Observations}

In 1995-2007 spectra of 3C 390.3 were taken with the 6 m and 1 m
telescopes of the SAO RAS and with INAOE's 2.1 m telescope of the
''Guillermo Haro Observatory'' (GHO) at Cananea, Sonora, M\'exico
and with a long slit spectrographs, equipped with CCD detector
arrays. The typical wavelength interval covered was from 4000\,\AA\,
to 7500\,\AA, the spectral resolution varied between (4.5-15) \ \AA.
Spectra  were scaled using the [\ion{O}{iii}]
$\lambda\lambda$4959+5007 (for blue spectra) and the [OI]6300A (for
red spectra) integrated line flux under the assumption that fluxes
of this lines did not change during the time interval covered by our
observations (1995--2007). The narrow components of H$\alpha$ and
H$\beta$ were removed by applying the modified method of
\citet{vw92} \citep[see also][]{sh04} using the spectral template
for narrow components as reference spectrum. More details about
observations can be found in Paper I.

\section{Analysis of the broad line profiles}

\begin{figure}
\centering
\includegraphics[width=4cm]{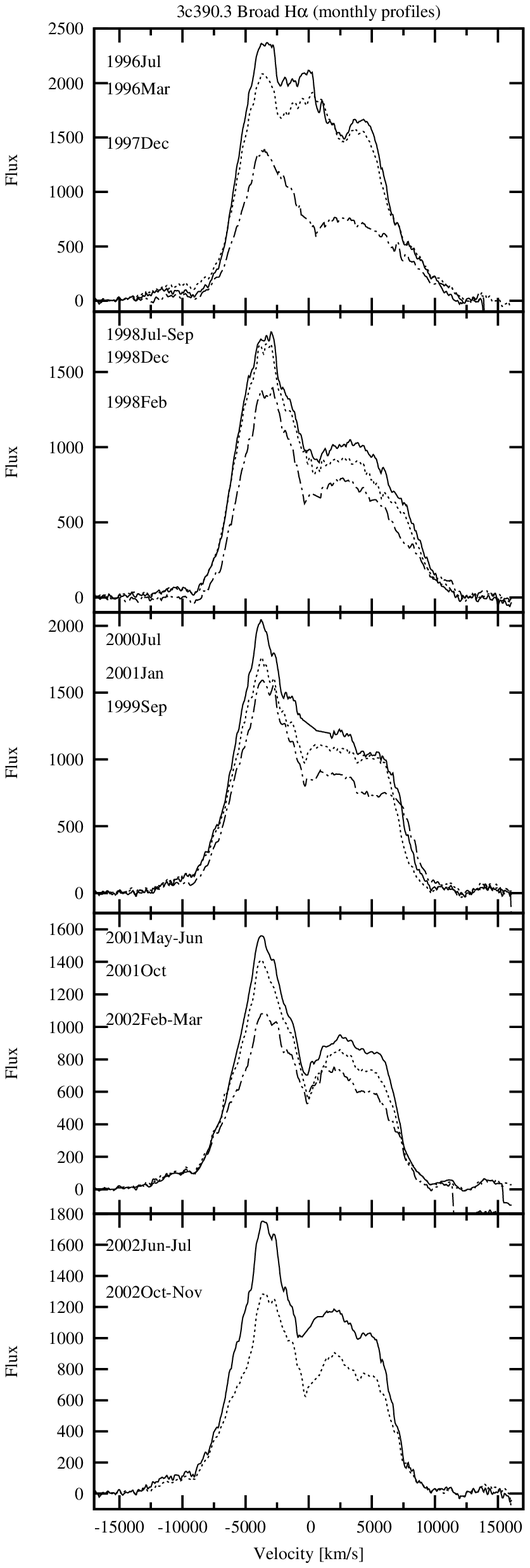}
\includegraphics[width=4cm]{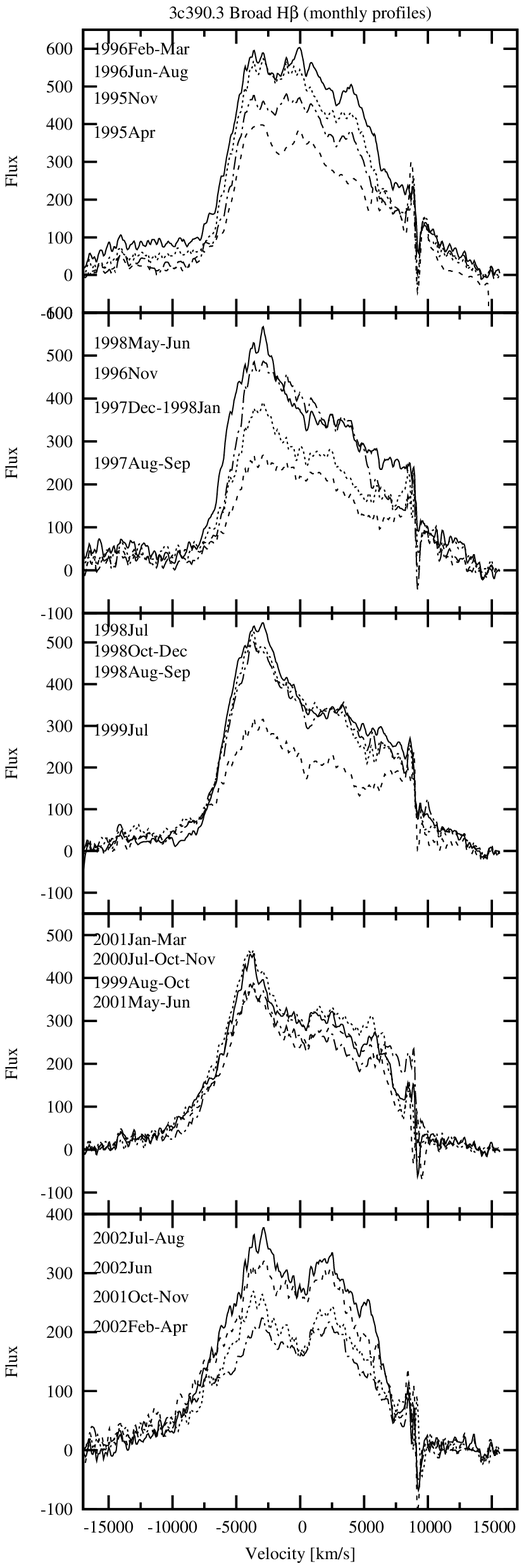}
\includegraphics[width=4cm]{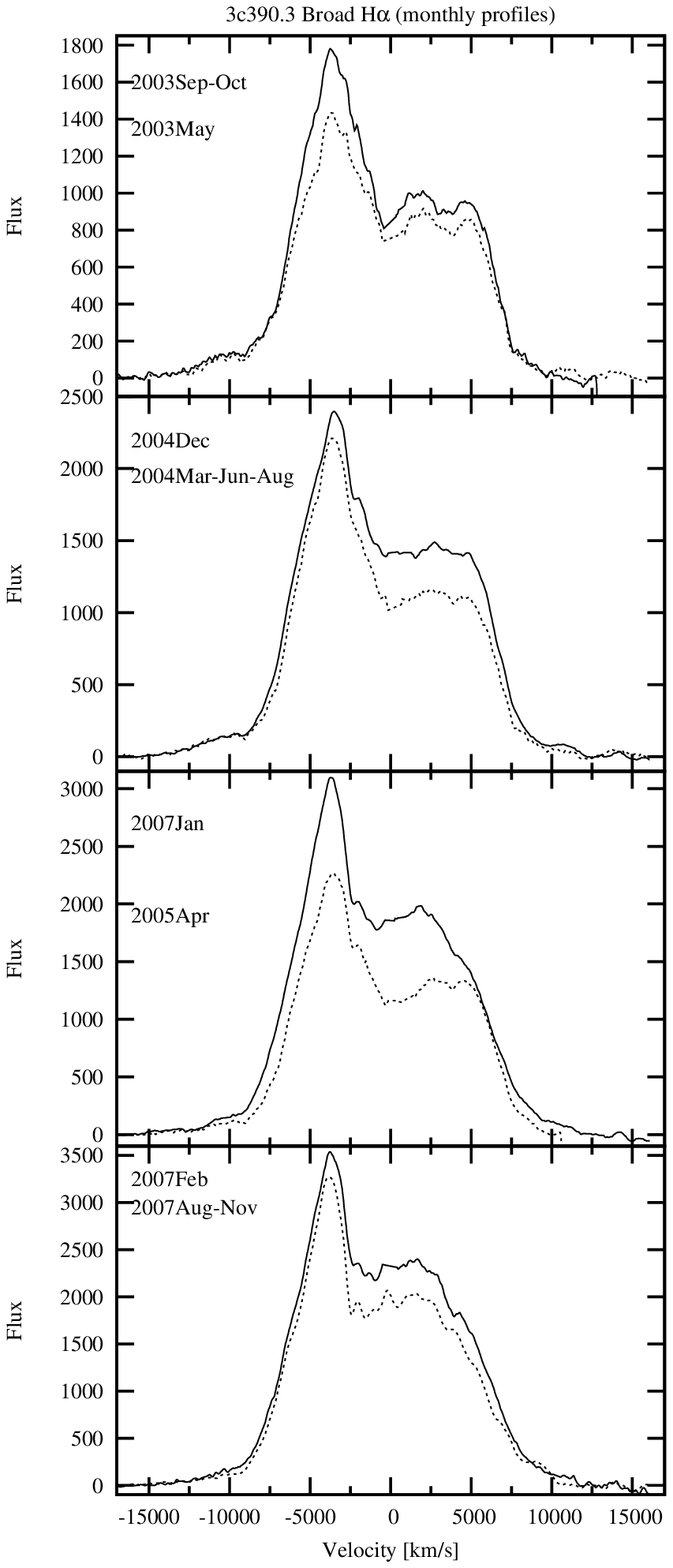}
\includegraphics[width=4cm]{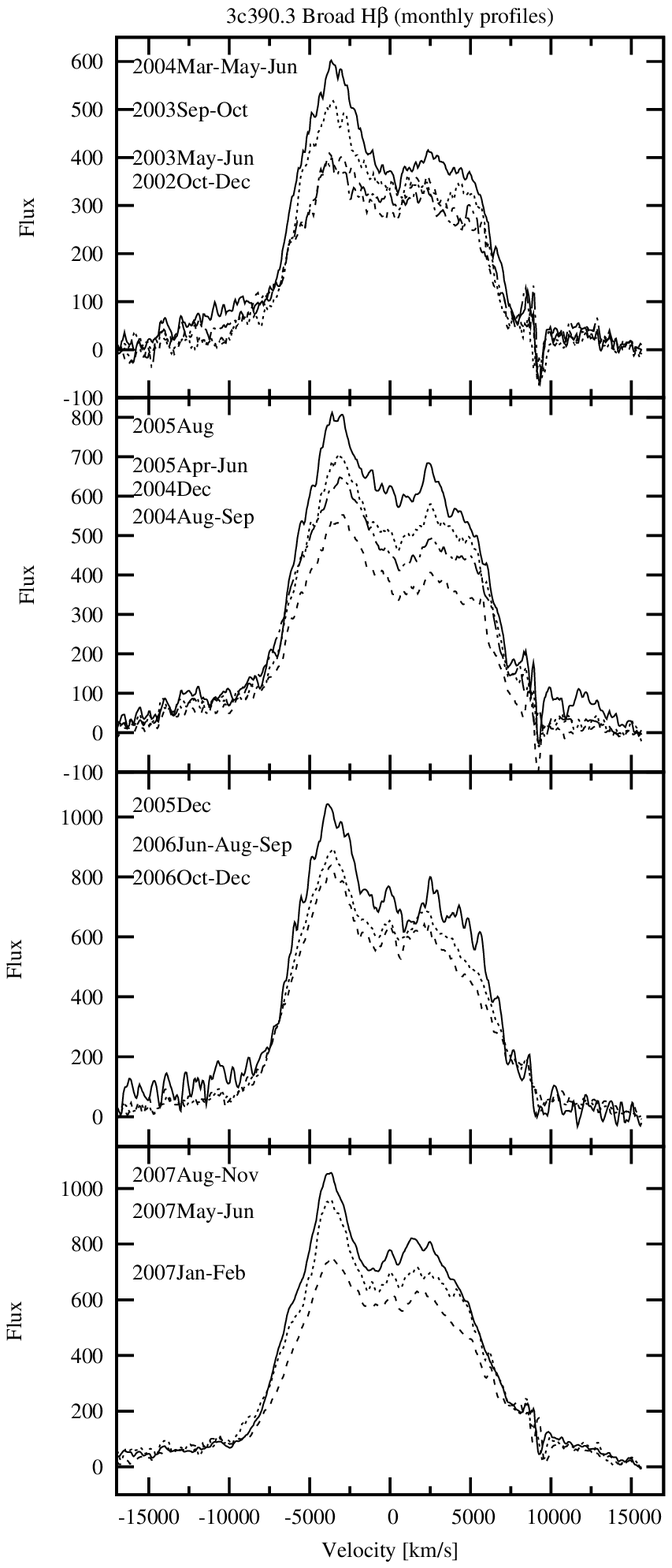}
\caption{The month-averaged profiles of the H$\alpha$ and H$\beta$
        broad emission lines in the period 1995--2007. The abscissa (OX) shows
        radial velocities relative to the narrow component of H$\alpha$
        or H$\beta$ line. The ordinate (OY) shows the line flux in units of
         $10^{-16} {\rm erg\ cm^{-2}s^{-1}}$\AA$^{-1}$.}\label{f01}
\end{figure}

In Paper I we explained the continuum and narrow line subtraction in
order to obtain only the broad H$\alpha$ and H$\beta$ line profiles,
and here that procedure will not be repeated. Using only broad line
component we analyze the line parameters: month- and year- averaged
line profiles (see Figs. \ref{f01}--\ref{f03}, Figs.
\ref{f02}--\ref{f03}- available electronically only), the position
(and separation) of prominent peaks, the fluxes of the line segments
and the H$\alpha$/H$\beta$ ratios (or Balmer decrement -- BD). In
this section we first give the analysis of the observed broad line
parameters, and after that we analyze the corresponding modeled line
parameter, obtained using an accretion disk model, in order to learn
about changes in the disk structure which may cause the line
parameter variability.

\onlfig{2}{\begin{figure*}
\centering
\includegraphics[width=13cm]{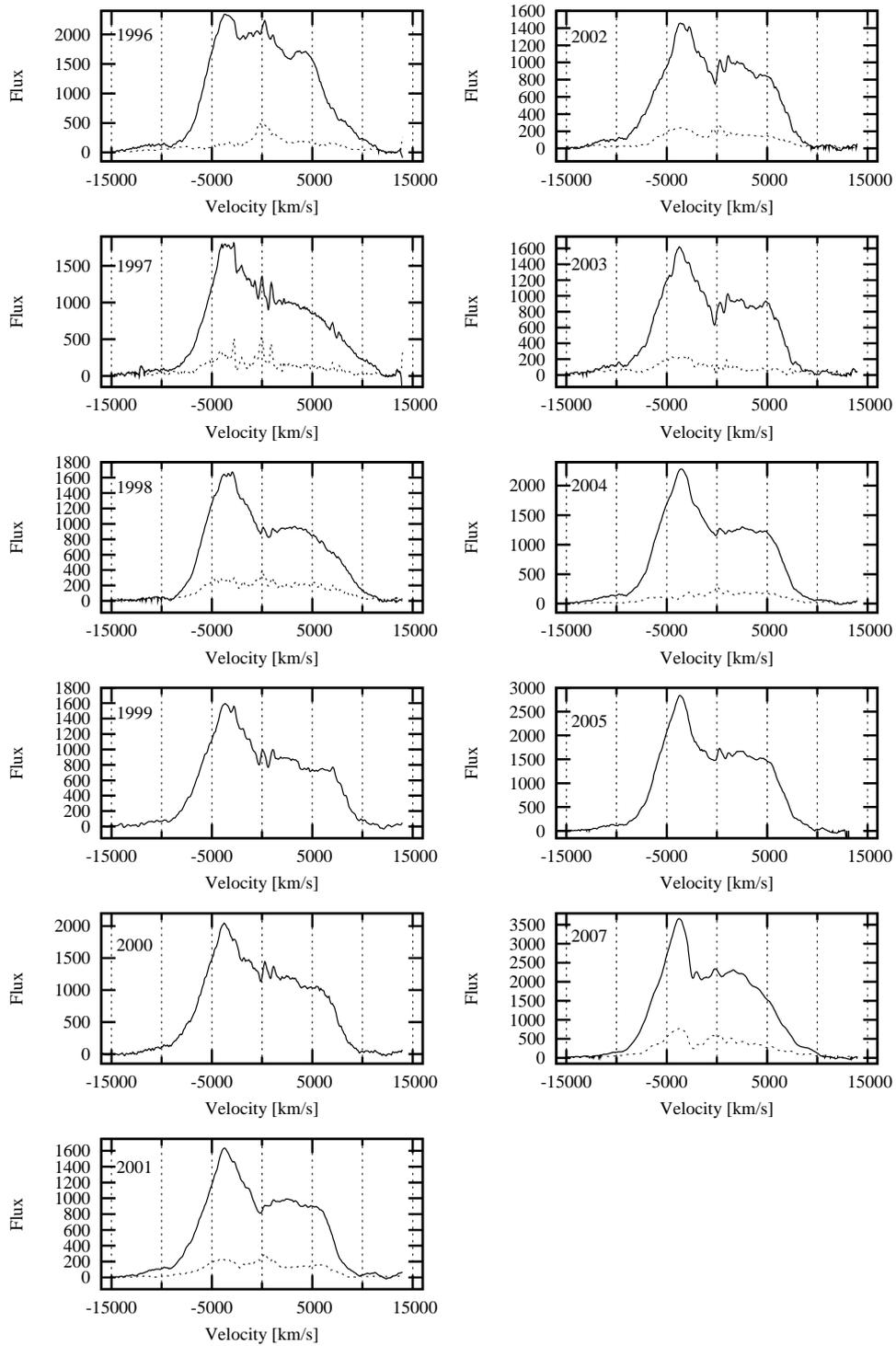}
\caption{The year-averaged  profiles (solid line) and their rms (dashed line)
       of the H$\alpha$ broad emission line in 1995--2007. The abscissa (OX)
       shows the radial velocities relative to the narrow component of the H$\alpha$ line.
       The ordinate (OY) shows the flux in units of $10^{-16} {\rm erg\ cm^{-2}s^{-1}}$\AA$^{-1}$.  }\label{f02}
\end{figure*}}

\onlfig{3}{\begin{figure*}
\centering
\includegraphics[width=13cm]{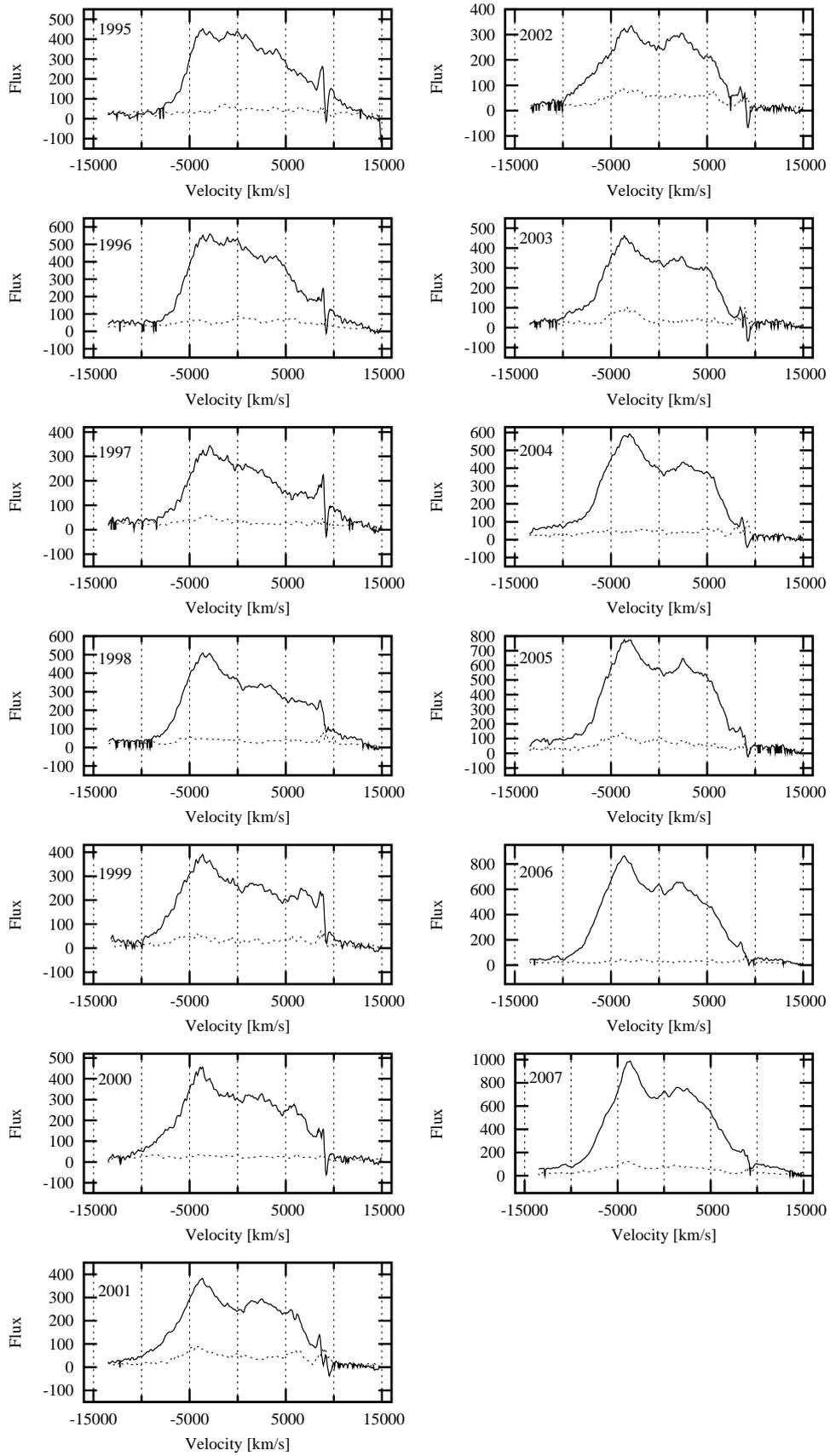}
\caption{The same as in Fig. \ref{f02}, but for H$\beta$ broad emission line.}\label{f03}
\end{figure*}}

\begin{figure}
\centering
\includegraphics[width=8cm]{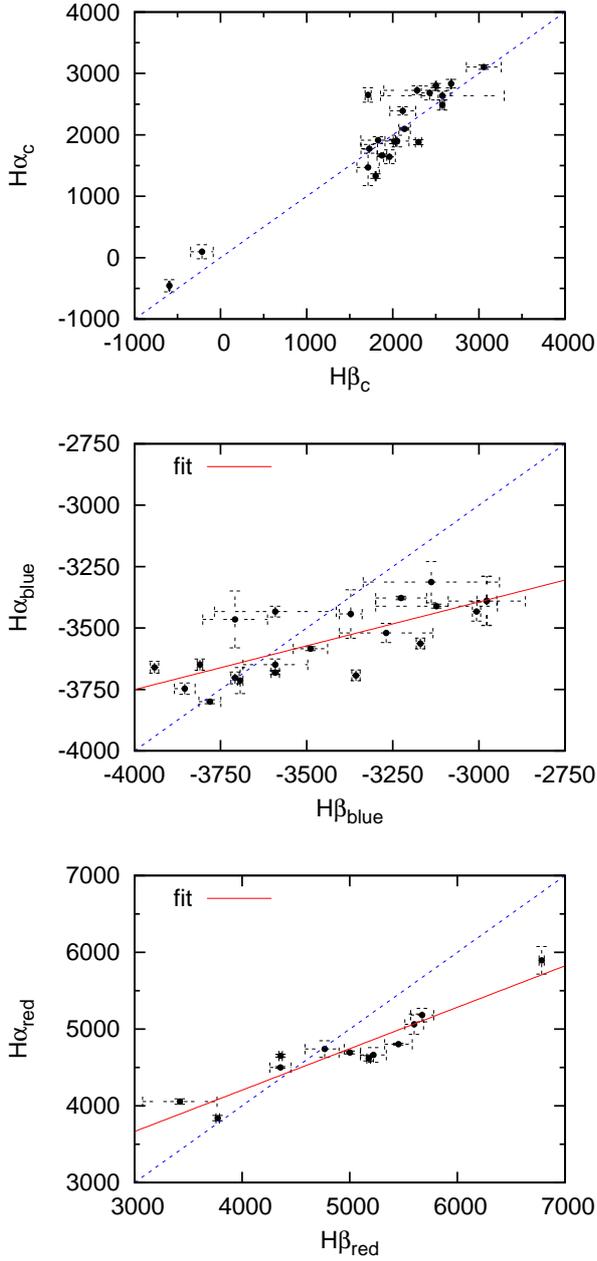}
\caption{The H$\alpha$ vs. H$\beta$ peak velocity for the central
(top), blue (middle) and red (down) peak. The dashed line represents
expected function for equal peak position in both lines, while solid
line represents the best fit of the measured data.}\label{f04}
\end{figure}

\begin{figure}
\centering
\includegraphics[width=8.5cm]{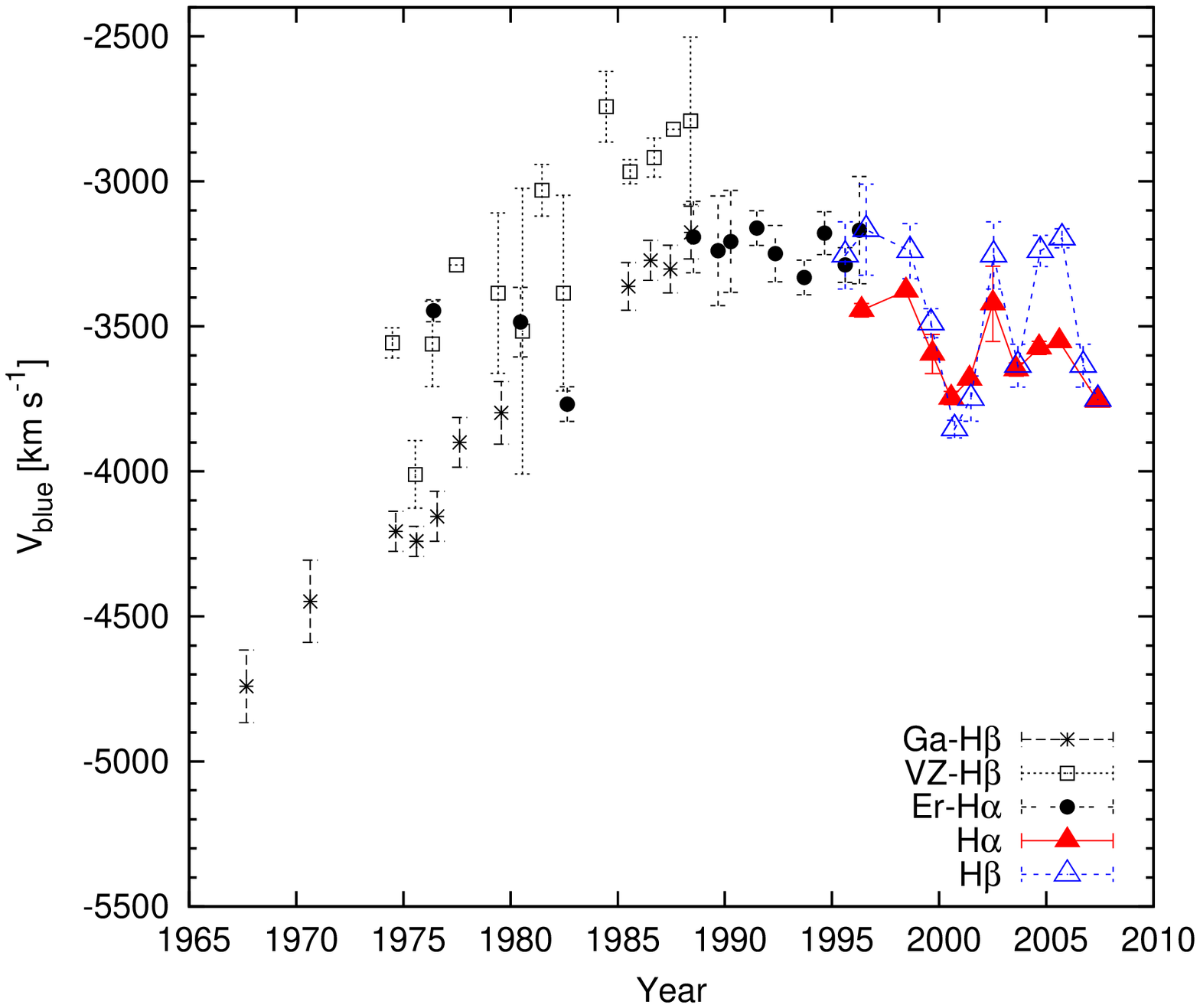}
\caption{The annual averaged radial velocities of the blue peak. The
different symbols represent: stars - the H$\beta$ data from
\citet{gas96}, open squares - the H$\beta$ measurements given in
\citet{vz91}, year-averaged to match other data, full circles - the
H$\alpha$ data from \citet{er97}, open and full triangles - our data
for H$\beta$ and H$\alpha$, respectively. }\label{f05}
\end{figure}

As reported in Paper I there is a differences in the H$\alpha $ and
H$\beta $ line profiles before and after the beginning of the
activity phase in 2002. Therefore, we consider separately line
profiles obtained in the period before March 05, 2002 (Period I, JD
2 452 339.01 according the minimum in H$\beta$) and after that date
(Period II, see Paper I). Also, we found that observations in 2001
and 2002 have sometimes closer characteristics as the data in Period
II, therefore in some plots we marked separately these observations
as 2001--2002.

\subsection{The averaged profiles of the H$\alpha$ and H$\beta$ broad emission lines}

It was also reported in Paper I that the line profiles of H$\alpha$
and H$\beta$ were changing during the monitoring period. In Fig.
\ref{f01}, we show the month- and in Figs. \ref{f02} and \ref{f03}
(available electronically only) year-averaged profiles of the
H$\alpha$  and H$\beta$. As it can be seen in Fig. \ref{f01} the
line profiles of H$\alpha$ (left) and H$\beta$ (right) vary, showing
clearly two, and sometimes three peaks. Similar variations can be
seen from the year-averaged profiles (see Fig. \ref{f02} for
H$\alpha$ and Fig. \ref{f03} for H$\beta$). Note here that in both
periods the blue peak is higher as it is expected in the case of a
relativistic accretion disk. {The most interesting is the
central peak at the zero velocity (clearly  seen in 1995 and 1996)
that may indicate perturbation in the disk or an additional emitting
region.} We found that this central peak (sometimes weaker) exists
almost always during both periods, therefore we measured the
position of the blue, central and red peak. Our measurements are
given in Table \ref{t01}. Sometimes the red peak could not be
properly measured (since it was too weak or even absent), thus in
that case the measurements are not given in Table \ref{t01}. The
measurements show that the position of the peaks are changing, and
as can be seen in Fig. \ref{f04} changes are higher in the H$\beta$
than in H$\alpha$ line. It is interesting that the position of the
blue peak varies around several hundreds km s$^{-1}$ (e.g. in
H$\alpha$ it is -3550$\pm$250 km s$^{-1}$, in H$\beta$ -3500$\pm$400
km s$^{-1}$), while the position of the red peak has higher
variation (for H$\alpha\ \sim$ +5000$\pm$ 1000 km s$^{-1}$, and for
H$\beta\ \sim$ 5250$\pm$1700 km s$^{-1}$). This may indicate that
the emission is coming relatively close to the central black hole.

There is a difference between the H$\alpha$ and H$\beta$ red and
blue peak velocities (see Fig. \ref{f04}), that can be expected due
to the stratification of the disk, i.e. the different dimensions of
the disk region that emits H$\alpha$ or H$\beta$. It is interesting
that the central peak seems to stay at similar position in both
lines, i.e. as it can be seen in Fig. \ref{f04} (panel top) the
points follow well the linear bijection function (represented as
dashed line). This could indicate that the central peak is connected
with some kind of perturbation in a disk or an additional region
that in the same way affects the H$\beta$ and H$\alpha$ line
profiles. Note here that  the so-called central peak was slightly
shifted to the blue side (or close to the zero shift) in 1996--1997
(see Fig. \ref{f01}), but after that this peak has been redshifted
(between 1500 and 2500 km s$^{-1}$).

In order to compare the variability of the blue peak position during
a longer period, in Fig. \ref{f05} we presented our measurements for
the blue peak position and measurements obtained by  \citet{vz91},
\citet{gas96}  and \citet{er97}. {Note here that the estimates
of the blue peak position in the H$\alpha$ line is more uncertain
than in H$\beta$, due to the substraction of the atmosphere B-band
(Paper I), while H$\beta$ is more uncertain in the red part due to
subtraction of the [OIII] lines (see \S 4.1).} As it can be seen in
Fig. \ref{f05} the measurement follow well the previous
observations, and there is an increase of the blue peak position
velocity from 1970 to 1990 (1995), while after 1995, the the blue
peak position velocity is slightly decreasing.

\begin{table*}
\caption[]{Measurements of the peak position velocities (blue, central, red) from  month-averaged
 H$\alpha$ and H$\beta$ profiles. }\label{t01}
\centering
\resizebox{18.5cm}{!}{
\begin{tabular}{llcccclcccc}
\hline \hline
     & \multicolumn{5}{c}{H$\alpha$} &       \multicolumn{5}{c}{H$\beta$}    \\
Year &Month  &  blue  & central & red  & $V_{\rm red}-V_{\rm blue}$ & Month  &blue  & central & red  & $V_{\rm red}-V_{\rm blue}$    \\
     &  &  km/s  & km/s  &  km/s  & km/s  &        &km/s  &  km/s  &  km/s  &  km/s   \\
\hline
\hline
1996& Mar     & -3520 $\pm$ 39  & 96   $\pm$ 115  &  3840 $\pm$ 35   &   7360 $\pm$ 74     &   Feb-Mar   &  -3269 $\pm$ 136  &  -216 $\pm$ 131  &  3773$\pm$  16   &  7042 $\pm$  152  \\
1996& Jul     & -3411 $\pm$ 8   & -457 $\pm$ 100  &  4056 $\pm$ 35   &   7467 $\pm$ 43     &   Jun-Aug   &  -3123 $\pm$ 177  &  -595 $\pm$ 7    &  3422$\pm$  346  &  6546 $\pm$  523  \\
1998& Feb     & -3390 $\pm$ 100 & 2639 $\pm$ 71   &  -               &   -                 &   May-Jun   &  -2977 $\pm$ 112  &  2575 $\pm$ 718  &  -               &  -                \\
1998& Jul-Sep & -3378 $\pm$ 7   & 3104 $\pm$ 35   &  -               &   -                 &   Jul       &  -3226 $\pm$ 74   &  3057 $\pm$ 202  &  -               &  -                \\
1998& Dec     & -3433 $\pm$ 22  & 2726 $\pm$ 71   &  -               &   -                 &   Oct-Dec   &  -3591 $\pm$ 177  &  2283 $\pm$ 387  &  6359 $\pm$ 162  &  9950  $\pm$ 340  \\
1999& Sep     & -3465 $\pm$ 116 & 1914 $\pm$ 56   &  5896 $\pm$ 180  &   9361 $\pm$ 296    &   Aug-Oct   &  -3708 $\pm$ 94   &  1830 $\pm$ 201  &  6783 $\pm$ 24   &  10491 $\pm$ 118  \\
2000& Jul     & -3747 $\pm$ 23  & 2649 $\pm$ 117  &  5063 $\pm$ 133  &   8810 $\pm$ 156    &   Jul-Nov   &  -3854 $\pm$ 30   &  1713 $\pm$ 35   &  5599 $\pm$ 87   &  9453  $\pm$ 116  \\
2001& Jan     & -3660 $\pm$ 24  & 1665 $\pm$ 21   &  4803 $\pm$ 11   &   8463 $\pm$ 35     &   Jan-Mar   &  -3942 $\pm$ 12   &  1874 $\pm$ 56   &  5453 $\pm$ 128  &  9395  $\pm$ 140  \\
2001& May-Jun & -3714 $\pm$ 53  & 2682 $\pm$ 112  &  5182 $\pm$ 88   &   8896 $\pm$ 141    &   May-Jun   &  -3693 $\pm$ 8    &  2429 $\pm$ 98   &  5673 $\pm$ 107  &  9366  $\pm$ 115  \\
2001& Oct     & -3693 $\pm$ 22  & 2390 $\pm$ 66   &  4803 $\pm$ 72   &   8496 $\pm$ 94     &   Oct-Nov   &  -3357 $\pm$ 12   &  2116 $\pm$ 151  &  -               &  -                \\
2002& Feb-Mar & -3389 $\pm$ 100 & 1881 $\pm$ 41   &  4750 $\pm$ 26   &   8139 $\pm$ 126    &   Feb-Apr   &  -2977 $\pm$ 29   &  2298 $\pm$ 36   &  -               &  -                \\
2002& Jun-Jul & -3313 $\pm$ 84  & 2098 $\pm$ 21   &  4662 $\pm$ 96   &   7975 $\pm$ 181    &   Jun       &  -3138 $\pm$ 198  &  2137 $\pm$ 67   &  -               &  -                \\
2002& Oct-Nov & -3443 $\pm$ 99  & 1892 $\pm$ 36   &  4619 $\pm$ 35   &   8062 $\pm$ 134    &   Oct-Dec   &  -3372 $\pm$ 33   &  2020 $\pm$ 15   &  5176 $\pm$ 17   &  8548  $\pm$ 50   \\
2003& May     & -3649 $\pm$ 23  & 1899 $\pm$ 95   &  4695 $\pm$ 19   &   8344 $\pm$ 42     &   May-Jun   &  -3810 $\pm$ 8    &  2049 $\pm$ 139  &  5001 $\pm$ 100  &  8811  $\pm$ 108  \\
2003& Sep-Oct & -3649 $\pm$ 23  & 1773 $\pm$ 71   &  4652 $\pm$ 19   &   8301 $\pm$ 42     &   Sep-Oct   &  -3591 $\pm$ 94   &  1727 $\pm$ 97   &  4358 $\pm$ 16   &  7948  $\pm$ 111  \\
2004& Mar-Aug & -3584 $\pm$ 8   & 2487 $\pm$ 82   &  4500 $\pm$ 11   &   8085 $\pm$ 19     &   Mar-Jun   &  -3489 $\pm$ 50   &  2575 $\pm$ 25   &  4357 $\pm$ 99   &  7846  $\pm$ 149  \\
2004& Dec     & -3433 $\pm$ 39  & 2833 $\pm$ 70   &  4740 $\pm$ 109  &   8173 $\pm$ 147    &   Dec       &  -3007 $\pm$ 12   &  2678 $\pm$ 5    &  4767 $\pm$ 182  &  7773  $\pm$ 194  \\
2005& Apr     & -3563 $\pm$ 22  & 2801 $\pm$ 35   &  4533 $\pm$ 35   &   8095 $\pm$ 57     &   Apr-Jun   &  -3170 $\pm$ 12   &  2502 $\pm$ 6    &  -               &  -          \\
2007& Jan     & -3682 $\pm$ 7   & 1643 $\pm$ 113  &  -               &   -                 &   Jan-Feb   &  -3591 $\pm$ 12   &  1961 $\pm$ 68   &  -               &  -          \\
2007& Feb     & -3703 $\pm$ 24  & 1329 $\pm$ 37   &  -               &   -                 &   May-Jun   &  -3708 $\pm$ 12   &  1800 $\pm$ 6    &  -               &  -          \\
2007& Aug-Nov & -3800 $\pm$ 8   & 1470 $\pm$ 297  &  -               &   -                 &   Aug-Nov   &  -3781 $\pm$ 32   &  1713 $\pm$ 130  &  -               &  -          \\
\hline
\end{tabular}}
\end{table*}

\begin{figure}
\centering
\includegraphics[width=9cm]{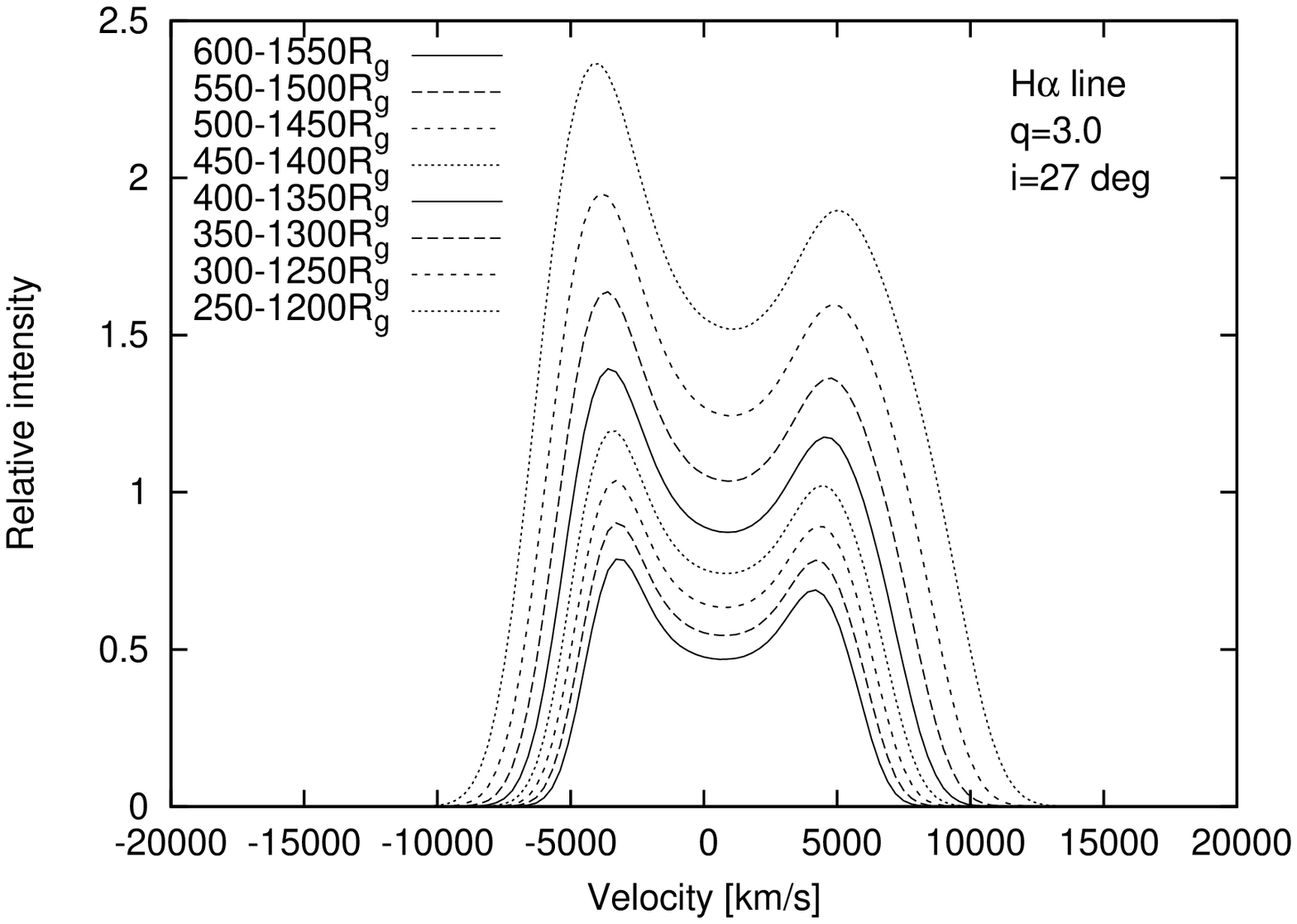}
\includegraphics[width=9cm]{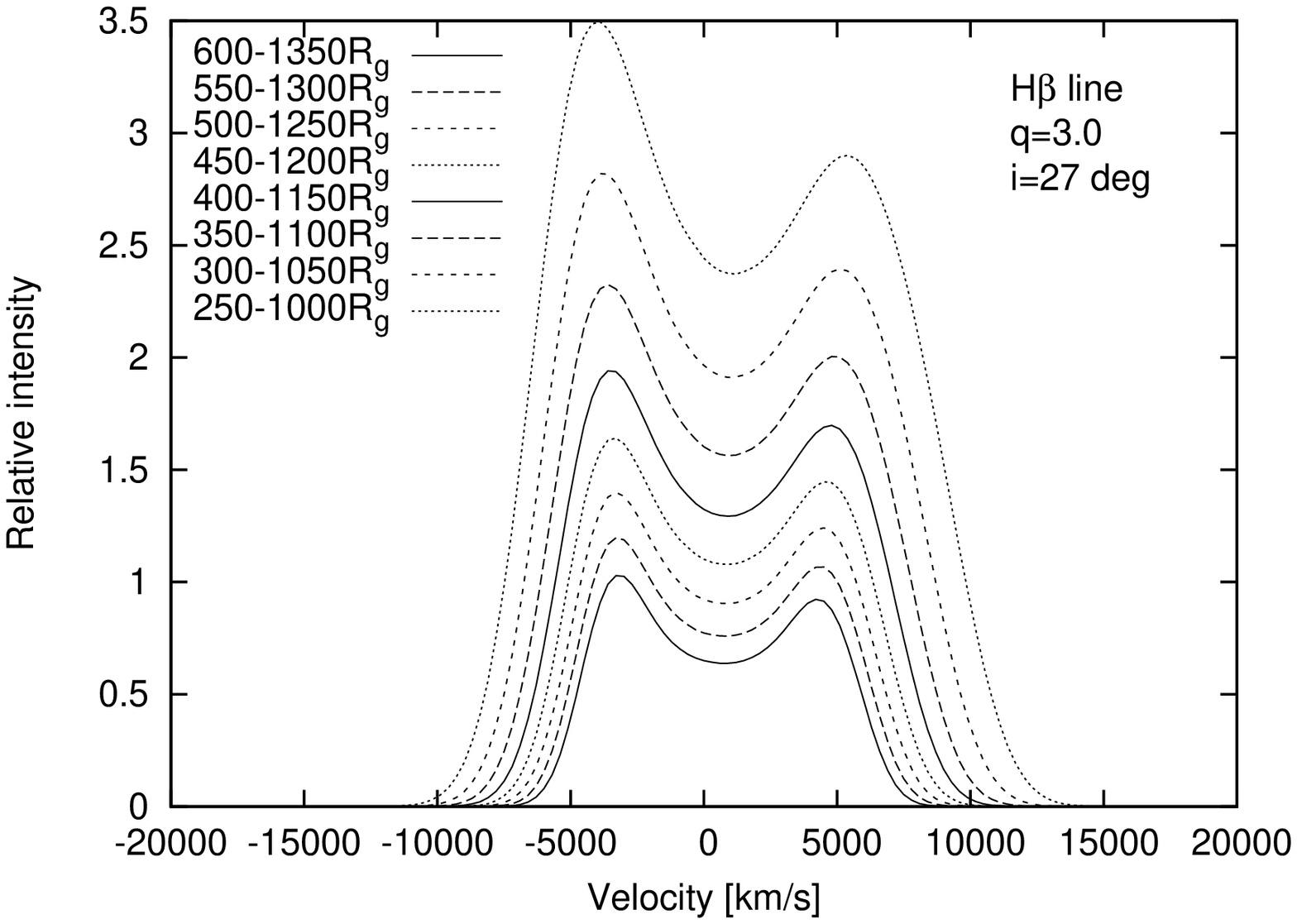}
\caption{The modeled line profiles for the H$\alpha$ (top) and H$\beta$ (bottom)
lines for different disk parameters (noted in the plots).}\label{f06}
\end{figure}

\begin{figure}
\centering
\includegraphics[width=9cm]{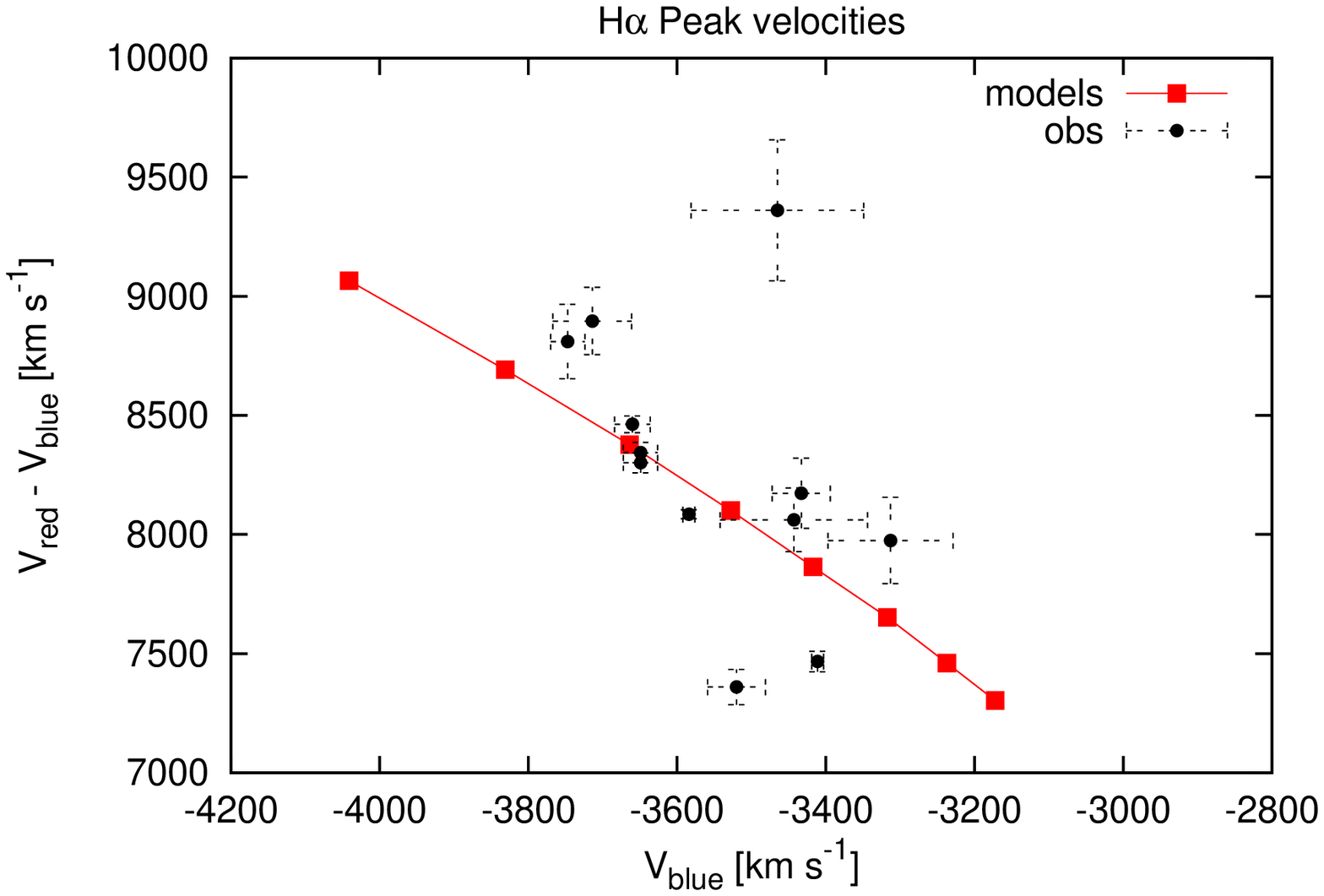}
\includegraphics[width=9cm]{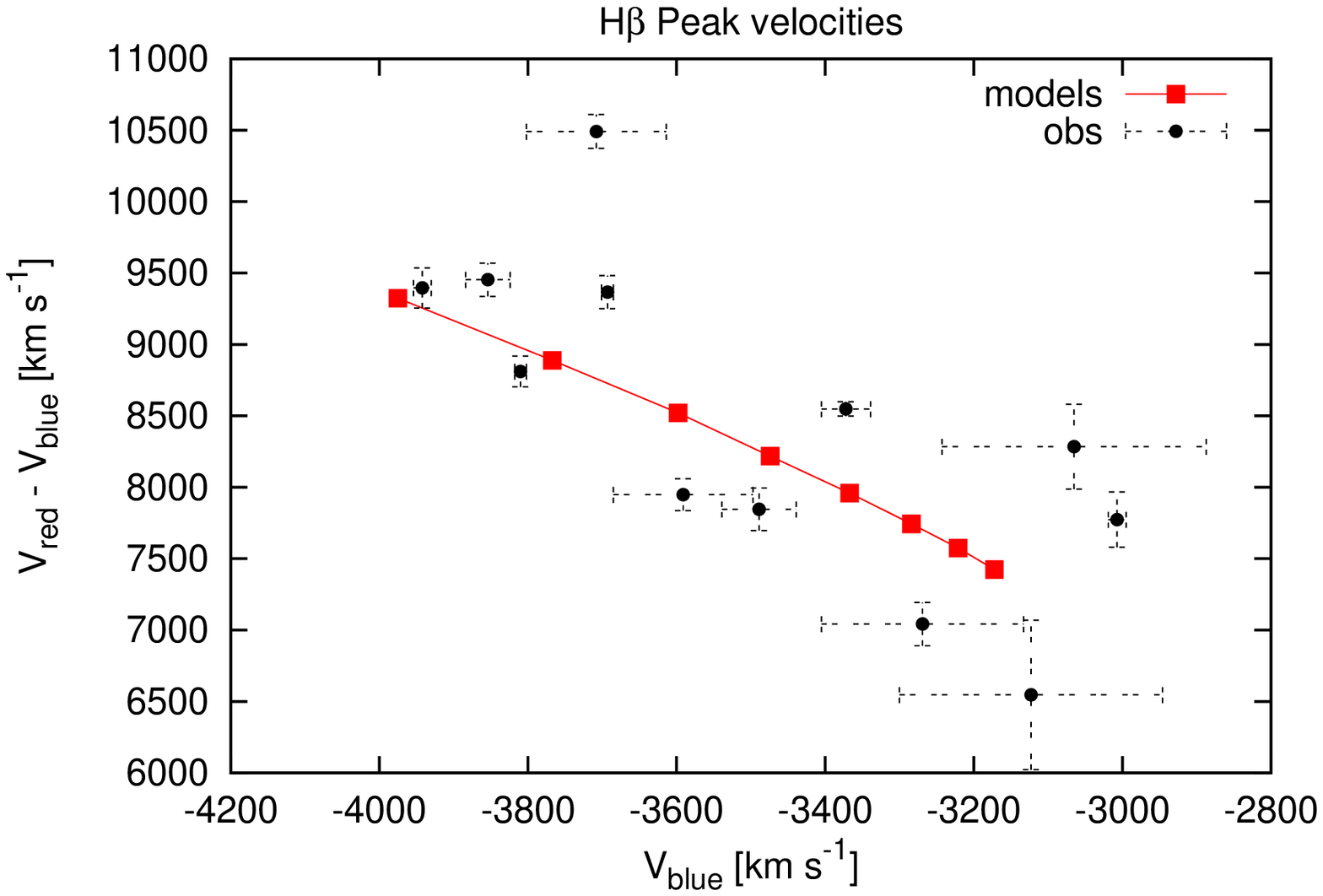}
\caption{The difference between the blue and read peak velocities
vs. the blue peak velocity for H$\alpha$ (top) and H$\beta$ line
(bottom).  The model parameters are marked with full squares, while
observations are denoted with full circles. }\label{f07}
\end{figure}

\subsection{Simulation of the line parameters  variations}

One can fit the broad double peaked line profiles to extract some
disk parameters \citep[see e.g.][etc.]{eh94, eh04,fl08,le10,jov10},
but here, since we have a large set of observational data, we will
use a simple model to simulate variations in different disk
parameters and give some qualitative conclusions.

To qualitatively explain the changes in the disk structure that can
cause the line parameter variations, we simulate the disk emission,
using a disc model given by \citet{chp89} and \citet{ch89}. {The
model assumes a relativistic, geometrically thin and circular disk
\citep[see in more details][]{chp89,ch89}. Note here  that in this
model relativistic effects are approximatively included and the
limit of the inner radius is $R_{inn}>100$ R$_g$ ($R_g=GM/c^2$ --
gravitational radius), but the estimated inner radius in the case of
3C390.3 is significantly larger \citep[see][]{fl08}, consequently,
the model given by \citet{chp89} and \citet{ch89} can  be  properly
used, i.e. it is not necessary to include a full relativistic
calculation \citep[as e. g. in][]{jov10}.}

To obtain the double-peaked line profiles we generated the set of
different disk-like line profiles. For the starting parameters of
the H$\alpha$ line we used the results of \citet{fl08}. They
obtained the following disk parameters from fitting the observed
H$\alpha$ line of 3C 390.3: the inner radius $R_{\rm inn}=450\
R_{\rm g}$, the outer radius $R_{\rm out}=1400\ R_{\rm g}$, the disk
inclination $i=27 \deg$, the random velocity in the disk $\sigma =
1300\ {\rm km\ s^{-1}}$, with the emissivity $r^{-q}$, where $q=3$
is assumed. To model the parameters for the H$\beta$ line, we took
into account the results from Paper I that the dimension of the
H$\alpha$ emitting region is $\sim$ 120 light days, and of H$\beta$
is $\sim$ 95 light days, consequently we proportionally (95/120)
rescaled the H$\beta$ disk parameters for outer radius (i.e. $R_{\rm
out}=1200\ R_{\rm g}$) keeping other parameters as in the case of
the H$\alpha$ disk. {Note here that we assumed that the inner
radius for both, H$\alpha$ and H$\beta$ is the same. It is an
approximation, but one  may expect that the plasma conditions
(temperature and density) are probably changing fast in the inner
part of disk, therefore at some distance from the black hole, the
plasma has such conditions that can emit Balmer lines (i.e.
recombination stays effective). On the other side, at larger
distances from the black hole, the conditions in plasma are slowly
changing (as a function of radius) and from  some distance, due to
the probability of transitions, there may be significant emission in
the H$\alpha$ line and very weak (or absent) H$\beta$ emission.}

Also, we tested what will be the consequence when changing disk
emissivity in the models and found that it has smaller influence on
the double peaked line profiles \citep[see also][]{b09}.

To probe different type of variability, first we consider that in
different time, different part of disk can contribute to the
emission of the broad H$\alpha$ and H$\beta$ lines, i.e. {we
keep constant the relative size of the disk regions responsible for
the emission of H$\alpha$ and H$\beta$ lines but we change the
locations of these regions with respect to the central black hole.}
Then the emitting region can be closer, i.e. inner and outer radius
closer to the black hole. For obtaining such models, we have been
shifting the emitting region closer to  the black hole for a step of
$\Delta R=50\ R_{\rm g}$. This was performed by varying the inner
and outer radius, keeping all other parameters as constant (except
of the random velocity that was changed accordingly, but this has
also small influence on the line profiles). In Fig. \ref{f06} we
presented simulated line profiles. The range of the inner radius was
$R_{\rm inn} = 250 - 600\ R_{\rm g}$, where the broadest line in
Fig. \ref{f06} corresponds to the $R_{\rm inn}=$250 R$_{\rm g}$.
Consequently, the outer radius was in the range of $R_{\rm out} =
1200-1550\ R_{\rm g}$ for H$\alpha$ and $R_{\rm out} = 1000-1350\
R_{\rm g}$ for H$\beta$ (see Fig. \ref{f06}).

From the models we have measured the velocities of the blue and red
peak ($V_{\rm blue}$ and $V_{\rm red}$). Note here that we could not
obtain the central peak since, as it was mentioned above, it
probably has a different origin, either it originates outside of the
disk, or it is due to some disk perturbations.

To compare the modeled values and observed ones, we plot in Fig.
\ref{f07} the difference between the blue and read peak velocities
$V_{\rm red}- V_{\rm blue}$ as a function of the blue peak velocity
for H$\alpha$ (top) and H$\beta$ line (bottom).  The modeled values
are presented as full squares. {Fig. \ref{f07} shows that the
modeled values are in general agreement with the observed ones,
which however show a considerable scattering around the modeled
trend.} It seems that the emission of the different parts of the
disk in different epochs can explain the peak position variations.
In such scenario, the line profile variability {can be
explained} as following: when we have the inner radius closer to the
central black hole, the separation between peaks stays larger and
the blue peak is more blue-shifted, but when the inner radius is
farther from the central black hole, the distance between peaks is
smaller and the blue peak is shifted to the center of the line.
Note here that \citet{sh01} found the anticorrelation between the
continuum flux and $V_{\rm red}-V_{\rm blue}$ in the period
1995-1999.

\section{The  H$\alpha$ and H$\beta$ line-segments variation}

To see if there are any changes in the structure of the disk or disk
like-region, we investigated the light curves for the different
segments of the H$\alpha$ and H$\beta$ broad emission lines. First,
we divided the line profiles along the velocity scale into 9
segments. The size of intervals are {defined} as 2000 ${\rm km
\, s^{-1}}$  for segments from 0 to $\pm$3 and 3000 ${\rm km \,
s^{-1}}$ for segment $\pm$4 (the segments and corresponding
intervals are given in Table \ref{t02}).

The observational uncertainties were determined for each segment of
the H$\alpha$ and H$\beta$ light curves. In evaluating the
uncertainties, we account for errors due to the effect of the
subtraction of the template spectrum (or the narrow components and
continuum). We compare fluxes of pairs of spectra obtained in the
time interval from 0 to 2 days. In Table \ref{t03} (available
electronically only), we present the year-averaged uncertainties (in
percent) for each segment of H$\alpha$ and H$\beta$ and the
corresponding mean year-flux. {We adopted the years for each
line where frequency and number of observations were enough to
estimate error-bars.} The mean values of uncertainties for all
segments are also given.  As one can see from Table \ref{t03}, for
the far wings (segments $\pm$4) the errorbars are greater
($\sim$30\%-50\%) in H$\beta$ than in H$\alpha$ ($\sim$20\%). But
when comparing the errorbars in the far red and blue wings of each
lines, we find that the errorbars are similar. Larger errorbars can
also be seen in the central part of the H$\alpha$ due to the narrow
line subtraction.



Our measurements of line-segment fluxes for H$\alpha$ and H$\beta$
are given in Table \ref{t04} and \ref{t05} (available electronically
only), respectively. The light curves  for each segment of the
H$\alpha$ and H$\beta$ lines are presented in Fig. \ref{f09}
(available electronically only), where  we show blue part (crosses),
red part (pluses), core (full circles), and continuum (solid line)
in arbitrary units for comparison. As can be seen in Fig. \ref{f09},
the fluxes from 0 to $\pm$4 segments of both lines have similar
behavior during the monitoring period. As a rule, the blue segments
are brighter or equal to the red ones. Only in $\pm$4 in the period
of 1997--1999  the red segments are brighter than the blue ones. It
is interesting that the red and blue wings of H$\beta$ in 1995--1997
where with small difference in brightness, while in 2006 the blue
peak was around 30\% brighter than the red one.

The variation in line segments of the H$\alpha$ and H$\beta$ are
similar, and there are more observations of H$\beta$ than of
H$\alpha$, further in the analysis we will consider only the segment
variation of the H$\beta$ line, except in the case of segments
$\pm$4, where we consider both H$\alpha$ and H$\beta$.

\onlfig{8}{\begin{figure*} \centering
\includegraphics[width=12cm]{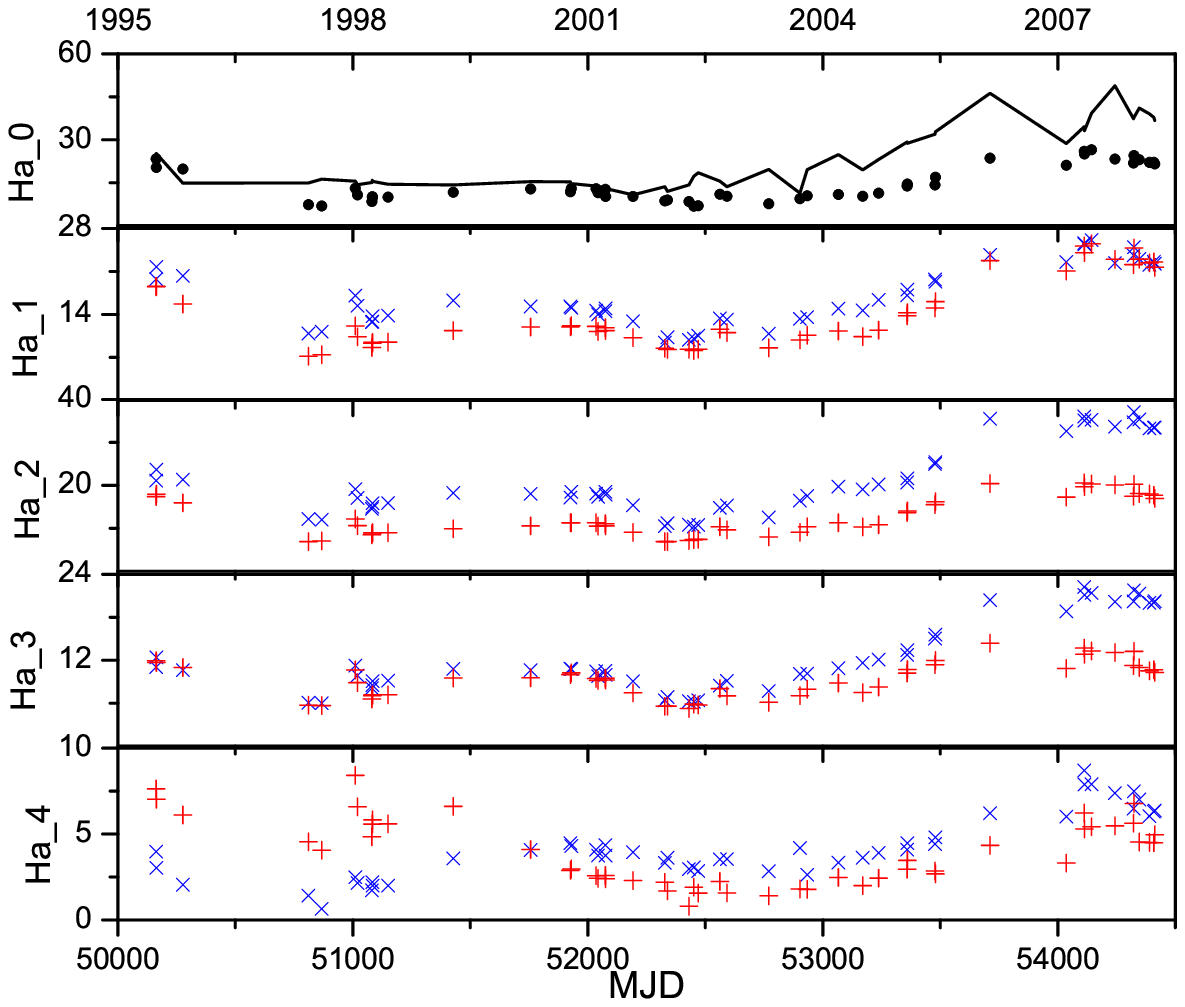}
\includegraphics[width=12cm]{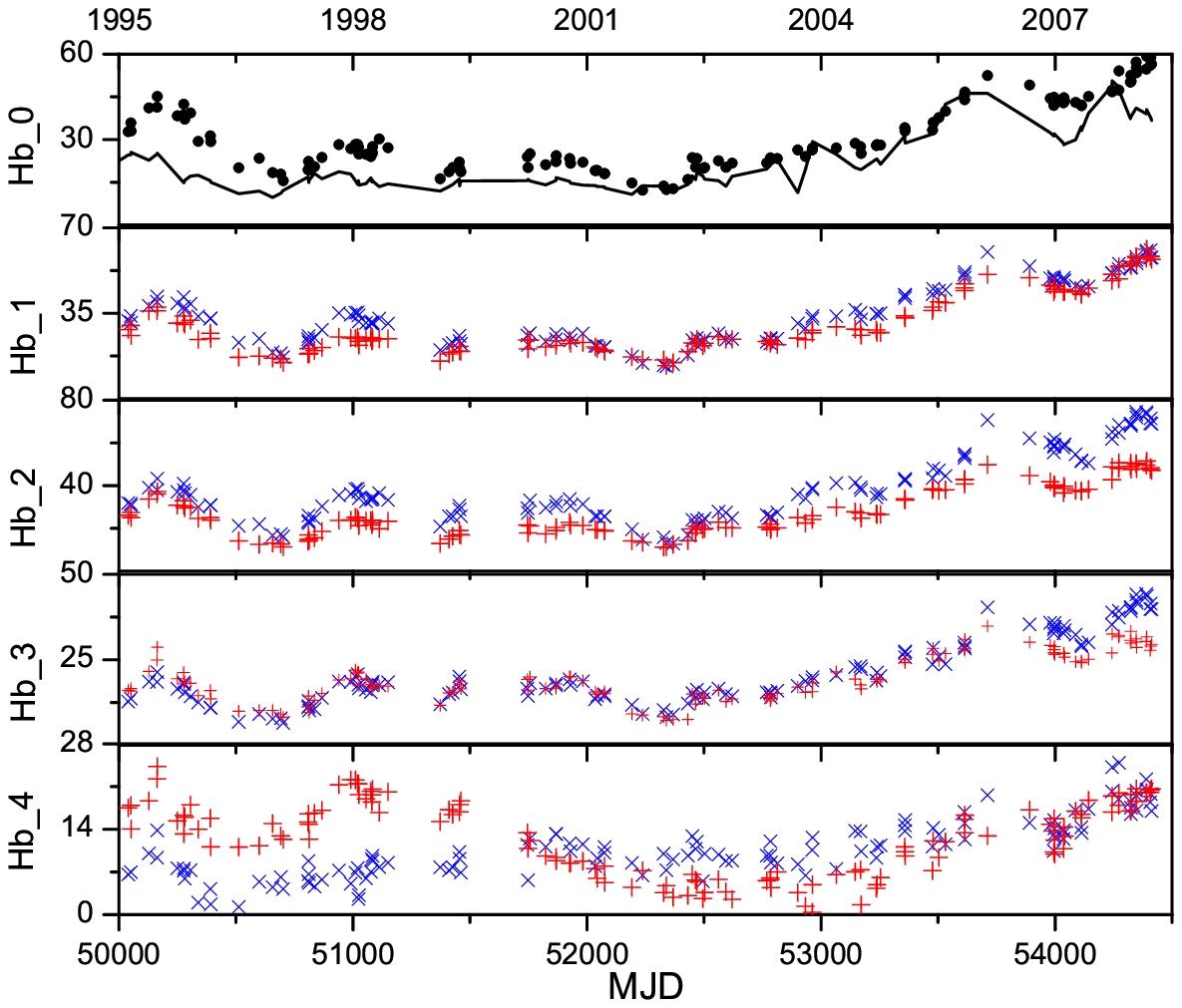}
\caption{The light curves for different segments of the H$\alpha$
(top) and H$\beta$ (bottom) broad emission lines. {The
line-segment fluxes are given in $10^{-14}$ and $10^{-15} {\rm
erg~cm^{-2}~s^{-1}}$ for H$\alpha$ and H$\beta$, respectively, while
the date is given in the Modified Julian Date as MJD=JD-2400000.5}
The mean velocity for segments number is as following: ($\pm$4)
correspond ($\pm$8500)${\rm km \ s^{-1}}$ ; ($\pm$3) correspond
($\pm$6000); ($\pm$2) correspond ($\pm$4000); ($\pm$1) correspond
($\pm$2000)${\rm km \ s^{-1}}$; 0  corespond (1000)${\rm km \
s^{-1}}$. The blue part is denoted with crosses, the red part with
pluses, and the line core with full circles. The solid line
represents the continuum flux in arbitrary units for comparison.}
\label{f09}
\end{figure*}}

\begin{figure*}
\centering
\includegraphics[width=8cm]{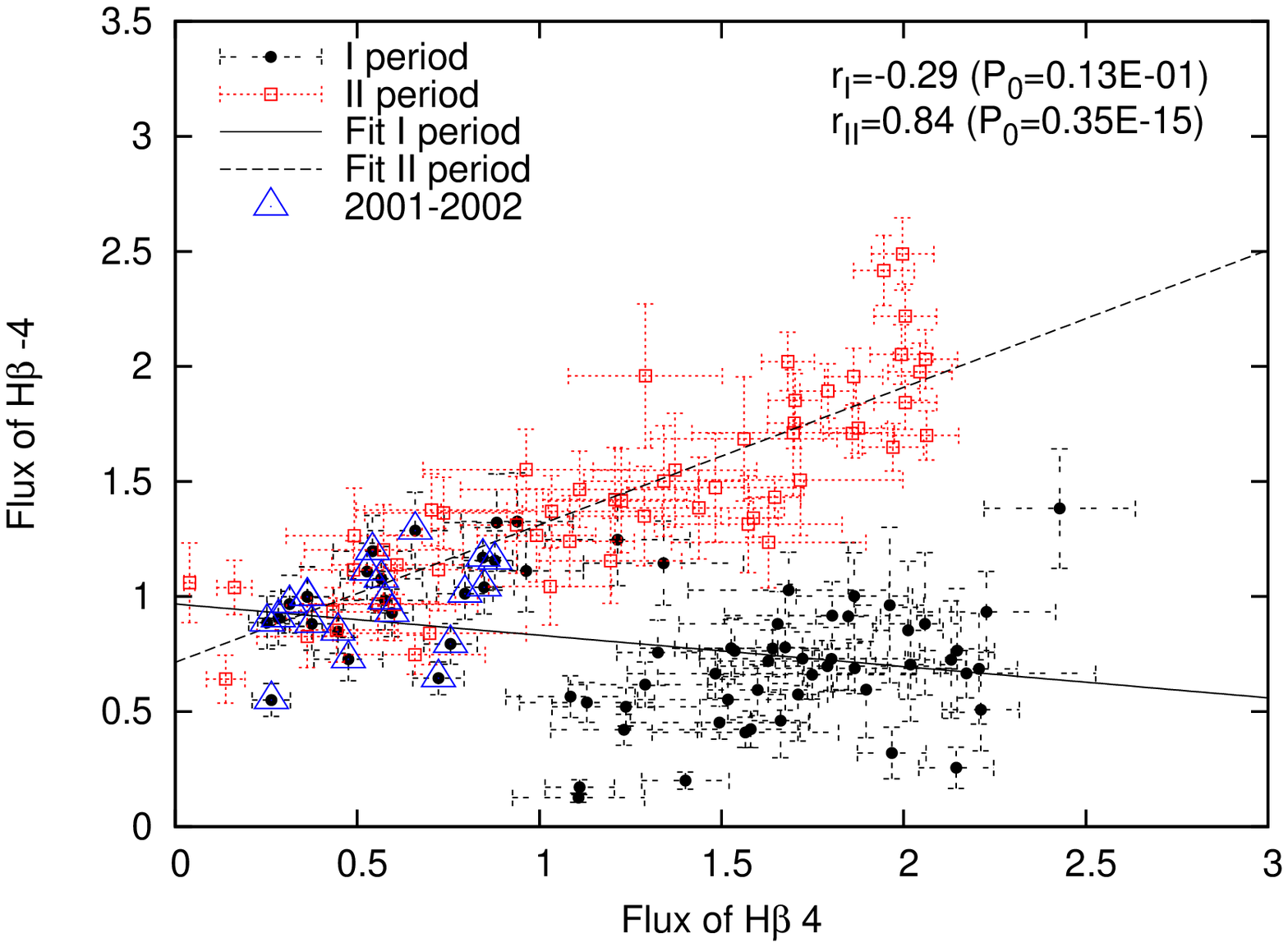}
\includegraphics[width=8cm]{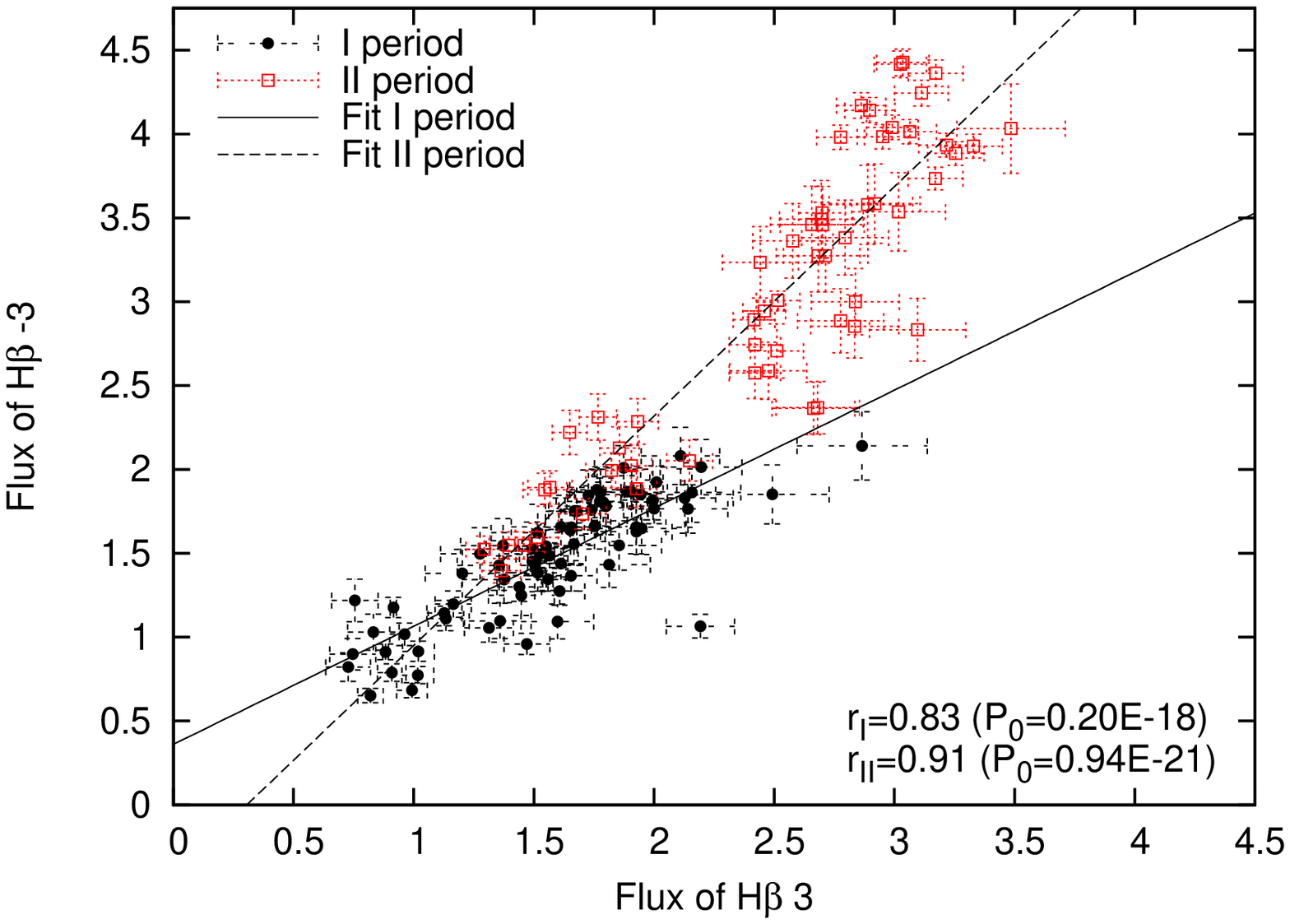}
\includegraphics[width=8cm]{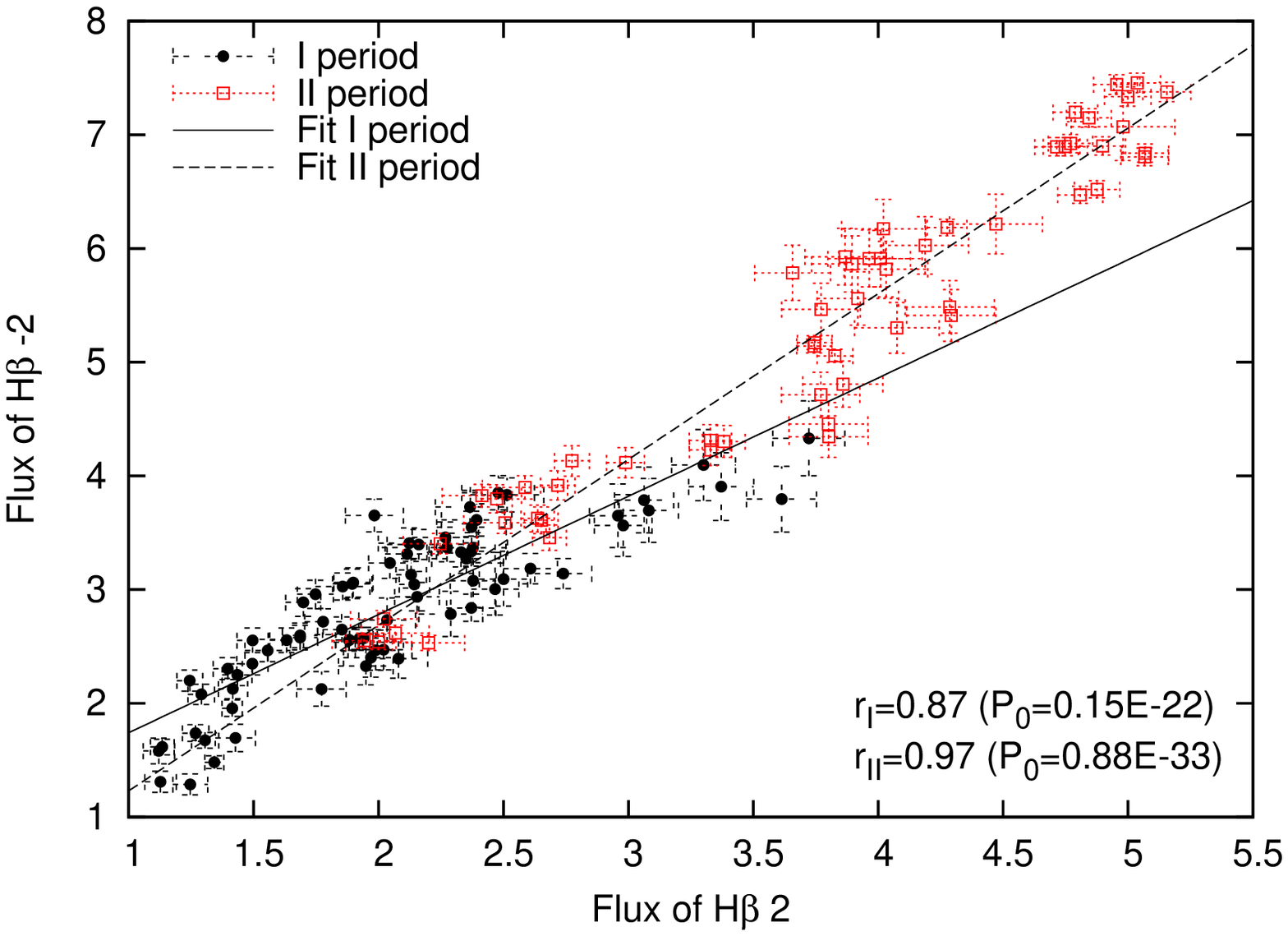}
\includegraphics[width=8cm]{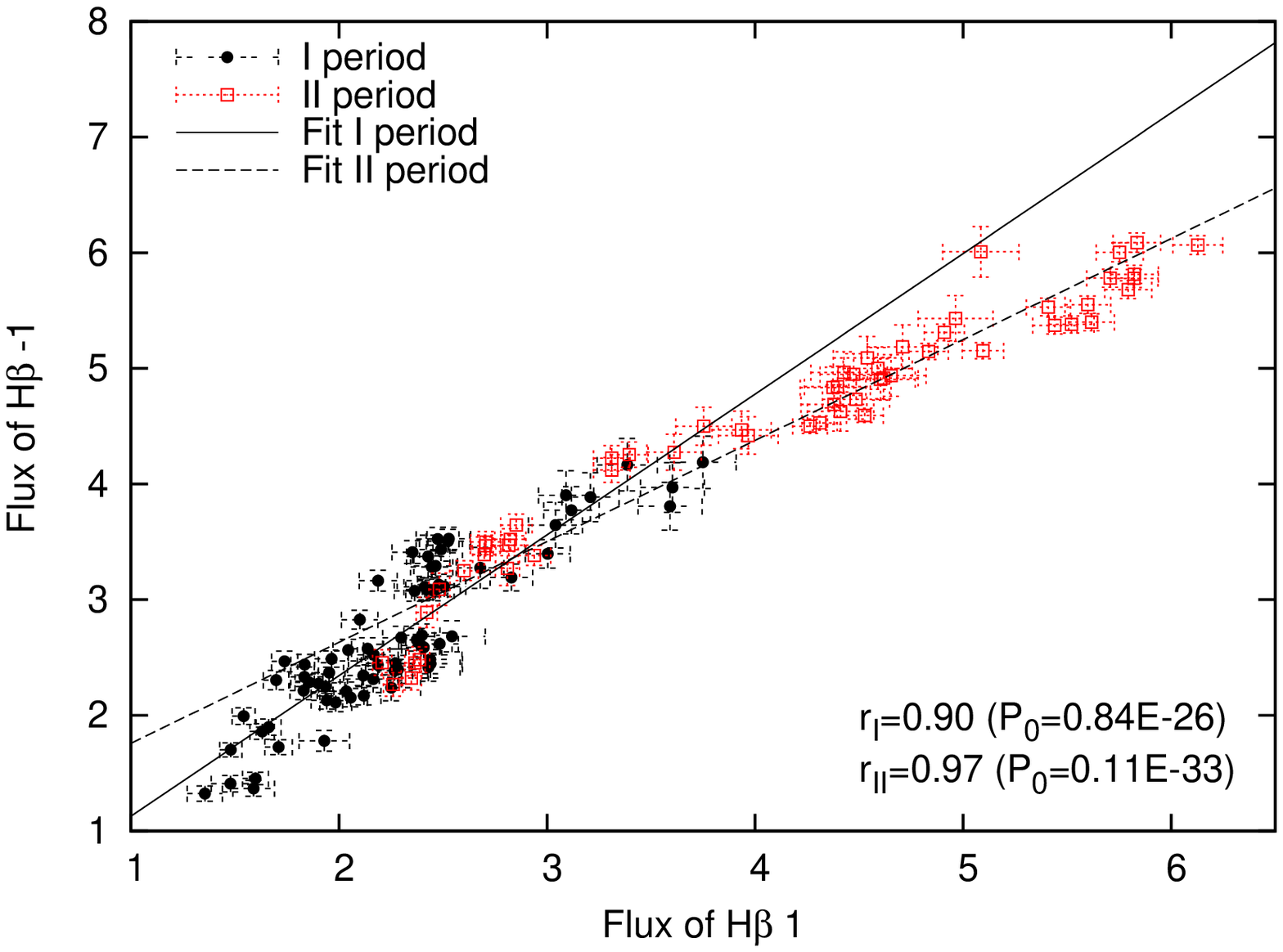}
\caption{The line segment vs. line segment response. The
observations in Period I are denoted with full circles and in Period
II with open squares. The line-segment flux is in units
$10^{-14}\,{\rm erg \, m^{-2}\,s^{-1}}$. {The Pearson correlation coefficient $r$
and the null hypothesis value $P_0$ are given on each plot for both periods.}}\label{f11}
\end{figure*}

\begin{table*}
\begin{center}
\caption[]{The beginning and ending radial velocities, $V_{\rm beg}$
and $V_{\rm end}$, for different segments in the line profiles of
H$\alpha$ and H$\beta$.}
\label{t02}
\begin{tabular}{llllllllll}
\hline
\hline
\backslashbox[0pt][l]{Vr}{segment}    &   -4  &  -3   &  -2   &  -1
&  0(C) & +1  &   +2  &   +3  &  +4   \\
\hline
$V_{\rm beg}$     & -10000 & -7000 & -5000 & -3000 & -1000 & 1000
& 3000 &  5000 &  7000    \\
$V_{\rm end}$     & -7000  & -5000 & -3000 & -1000 &  1000 & 3000
& 5000 &  7000 & 10000     \\
\hline

\end{tabular}
\end{center}
\end{table*}

\subsection{Line segment vs. line segment and continuum variations}

We study the response of symmetrical line segments (-4 vs. 4, etc.),
and as it shown in Fig. \ref{f11}, there is only a weak response
from the far blue to the far red segment in Period I.  We plot the
best linear fits for Period I and Period II, and find that the
correlation between the far wings (segment -4 vs. segment +4) has a
negative tendency {with low significance (the Pearson correlation
coefficient $r=-0.29$, the null hypothesis value $P_0$=0.13E-01)},
but in the Period II there is a linear response of the blue to the red
wing {with significant correlation ($r=0.84$, $P_0$=0.35E-15)}.
In other segments, one can see that the change in the blue wing in
Period I was smaller than in the red one (Fig. \ref{f11}).
It is interesting that in the Period I, the flux of  the far blue
wing (-4) stays nearly constant, while the flux of the red part (+4)
is increasing. {Even if we consider the Period I without the period
2001-2002, the correlation is weak and insignificant ($r=0.03$, $P_0$=0.84)
and in favor of the previous statement.}

\begin{figure*}
\centering
\includegraphics[width=8.5cm]{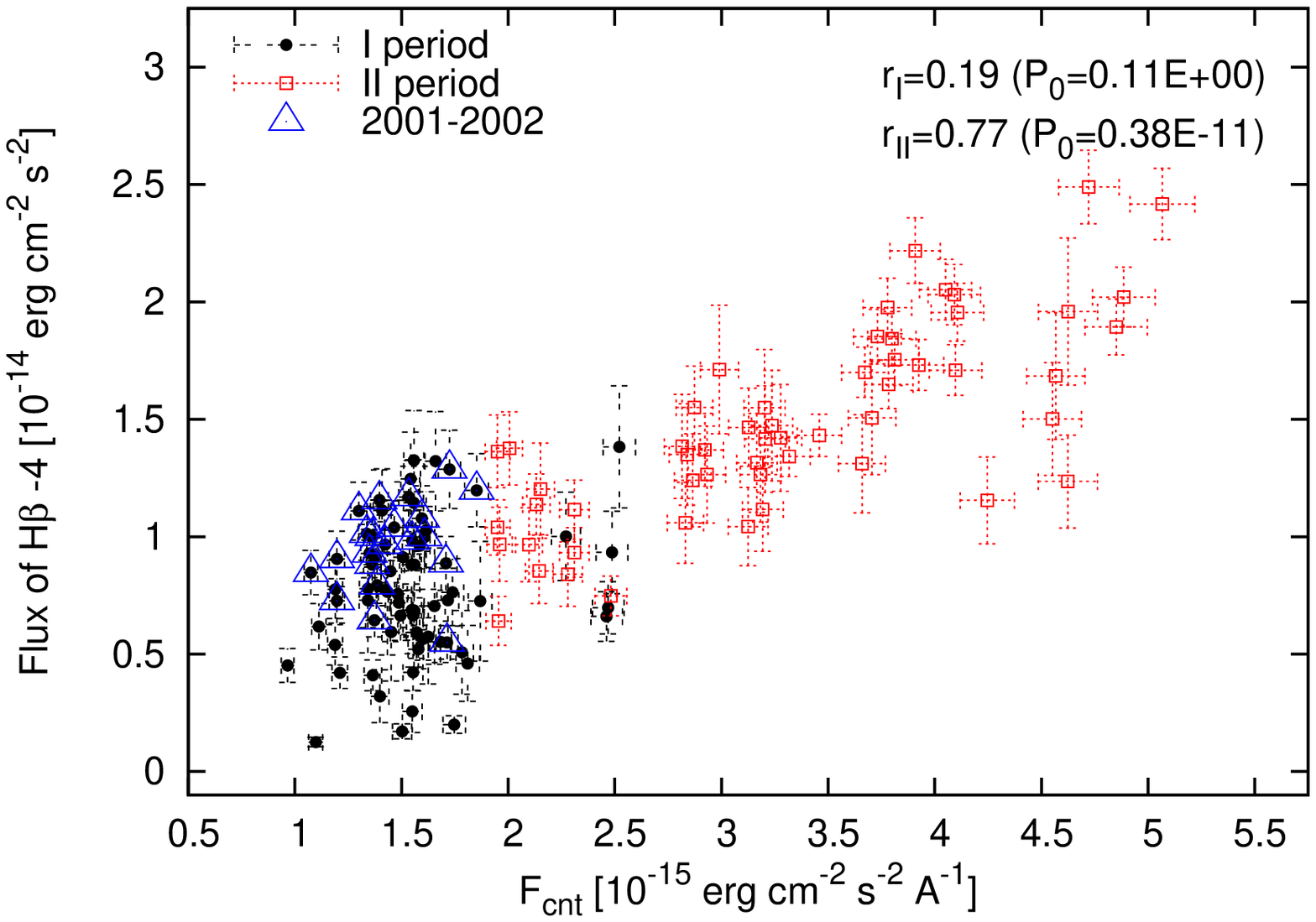}
\includegraphics[width=8.5cm]{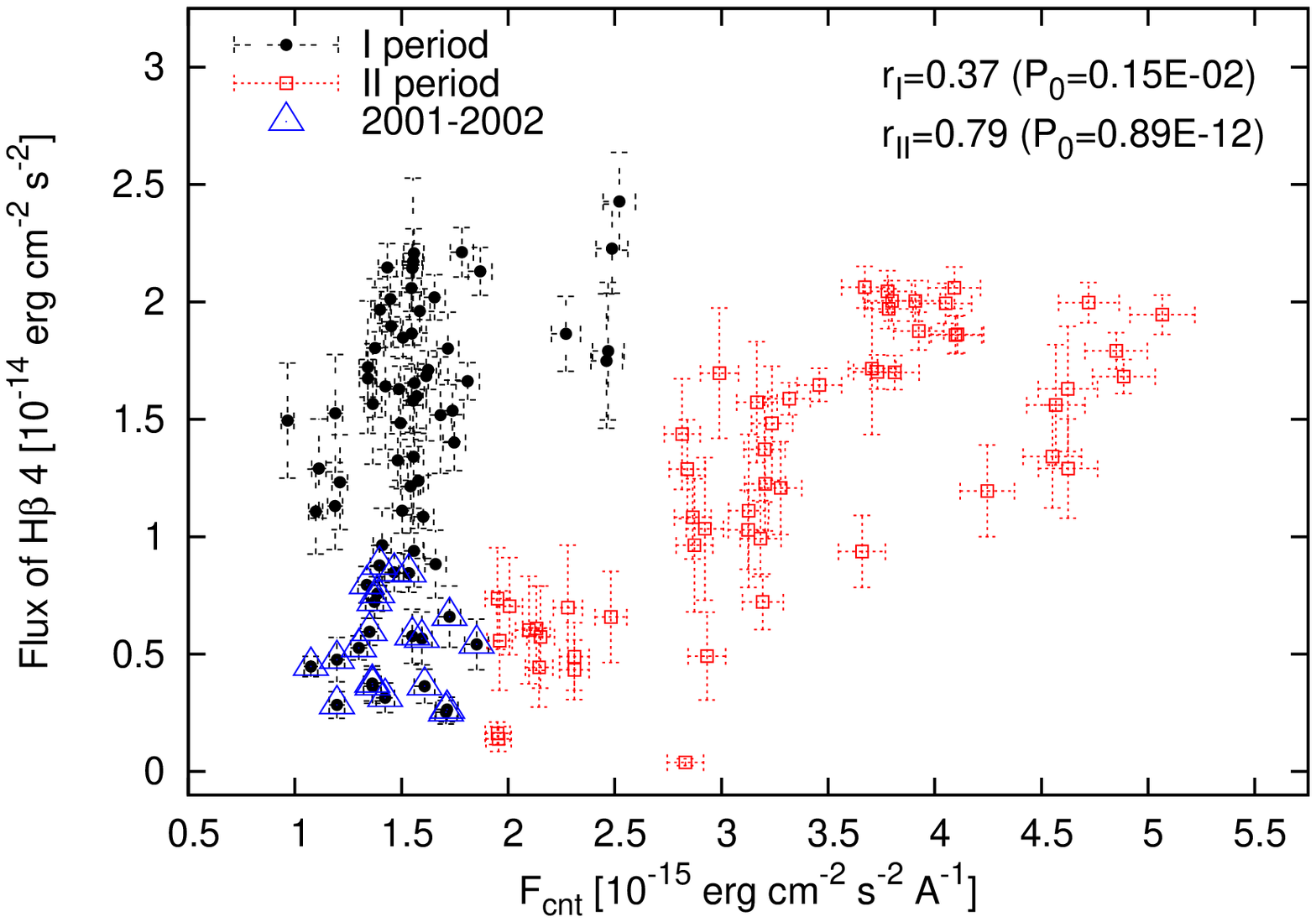}
\includegraphics[width=8.5cm]{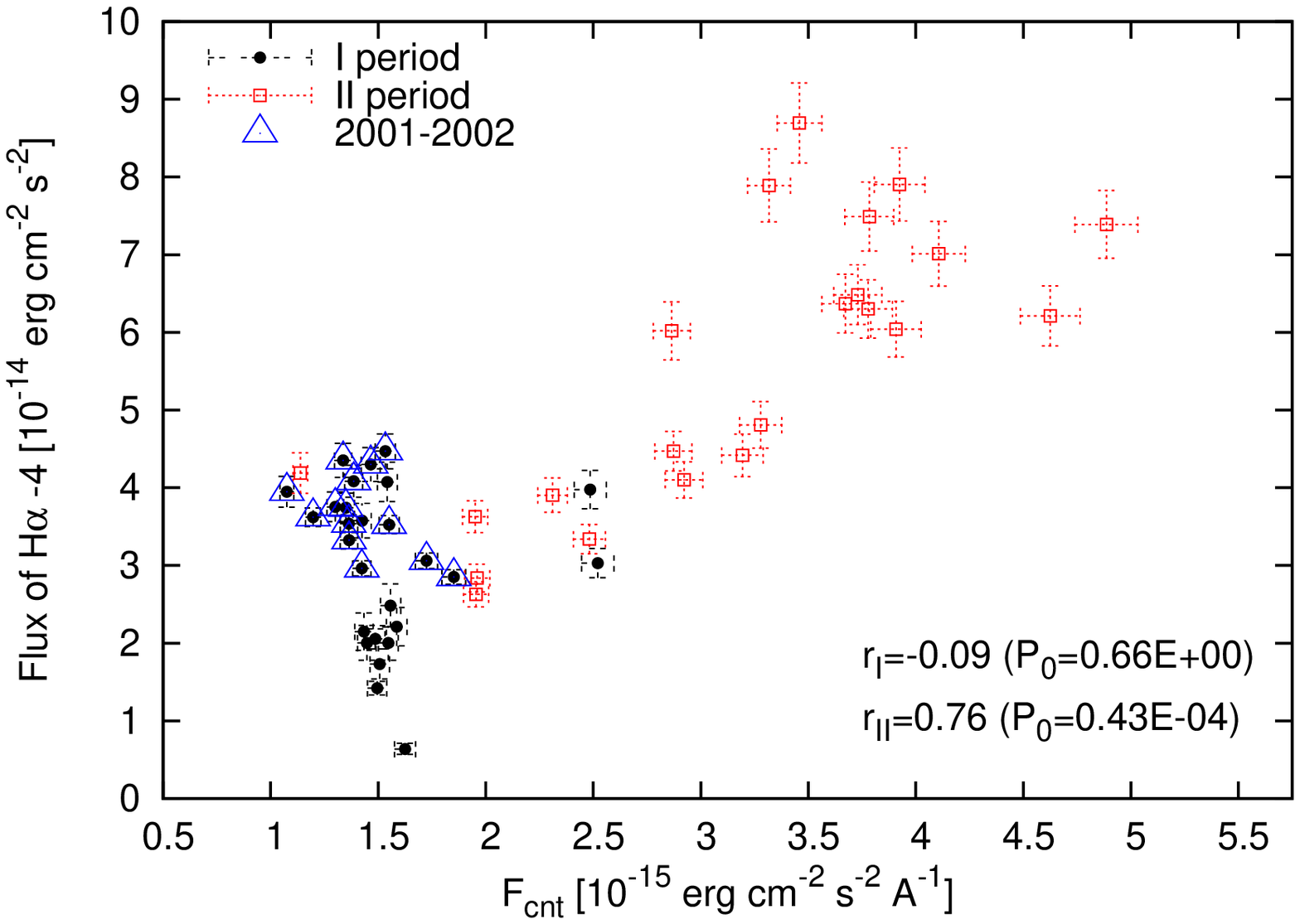}
\includegraphics[width=8.5cm]{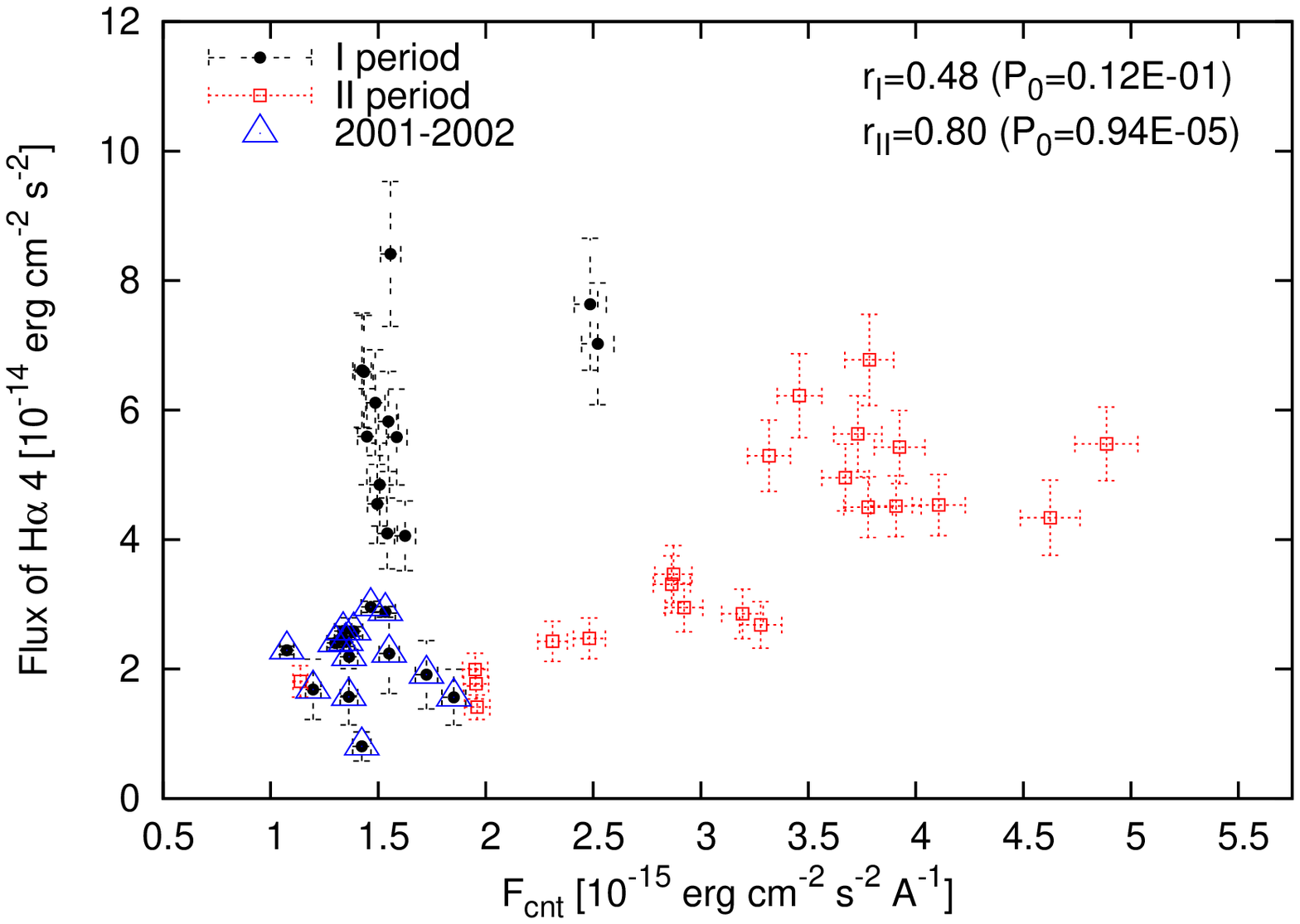}
\caption{The far blue (left) and far red (right) wings flux against
the continuum-flux variation for H$\beta$ (top) and H$\alpha$
(down). The notation is the same as in Fig. \ref{f11}.
{The Pearson correlation coefficient $r$
and the null hypothesis value $P_0$ are given on each plot for both periods.}}\label{f12}
\end{figure*}

To see whether there is a similar response of the line segments to
the continuum flux we construct plots in Figs. \ref{f12} and
\ref{f13} (Fig. \ref{f13} available electronically only), where the
flux of all segments against the continuum flux are presented. As it
can be seen in Figs. \ref{f12} and \ref{f13} there is a relatively
good response of the line segments to the continuum flux {(see
plots for the values of the Pearson correlation coefficient $r$ and
the null hypothesis $P_0$ for both periods)}. However, the line
segments ($\pm$4) {(especially in the case of the H$\beta$ line (Fig. \ref{f12}, top))} in
the Period I have a weak response (or even absent response, in the case segment
+4) to the continuum flux. For example, the flux in the far red
segment (H$\beta$ +4) increases around 4-5 times without the
increase of the continuum flux.

Note here that in the  H$\beta$ segment +4 one can expect higher
errors due to the subtraction of the bright [OIII] lines. {As a
consistency check}, we plot in Fig. \ref{f12} (bottom)  H$\alpha$
($\pm$4) segments fluxes vs. continuum flux. As it can be seen, the
trend is similar as in H$\beta$, and also in Period I  H$\alpha$
segment ($\pm$4) fluxes do not respond to the continuum flux. {This may indicate that there are periods when some processes} in the
disk that are not connected with the central continuum source.

\onlfig{11}{\begin{figure*} \centering
\includegraphics[width=8cm]{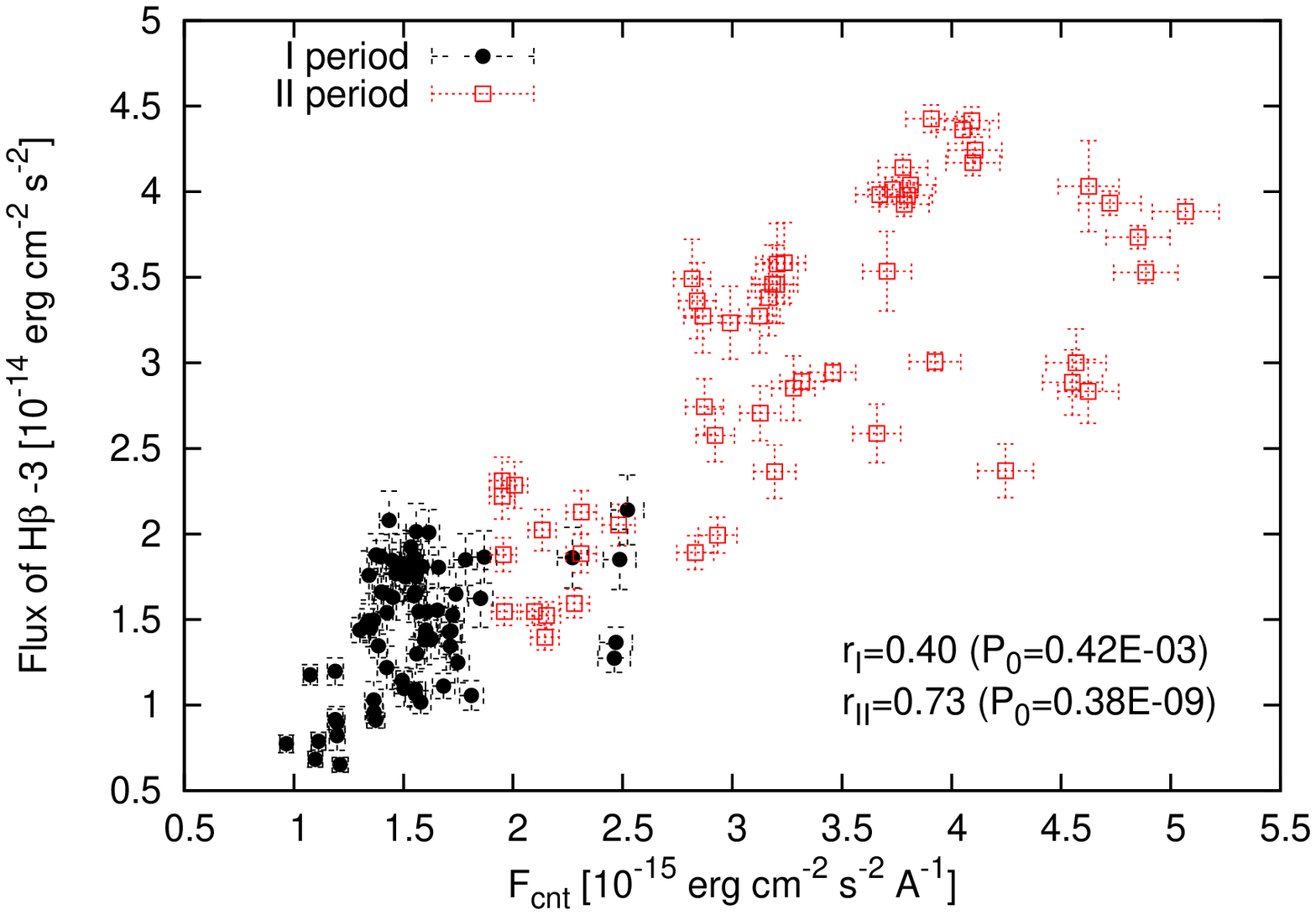}
\includegraphics[width=8cm]{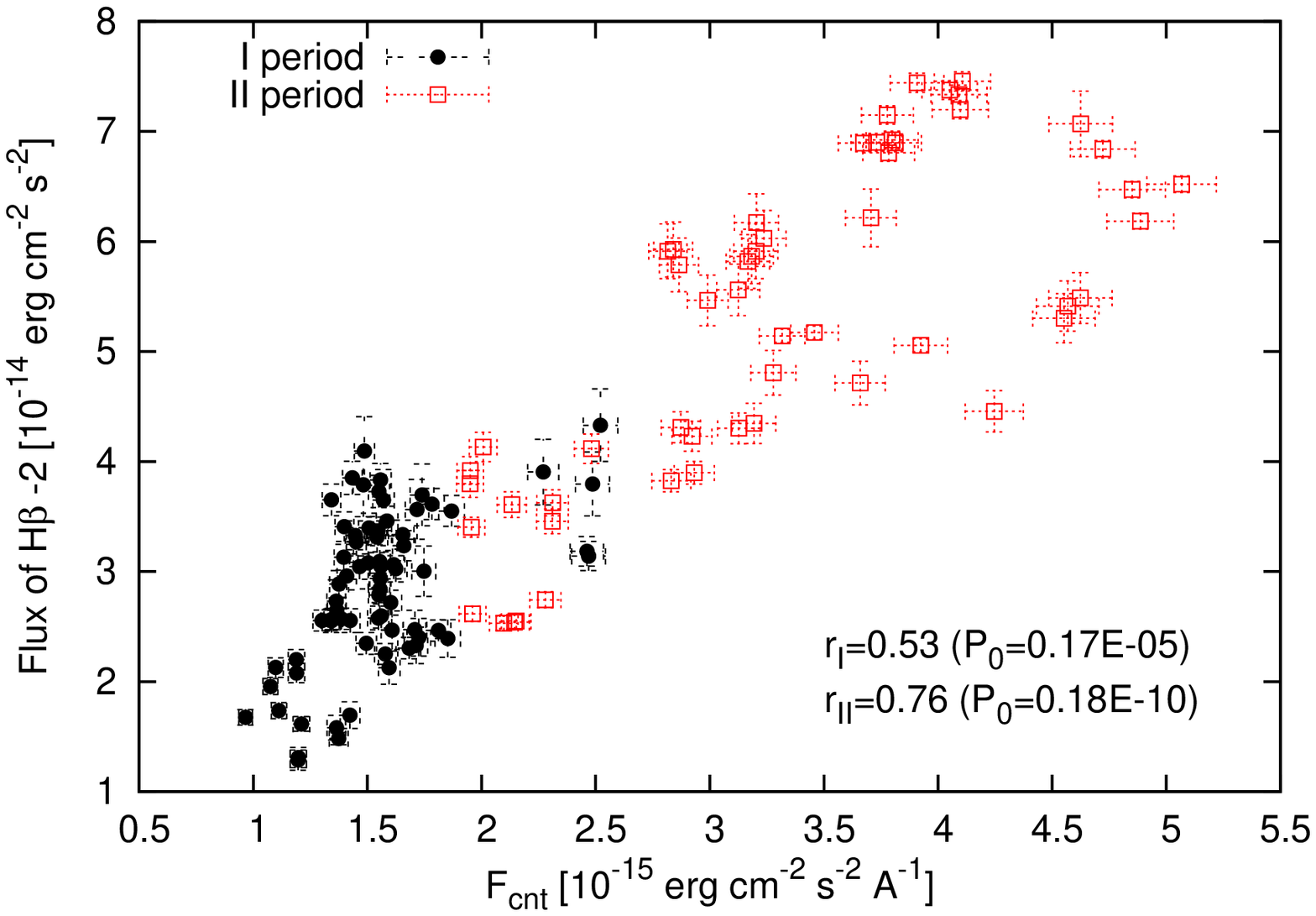}
\includegraphics[width=8cm]{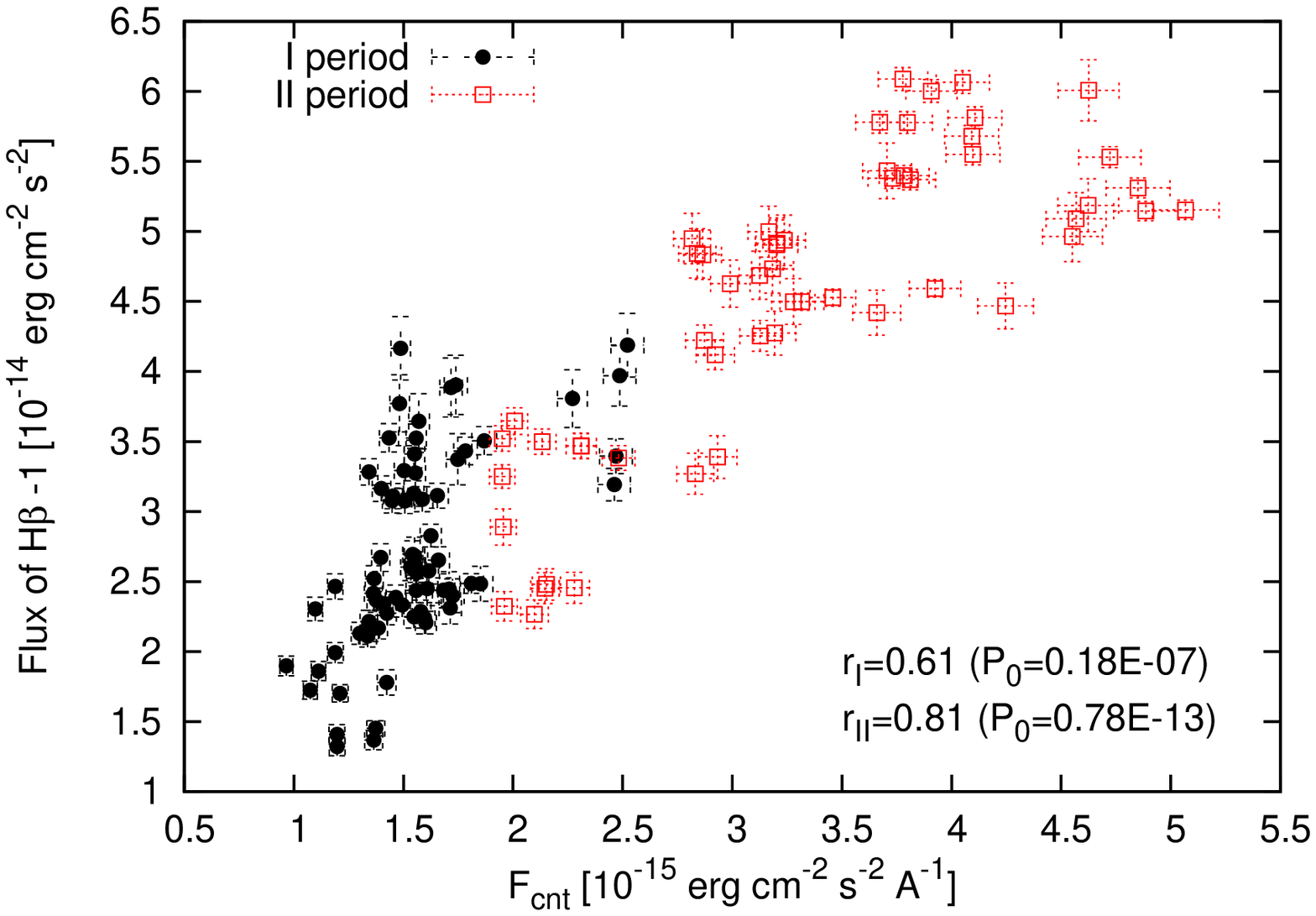}
\includegraphics[width=8cm]{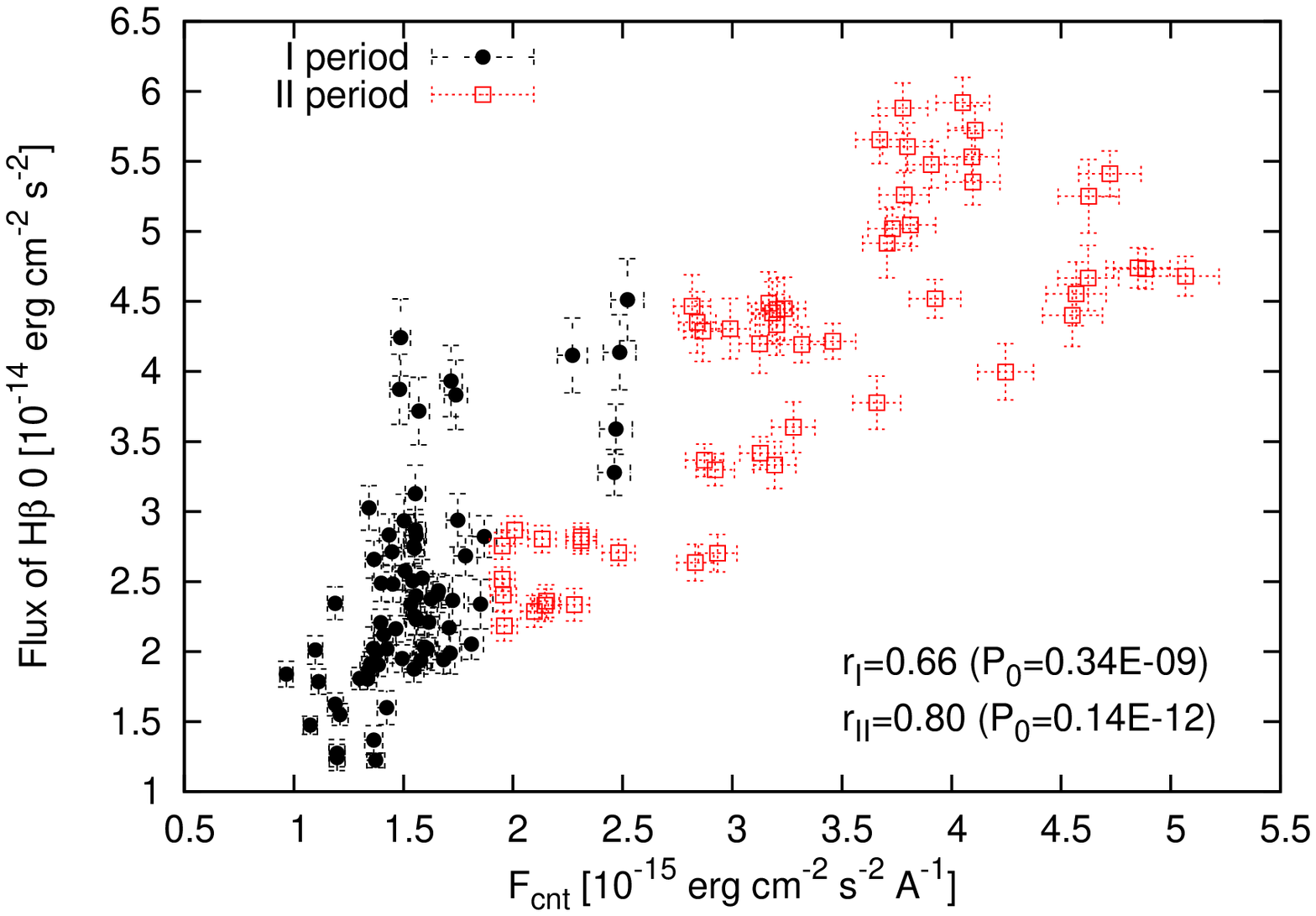}
\includegraphics[width=8cm]{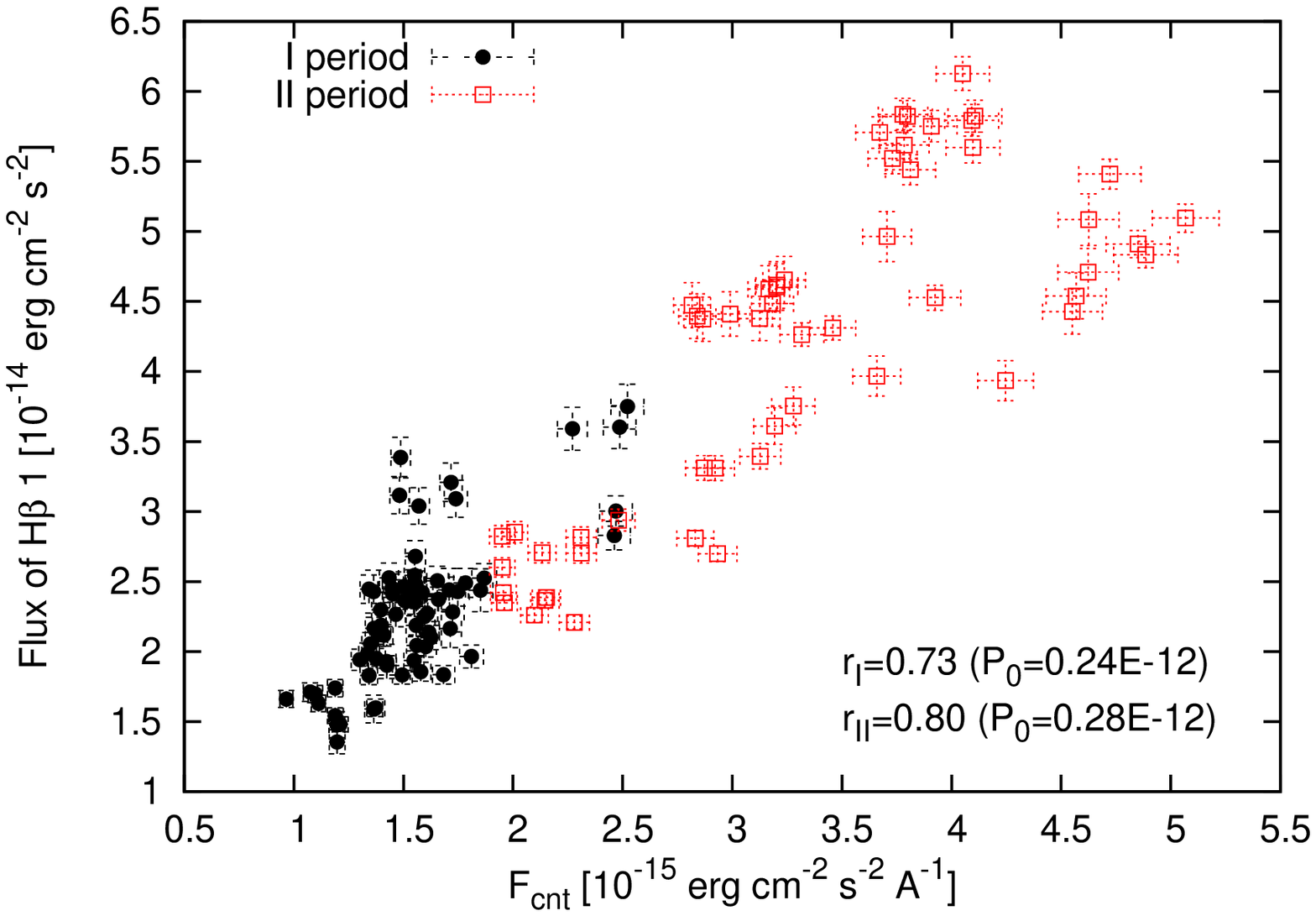}
\includegraphics[width=8cm]{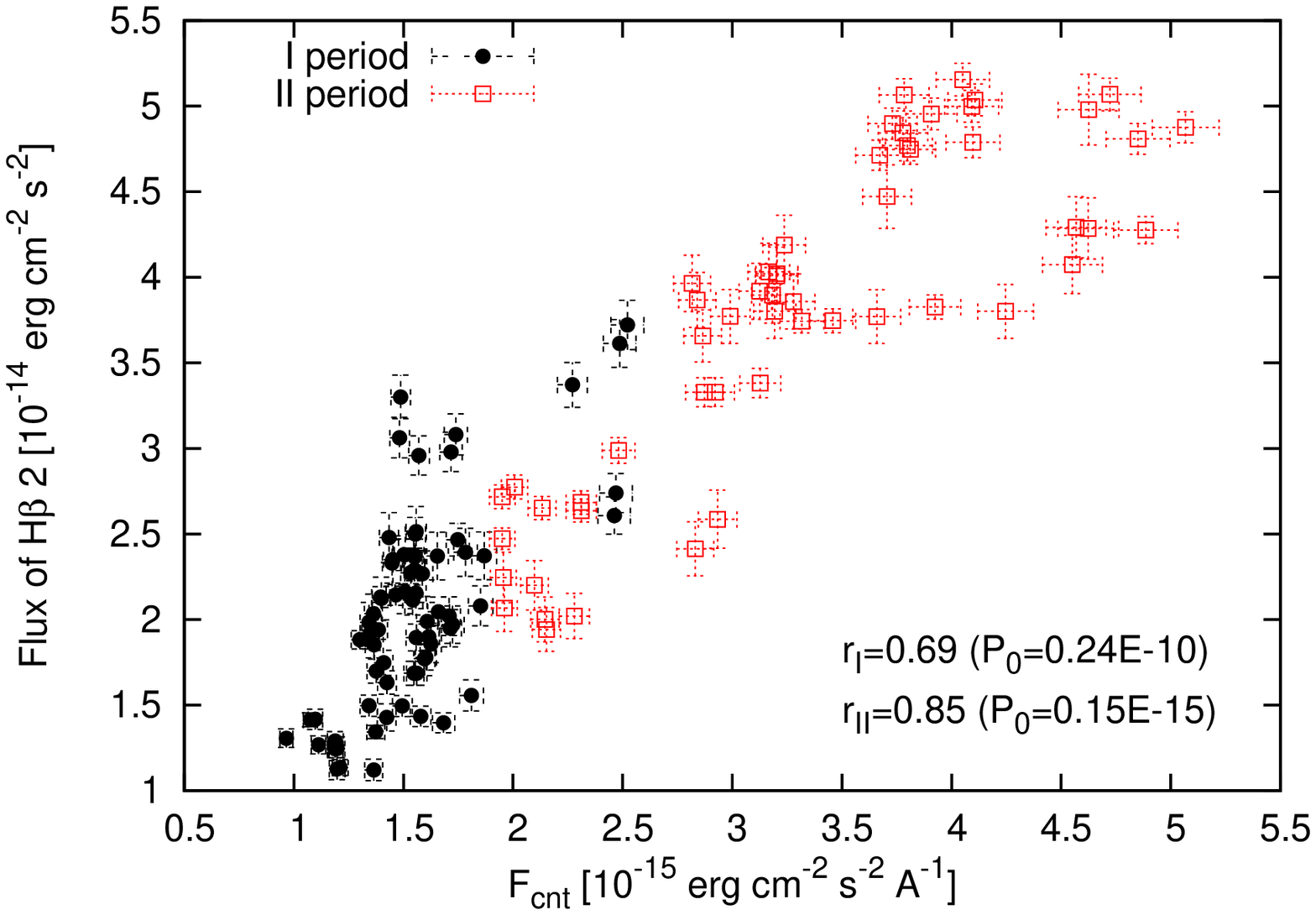}
\includegraphics[width=8cm]{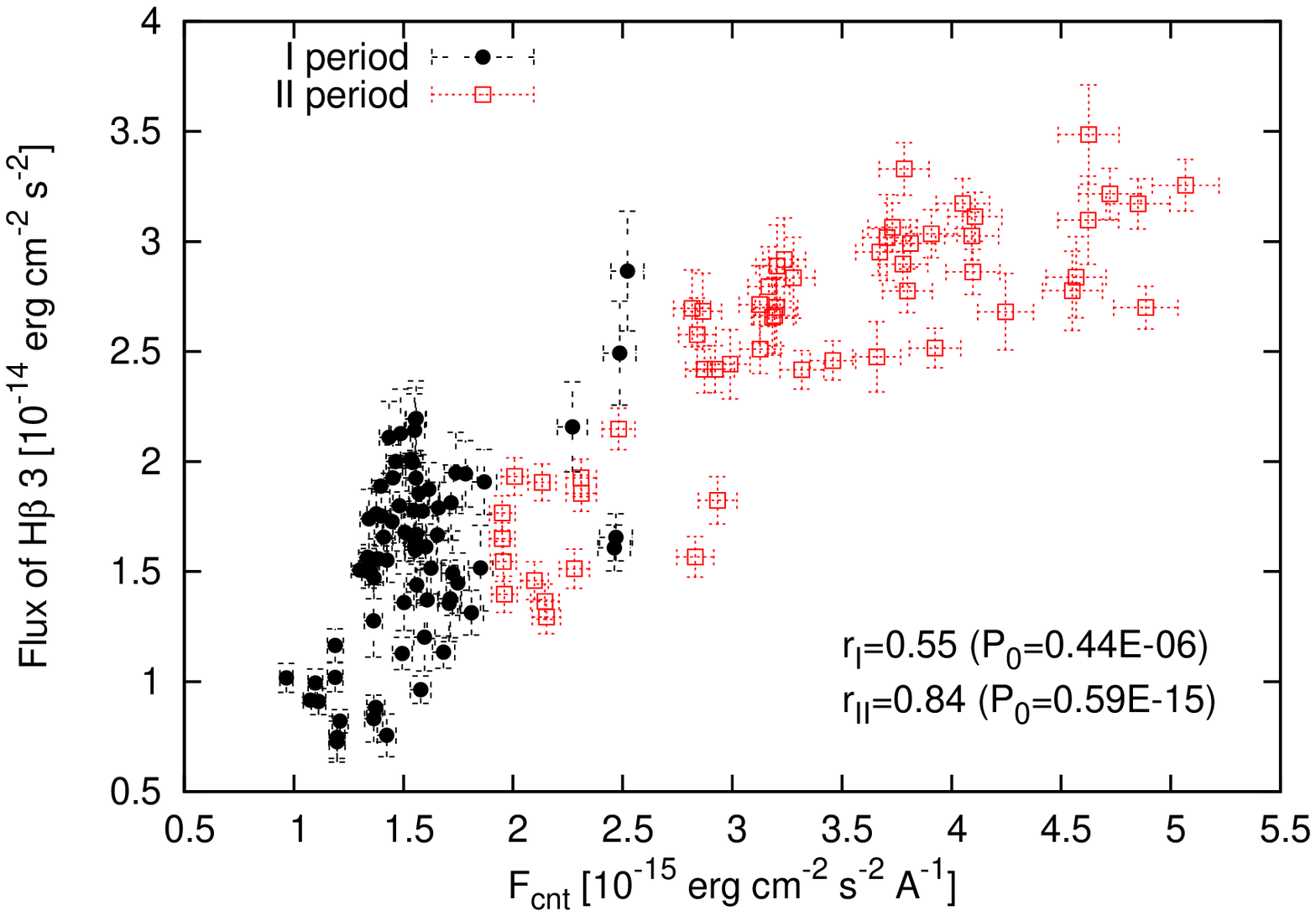}
\caption{The same as in Fig. \ref{f12}, but for the rest of
line-segments: -3, -2, -1, 0, 1, 2, 3.}\label{f13}
\end{figure*}}

\begin{figure*}
\centering
\includegraphics[width=4.5cm]{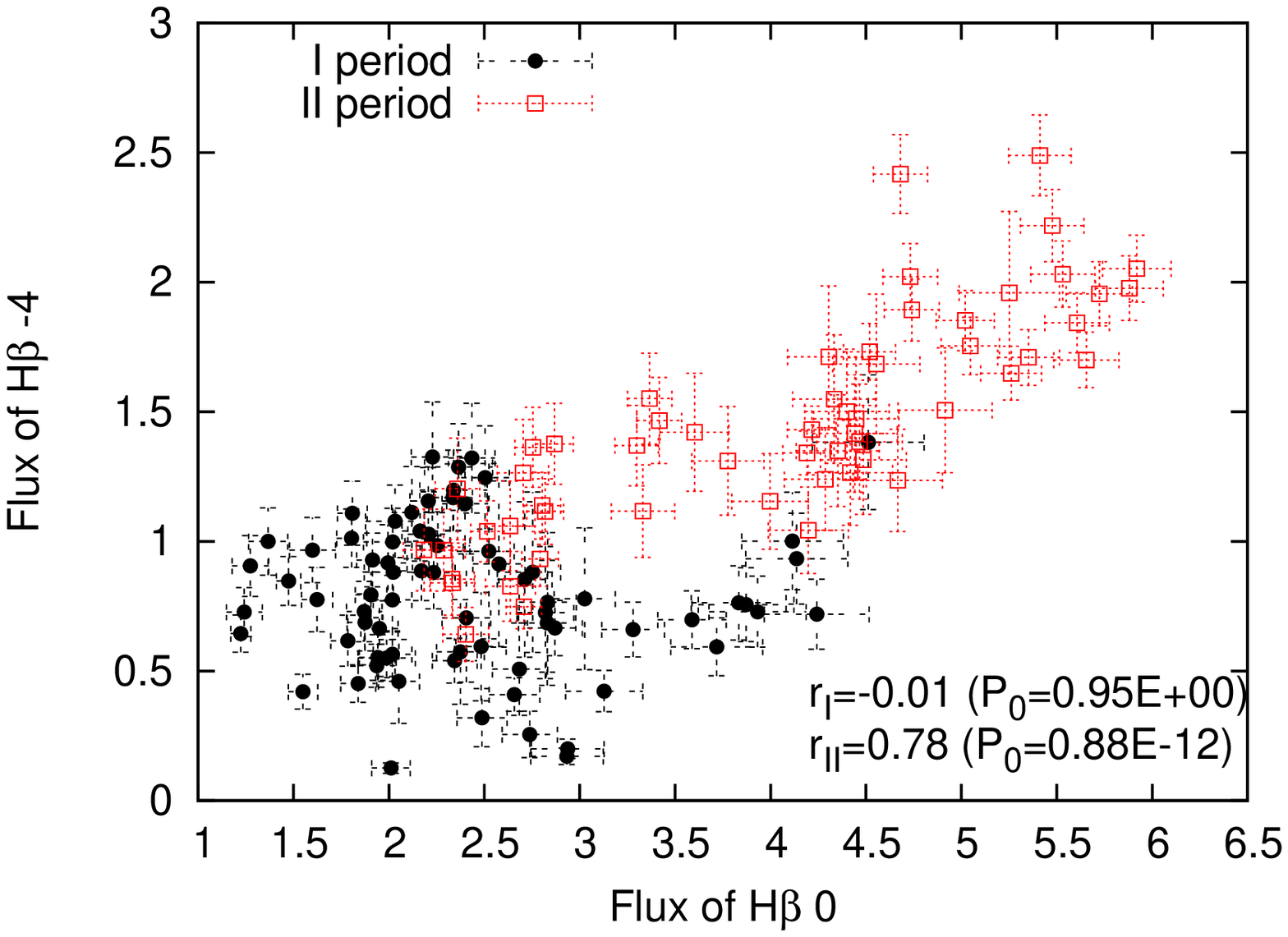}
\includegraphics[width=4.5cm]{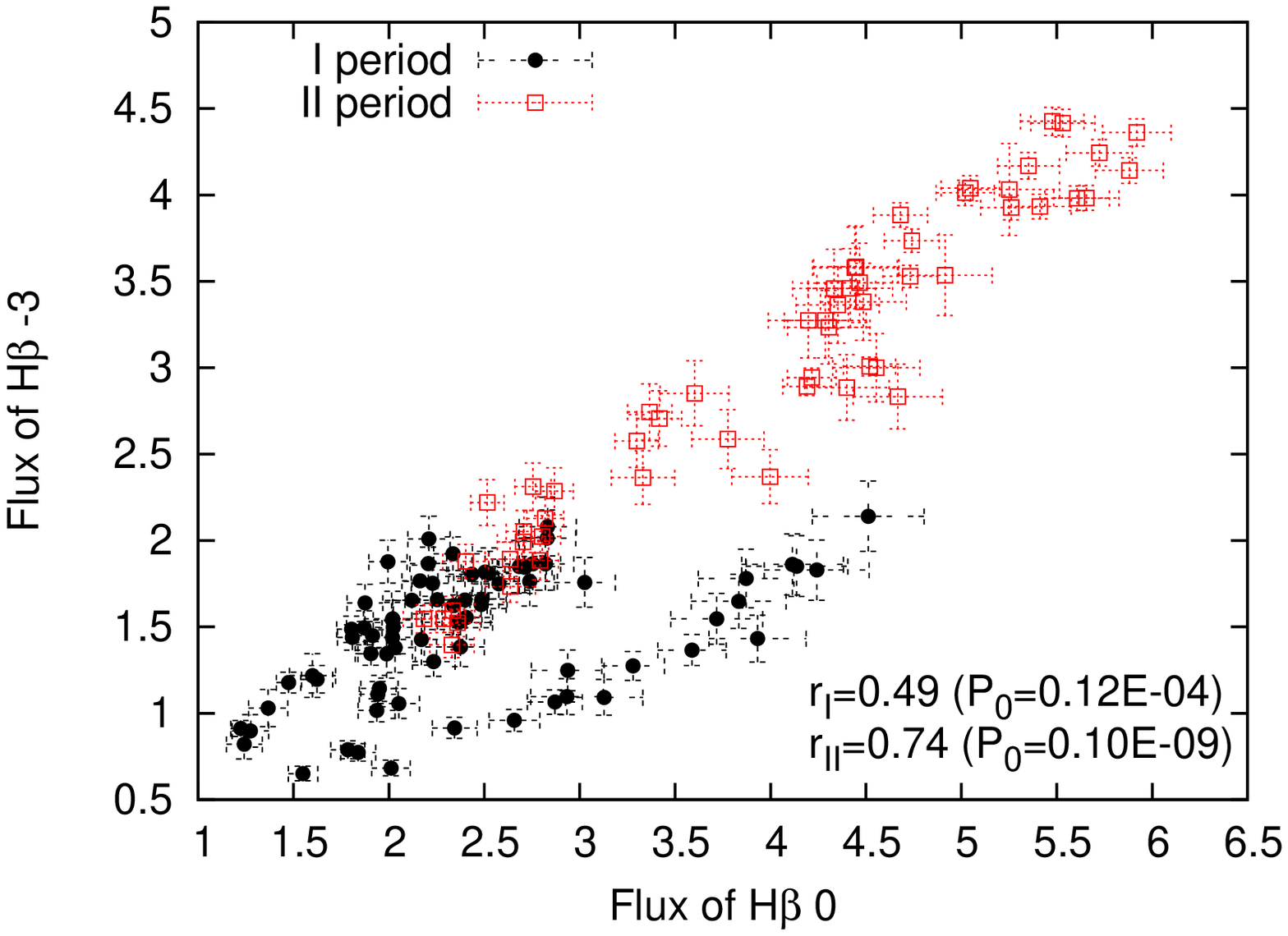}
\includegraphics[width=4.5cm]{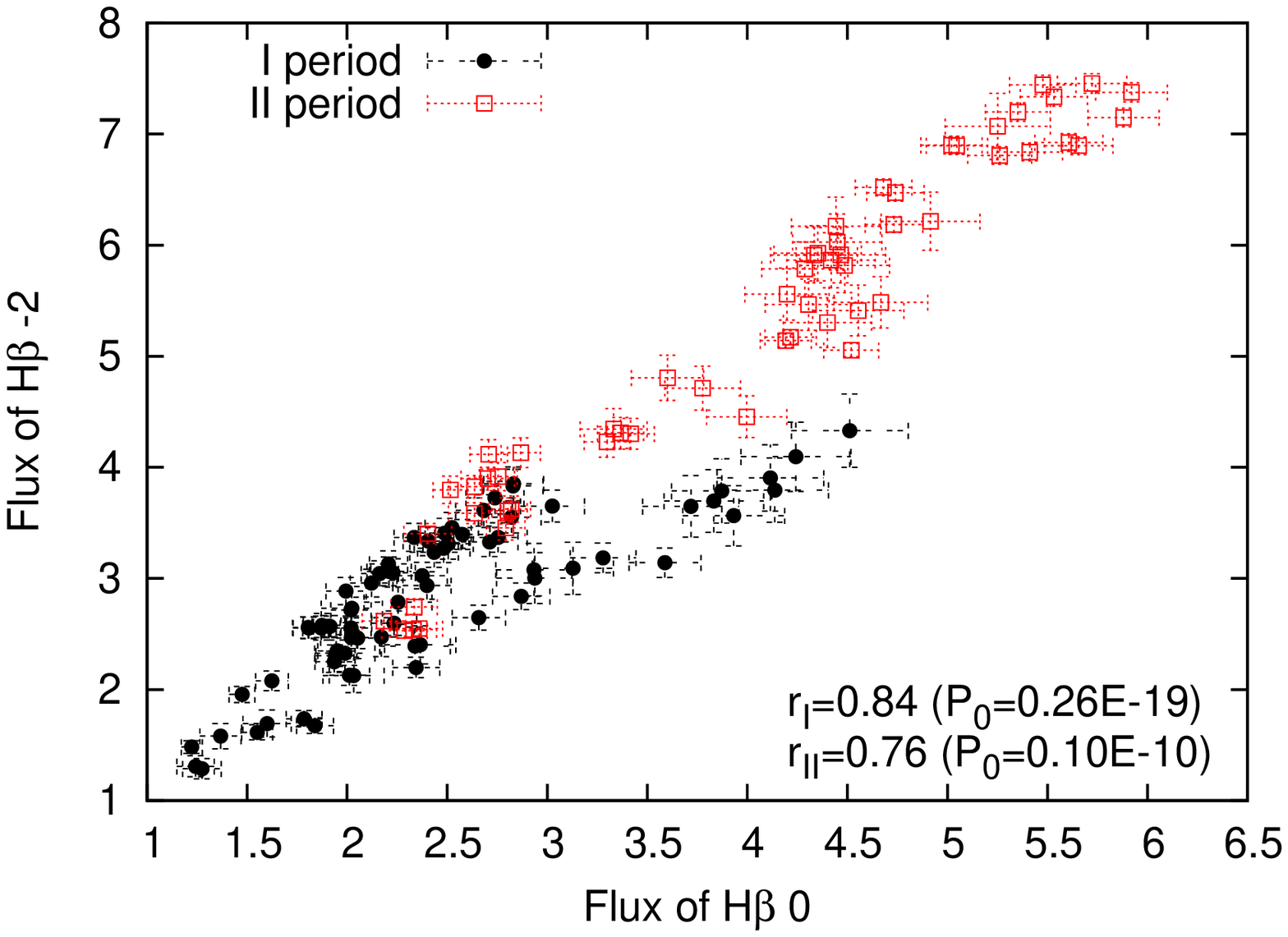}
\includegraphics[width=4.5cm]{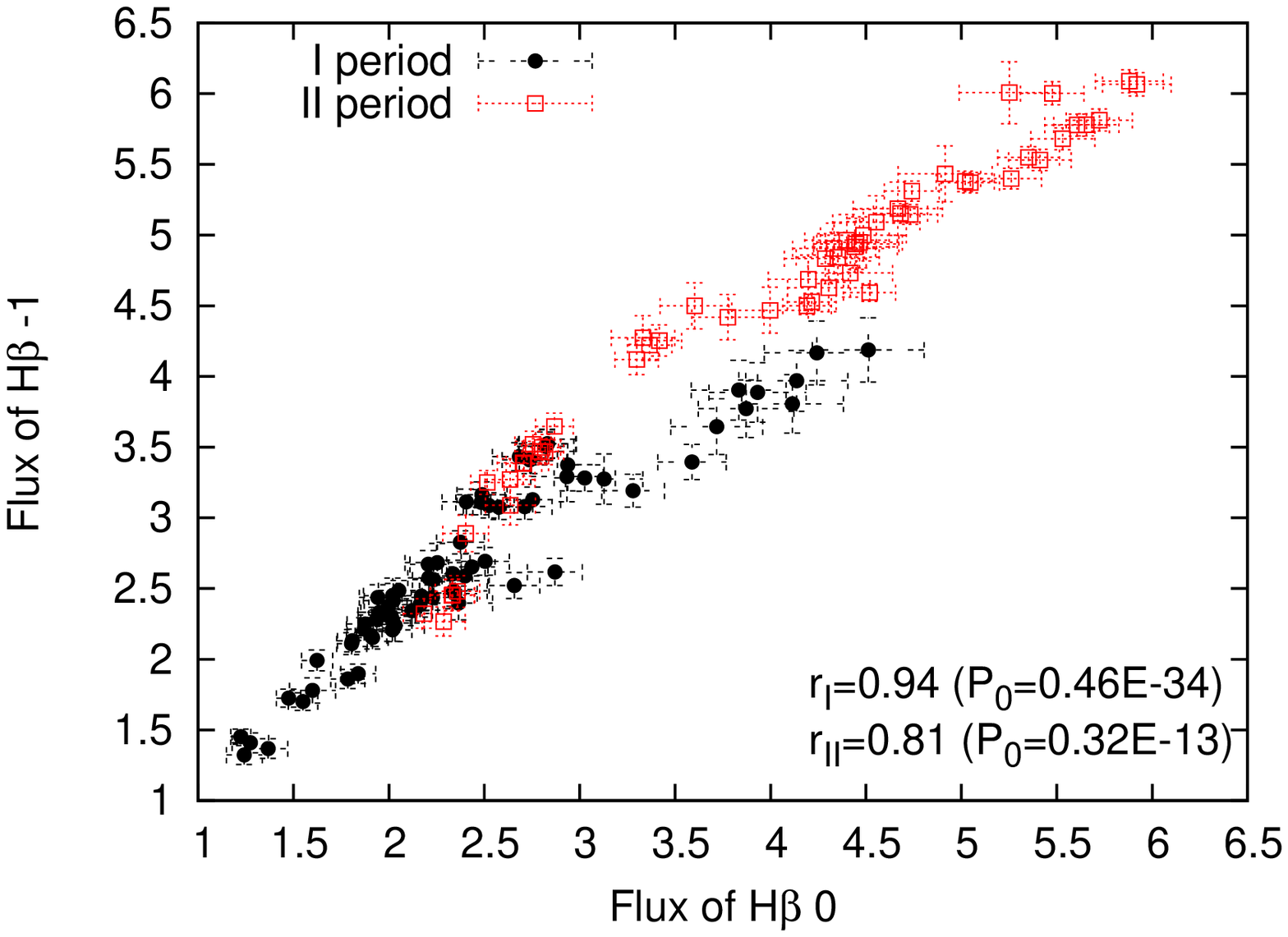}
\includegraphics[width=4.5cm]{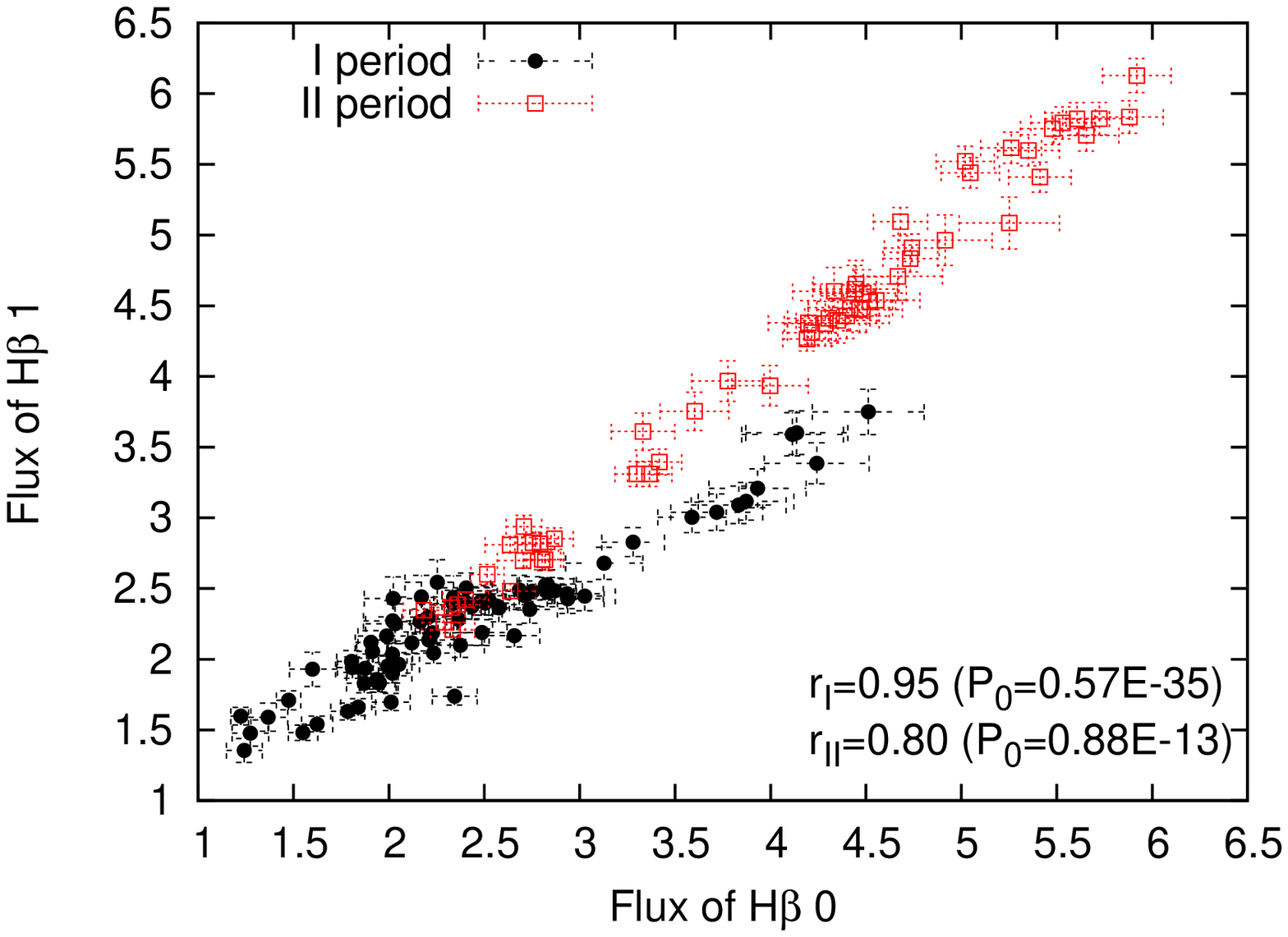}
\includegraphics[width=4.5cm]{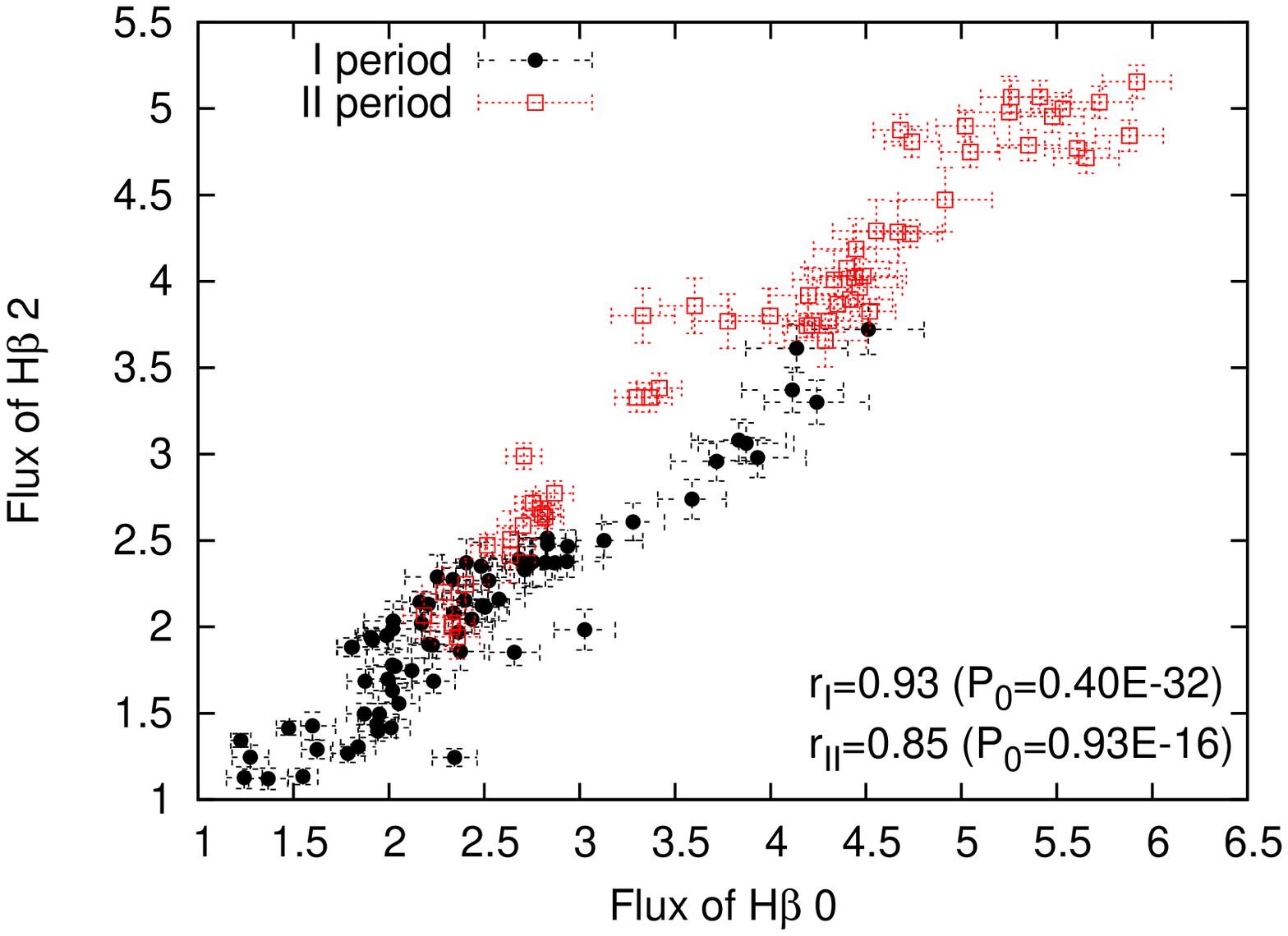}
\includegraphics[width=4.5cm]{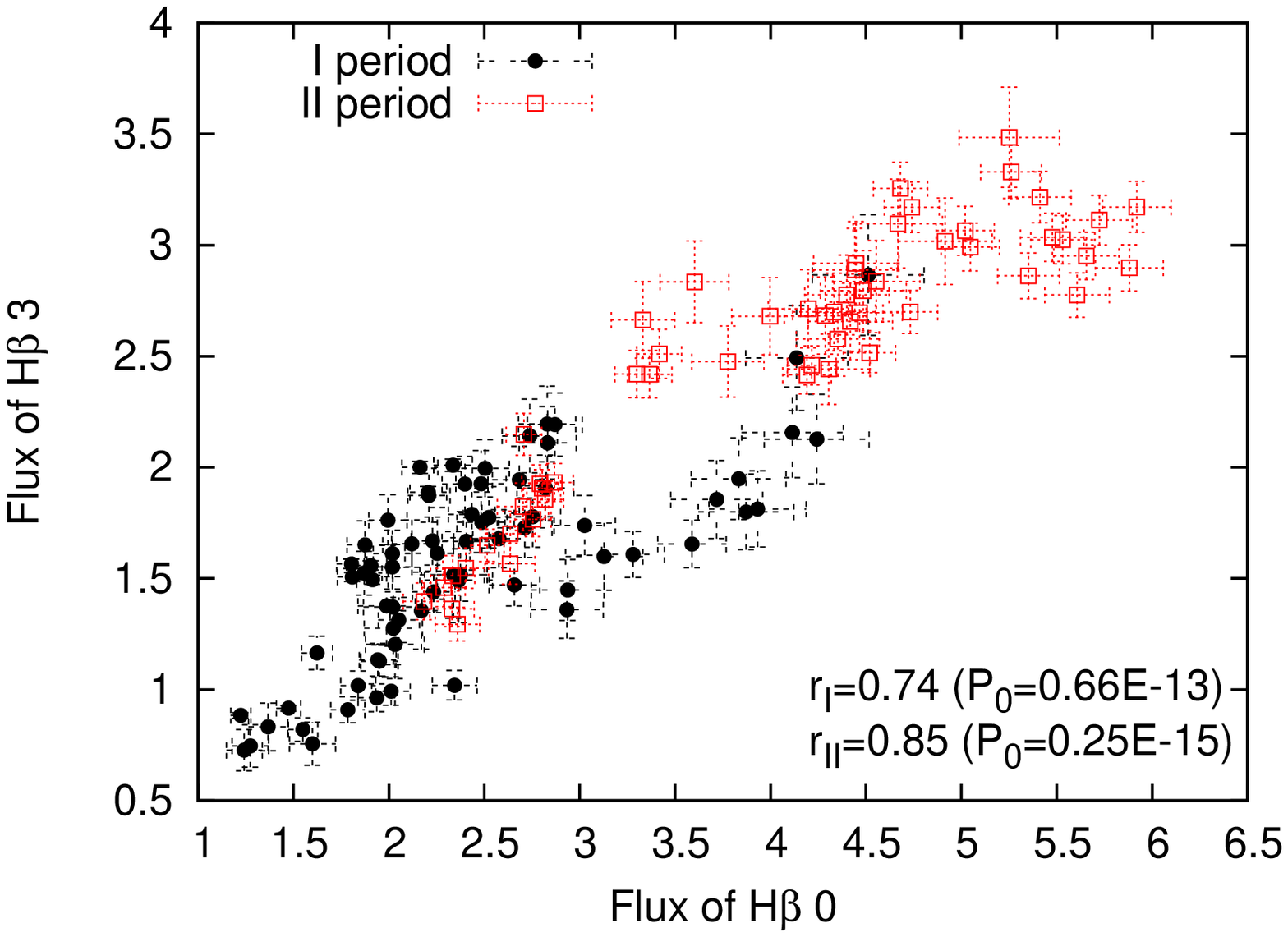}
\includegraphics[width=4.5cm]{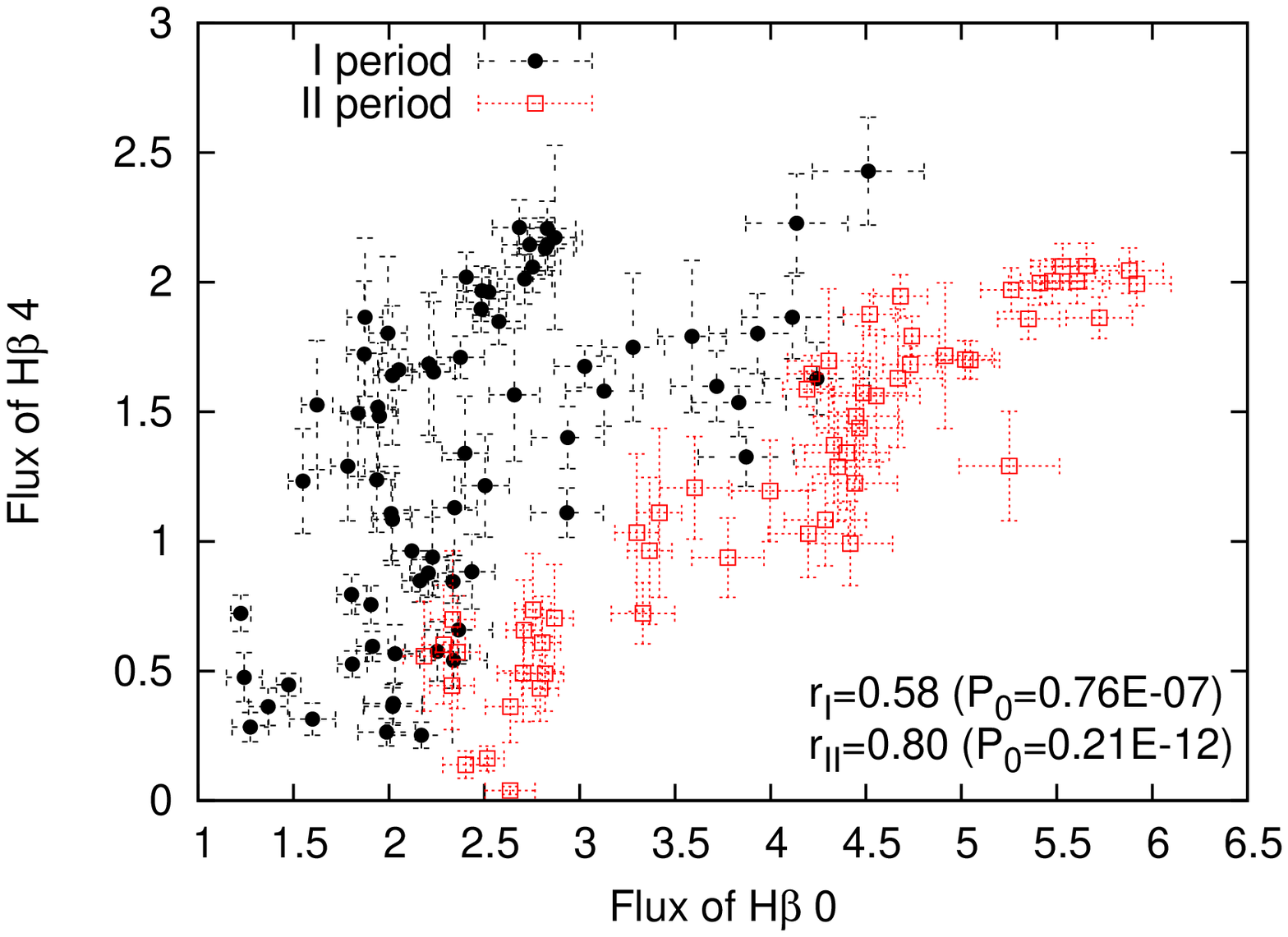}
\caption{The response of the line segments to the central segment 0
(from -1000 to 1000 km s$^{-1}$). The observations in Period I are
denoted with full circles and in Period II with open squares. The
line-segment flux is in units $10^{-14}\,{\rm erg \,
m^{-2}\,s^{-1}}$. {The Pearson correlation coefficient $r$
and the null hypothesis value $P_0$ are given on each plot for both periods.}}\label{f14}
\end{figure*}

As mentioned above, there is a central peak that is probably coming
from an additional emission region or is created due to some
perturbation in the disk. Therefore we plot in Fig. \ref{f14} the
line-segment fluxes vs. the flux of the central line-segment 0
(measured between -1000 and  +1000 km s$^{-1}$). As it can be seen
there is a good correlation between the central line-segment flux
and fluxes of other segments, only one can conclude that the
responses was better in the Period II. Thus, it seems that in this
period both the disk-line and additional component effectively
respond to the continuum variation. {Also, one can see that the
flux of the red part (segment +4) in Period I has a weak response to the central
line-segment 0 ($r=0.58$, $P_0$=0.76E-07), that is similar as in the case of
the segment to the continuum flux correlation (Fig. \ref{f12}).
This may have the same physical reason, that some kind of perturbation in disk were present
in Period I.}

\subsection{Modeling of the line segment variation of the H$\beta$ line}

\begin{figure}
\centering
\includegraphics[width=9cm]{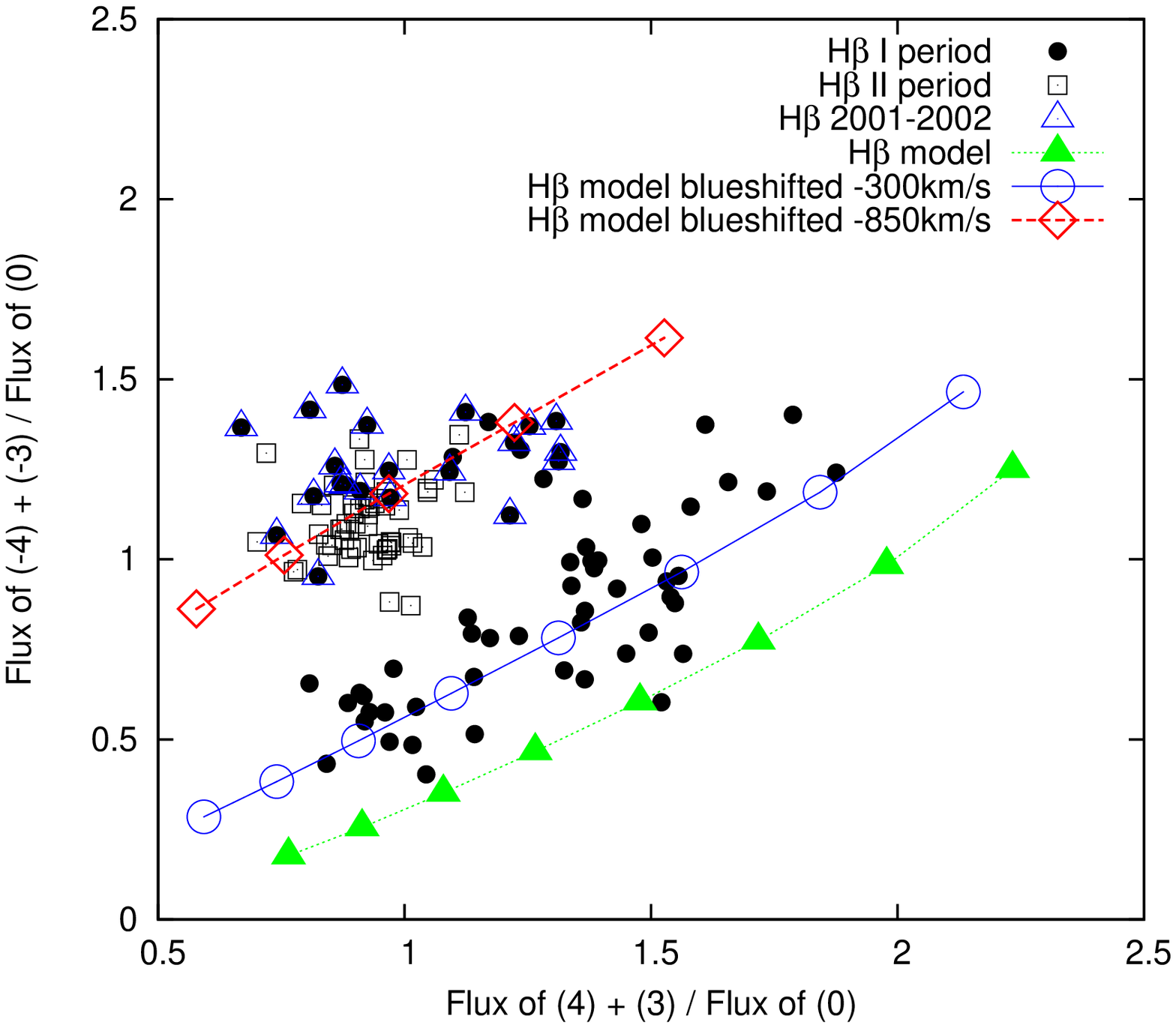}
\caption{The ratio {between} the blue-wing flux (two blue
segments, -4 and -3) and the central segment as a function of the
ratio of the red-wing (two red segments, +4 and +3) and the central
segment. Observations in Period I are denoted with full circles, in
Period II with open squares, while observations from 2001--2002 are
denoted with open triangles.}\label{f13a}
\end{figure}

We used the same models as described in \S 3.2, but now we have
measured the line-segments fluxes for the modeled lines, defined in
the same way as for the observed lines. Here we plot in Fig.
\ref{f13a} the sum of -4 and -3 segment fluxes as a function of the
sum of +4 and +3 segment fluxes normalized on the central 0 segment,
i.e. $F_{\rm blue \ wing}=[F(-4)+F(-3)]/F(0)$ vs. $F_{\rm red \
wing}=[F(+4)+F(+3)]/F(0)$.

Observations in Period I are denoted with full circles and in Period
II with open squares (see Fig. \ref{f13a}). Also we see that
observations in 2001 and 2002 are closer to the Period II, so we
denoted these with open triangles. It is interesting  that in the
first period  a good correlation between $F_{\rm blue \ wing}$ and
$F_{\rm red \ wing}$ is present. {The modeled $F_{\rm blue \
wing}$ vs. $F_{\rm red \ wing}$ relationship, marked with full
triangles, appears to follow the trend of the observed points in
Period I but is shifted below the observed correlation pattern.} To
find the causes for such disagreement, we measured the shift of the
H$\beta$ averaged profiles for two periods (given in Paper I) with
respect to the averaged profile constructed from the modeled
profiles (for different positions of the H$\beta$ emitting disk). We
measured the shift at the center of the width of the H$\beta$ line
at half maximum and at 20\% of the maximum. We found that the shift
of the modeled average profile is +900 km s$^{-1}$ at 20\% of the
maximum and +600 km s$^{-1}$ at half maximum, while the measurements
for Period I are $\sim$+700 km s$^{-1}$ at 20\% of the maximum and
$\sim$+110 km s$^{-1}$ at half maximum. In Period II the observed
averaged H$\beta$ line was more blue-shifted with respect to the
modeled one. We found $\sim$+310 km s$^{-1}$ at 20\% of the maximum
and $\sim$ -330 km s$^{-1}$ at half maximum. Therefore we simulated
the segment variation (Fig. \ref{f13a}) taking into account that the
whole disk line is blue-shifted by 300 km s$^{-1}$ (open circles in
Fig. \ref{f13a}), and 850 km s$^{-1}$ (diamonds in Fig. \ref{f13a}).
As can be seen in Fig. \ref{f13a} the modeled values well fit the
observations from Period I (line blue-shifted for 300 km s$^{-1}$)
and Period II (line blue-shifted for 800 km s$^{-1}$). {It is
interesting that in Period I there are changes in the location of
disk regions responsible for the lines emission (from R$_{\rm
inn}=250$ R$_{\rm g}$ to R$_{\rm inn}=550$ R$_{\rm g}$), while in
Period II, the inner disk radius appears to have been changed by a
smaller amount (from 350 R$_{\rm g}$ to 450 R$_{\rm g}$).} This
indicates that in the first period (excluding 2001--2002) the
variability can be well explained by variation of the disk position
with respect to the black hole, while in the Period II and
2001--2002 (when outburst is starting), probably the disk position
does not change so much. It seems that in this case we have that the
part of the disk emitting the broad lines has only become brighter.

\subsection{2-D CCF of the H$\beta$ line profiles}

We now investigate in detail the cross correlation function (CCF) of
the line-profile variations of H$\beta$. We proceed in the same way
as it was studied the line-profile variations in Mrk 110
\citep{ko02,ko03}.

\begin{figure}
\centering
\includegraphics[width=56mm,angle=-90]{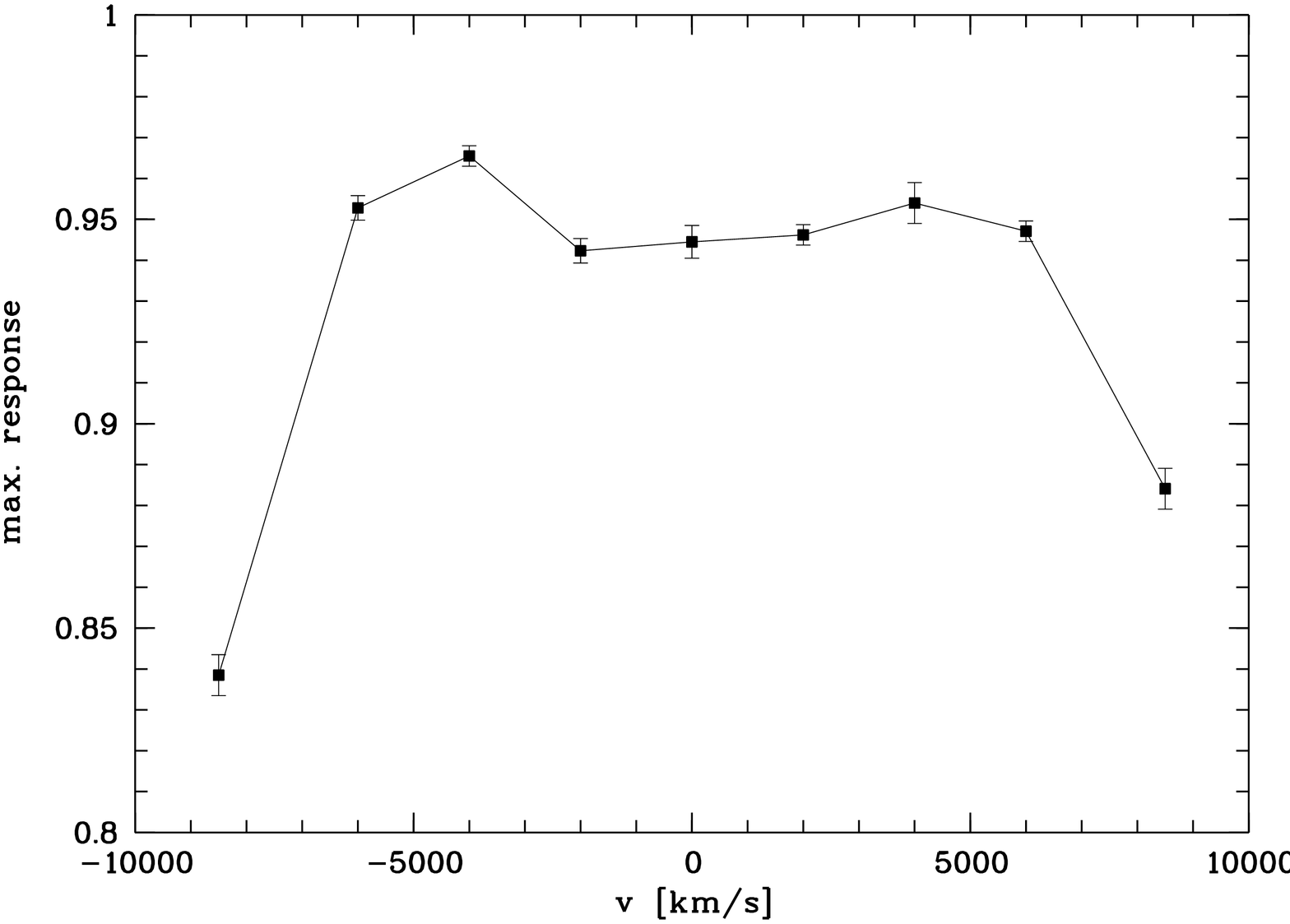}
\caption{Maximum response of the correlation functions of the H$\beta$ line
segment light curves with the continuum light curve. }\label{wk1}
\end{figure}

\begin{figure}
\centering
\includegraphics[width=63mm,angle=270]{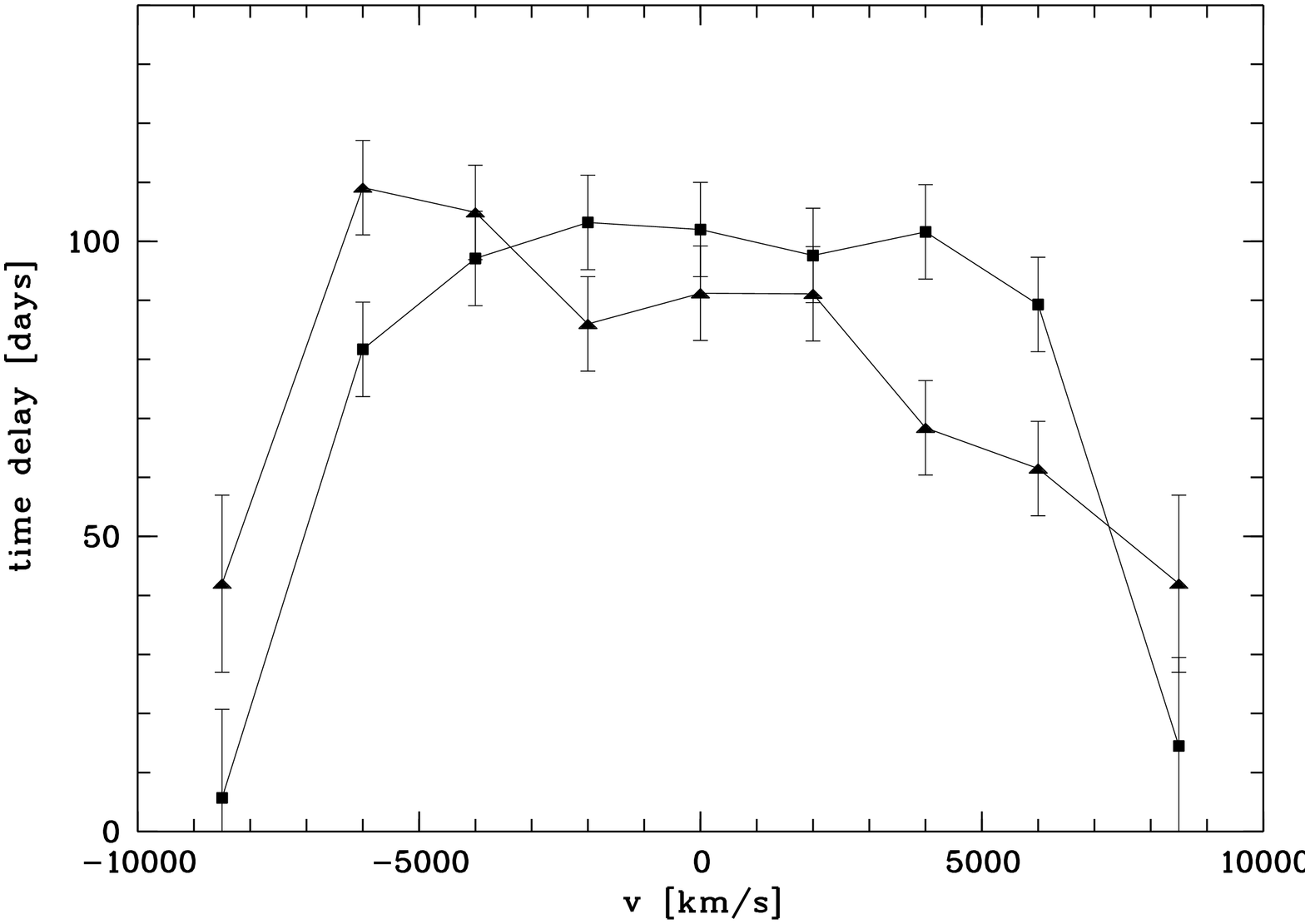}
  \caption{Time delay of the individual line segments of H$\beta$ for the
 first half of the observing period from 1995--2002 (filled squares) and for
  the second half from 2003--2007 (filled triangles) with respect to the
  continuum light curve in velocity space.}
  \label{wk2}
\end{figure}

{Fig. \ref{wk1} shows the maximum response of the correlation
functions of the H$\beta$ line segment light curves with the
continuum light curve and Fig. \ref{wk2} shows} the delays of the
individual H$\beta$ line segments in velocity space. Fig. \ref{wk2}
gives the delays of the individual line segments of H$\beta$
separately for the two observing periods: from 1995--2002 (filled
squares) and from 2003--2007 (filled triangles).

{The outer blue and red line wings show the shortest delay
clearly indicating that the line emitting region is connected with
an accretion disk \citep[e.g.][]{w01,h04}. A careful inspection of
the delays of the individual line segments in Fig. \ref{wk2}
indicates that a second trend is superimposed:  An earlier response
of the red line wing compared to the blue line wing is seen in Fig.
\ref{wk2} for the second half of the observing period. This behavior
is consistent with the prediction from disk-wind models
\citep{kk94}.}

It is intriguing that the very-broad line AGN 3C 390.3  shows the
same pattern in the velocity-delay maps as the narrow-line AGN Mrk
110 \citep[see][]{ko02,ko03}. {The outer red and blue wings
respond much faster to continuum variations than the central
regions.}

\begin{figure}
\centering
\includegraphics[width=8.5cm]{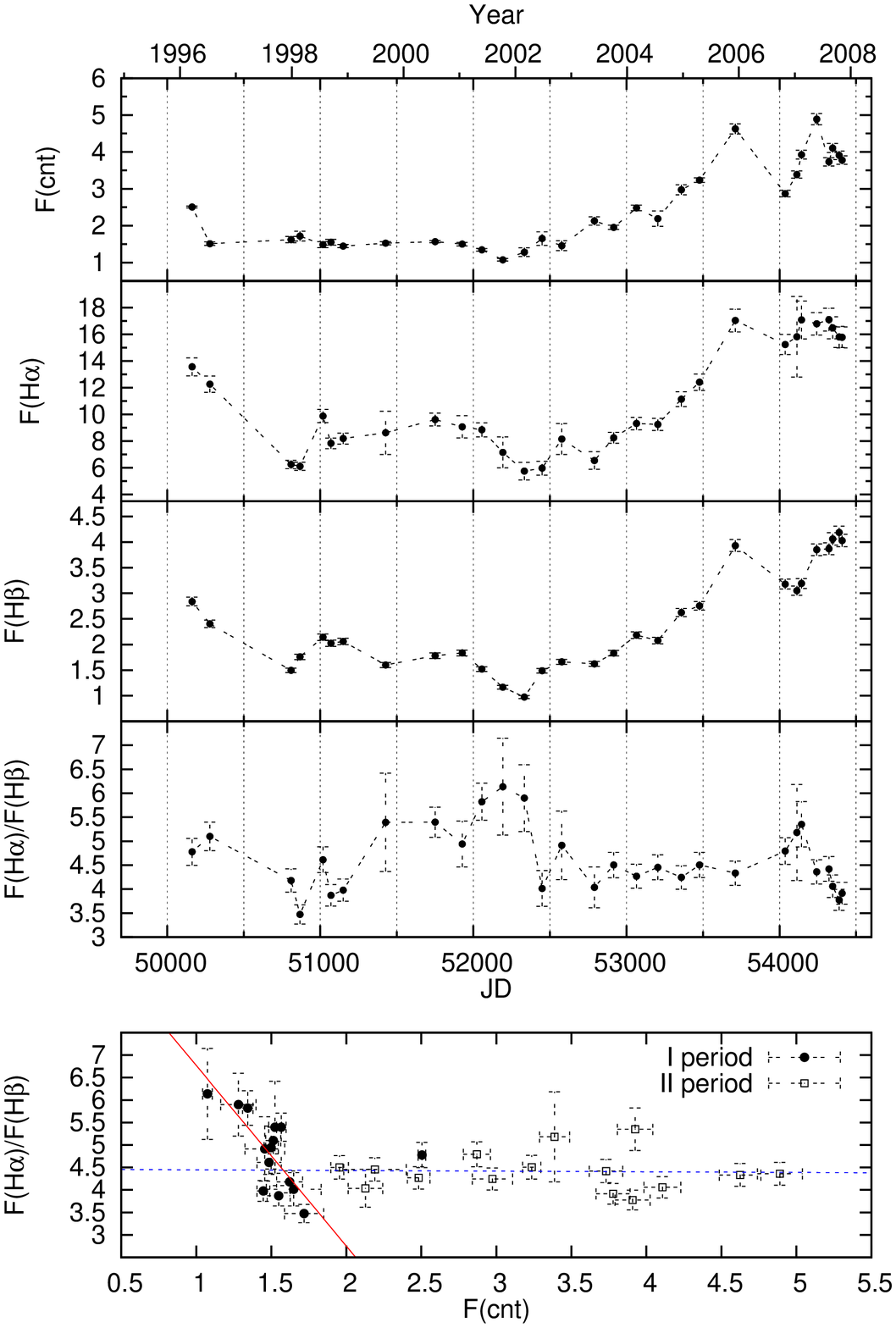}
\caption{Variations in the continuum flux at 5100 \AA\, (first
panel), in H$\alpha$ and H$\beta$ line fluxes (second and third
panel, respectively), and in the integral Balmer decrement
BD=$F(H\alpha)/F(H\beta)$ (fourth panel) of the month-averaged
profiles. The abscissa (OX) gives the Modified Julian date (bottom)
and the corresponding year (top). The continuum flux is in units
$10^{-15}\,{\rm erg \, m^{-2}\,s^{-1}\, \AA^{-1}}$ and line flux is
in $10^{-13}\,{\rm erg \, m^{-2}\,s^{-1}}$. The bottom panel gives
the BD as a function of the continuum flux, where full circles
denote the observations in Period I and open squares in Period II
(see text).}\label{f15}
\end{figure}

\section{Balmer Decrement (BD) variation}

{The ratio of H$\alpha$ and H$\beta$ depends on the physical
processes in the BLR and variations in the Balmer decrement (BD)
during monitoring time can indicate changes in physical properties
of the BLR.} Using the month-averaged profiles of the broad
H$\alpha$ and H$\beta$ lines, we determined their integrated fluxes
in the range between -10000 and +10000 ${\rm km~s^{-1}}$ in radial
velocity, i.e. the integrated flux ratio $F(H\alpha)/F(H\beta)$ or
integrated BD. In Fig. \ref{f15} we plot the time variation of the
continuum flux, H$\alpha$ and H$\beta$ line fluxes and the BD. The
BD reached its maximum in 2002 when the fluxes of lines and
continuum were in the minimum. As one can see from Fig. \ref{f15}
there is no correlation between BD and line/continuum variation.
Only, as it can be seen in the very bottom panel in Fig. \ref{f15},
which shows separate BDs for two periods (full circles represent
Period I, and open squares Period II), there is some tendency that
the BD is higher for weaker continuum flux. {When BD is plotted
versus the continuum flux (Fig.\ref{f15} bottom panel) a negative
trend seems to be present below 1.75 $\times 10^{-15}\rm erg\
cm^{-2}s^{-1}\AA^{-1}$. For higher values of the flux BD stays
roughly constant. Note that a similar BD behavior was observed by
\citet{se10}.}

\begin{figure*}
\centering
\includegraphics[width=15cm]{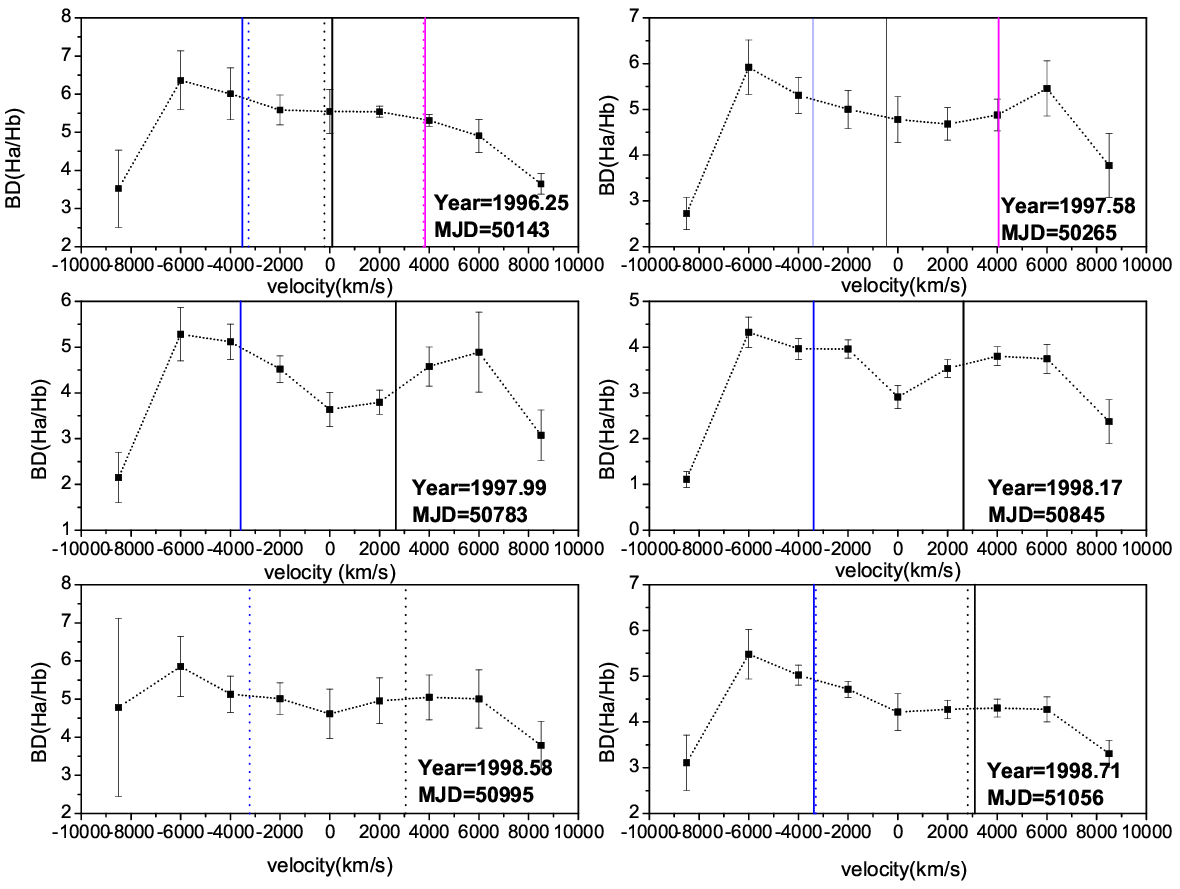}
\includegraphics[width=15cm]{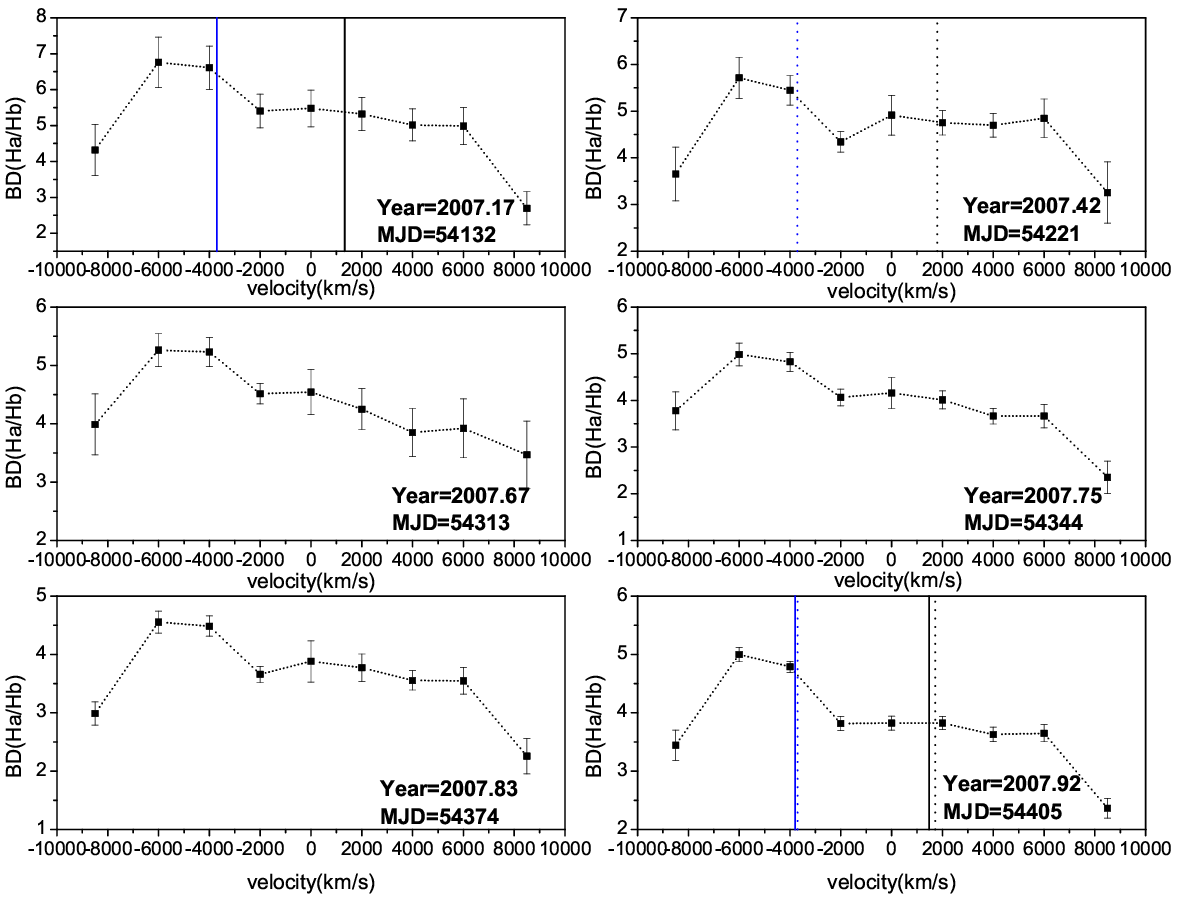}
\caption{Variations in the Balmer decrement (BD=H$\alpha$/H$\beta$)
         as a function of the radial velocity for month-averaged spectra
         of 3C 390.3. Upper panel represents BD from the period of 1995--1998,
         and bottom panel from 2007. {These two panels are chosen since they
         represent the characteristic behavior of BD vs. velocity}. The abscissa (OX)
         indicates the radial velocities
         relative to the narrow components. The position of the blue, central and
         red peaks are marked with blue, black and pink line, respectively for
         H$\alpha$ (solid line) and H$\beta$ (dashed line).
         {The rest of the panels are available electronically only.}}\label{f16}
\end{figure*}

\onlfig{17}{\begin{figure*} \centering
\includegraphics[width=16cm]{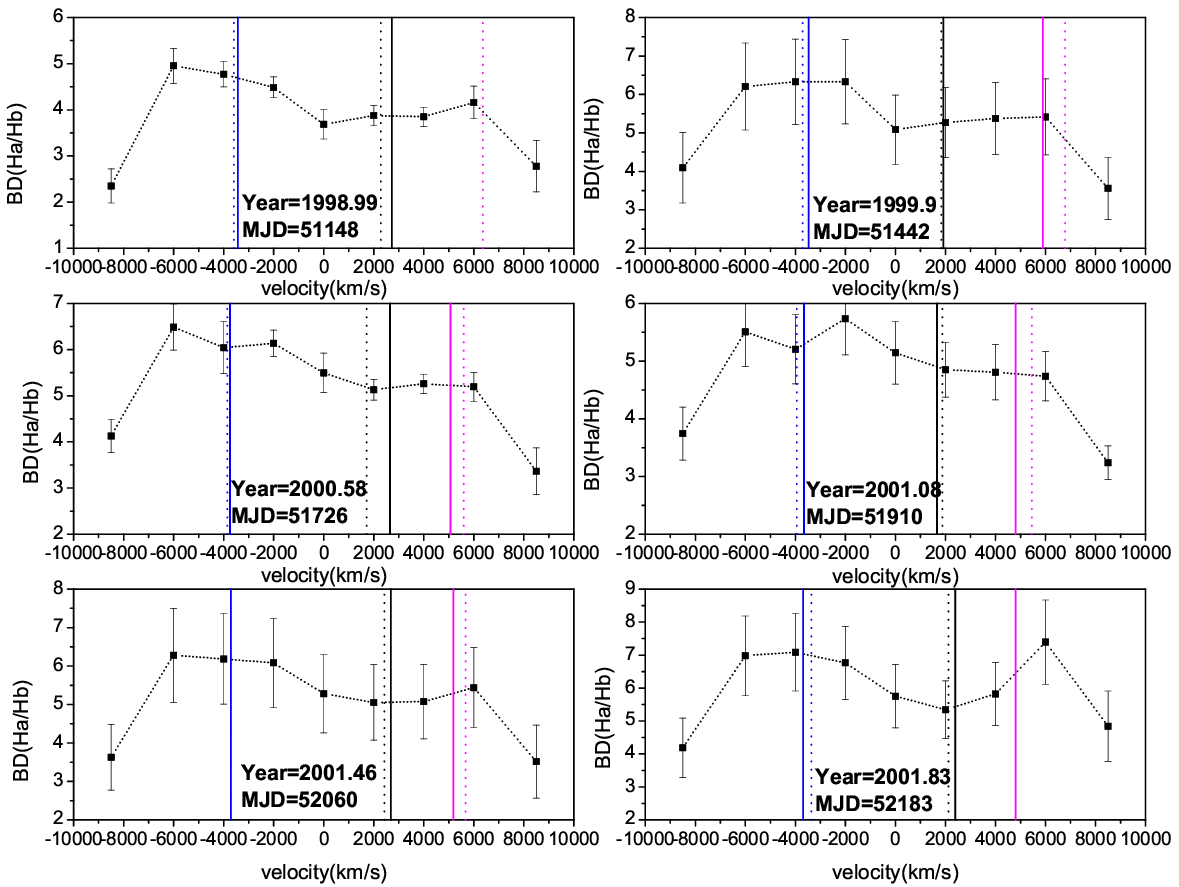}
\caption{Continued.}\label{f16}
\end{figure*}}

\onlfig{17}{\begin{figure*} \centering
\includegraphics[width=16cm]{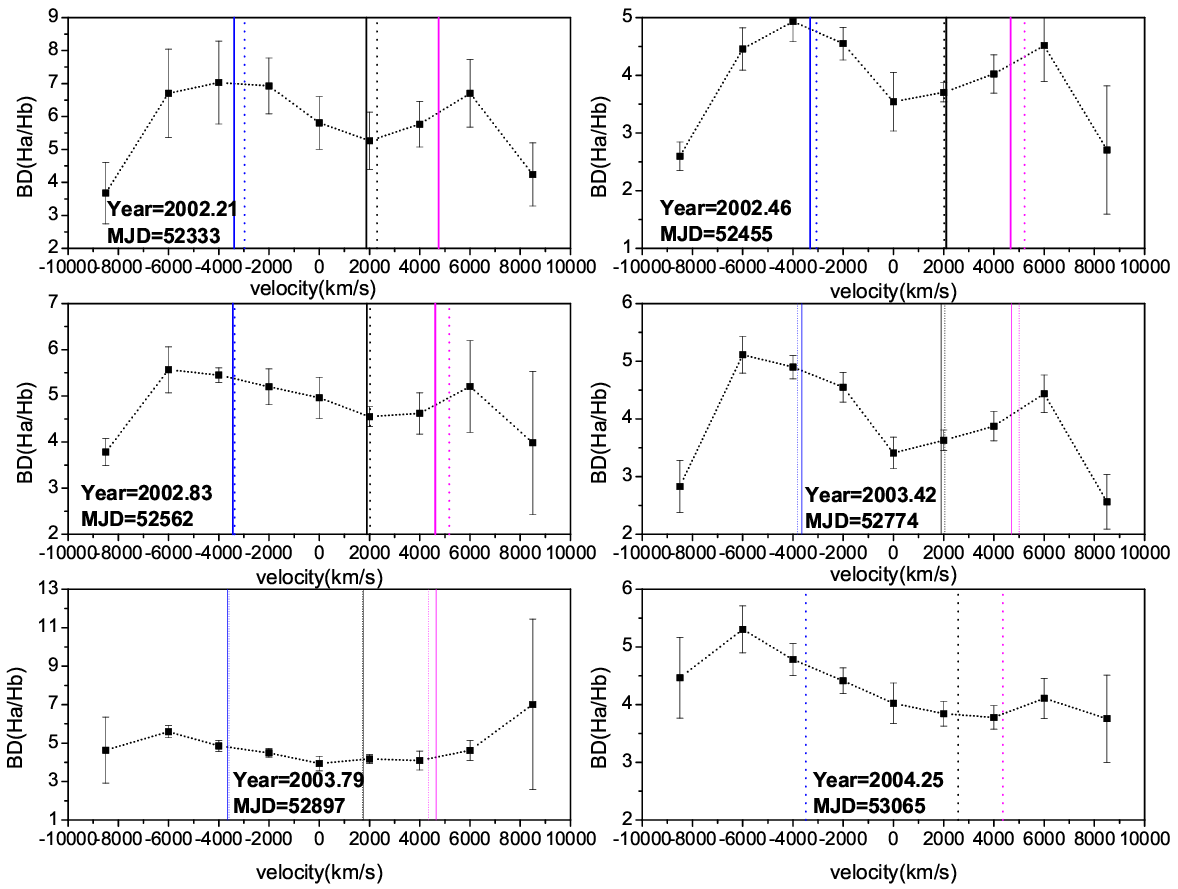}
\caption{Continued.}\label{f16}
\end{figure*}}

\onlfig{17}{\begin{figure*} \centering
\includegraphics[width=16cm]{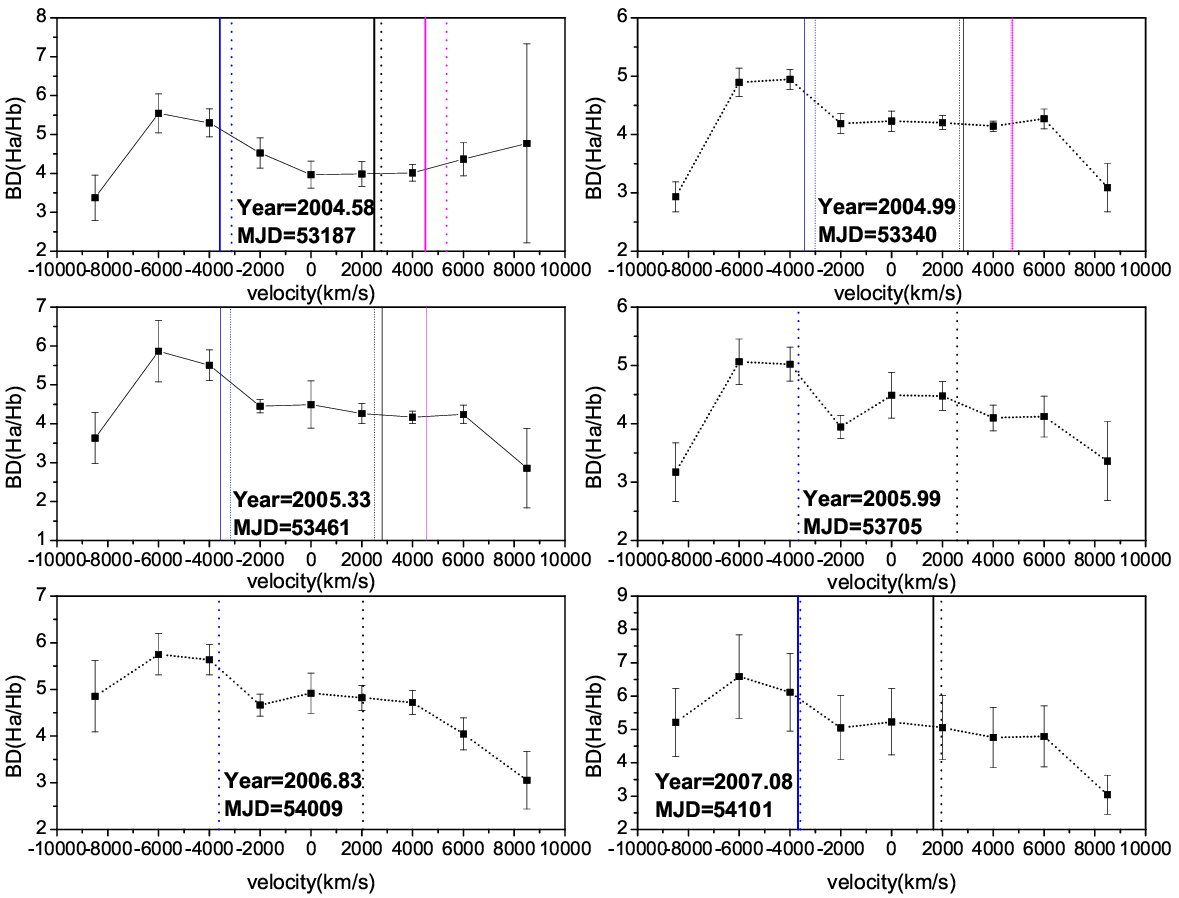}
\caption{Continued.}\label{f16}
\end{figure*}}

\subsection{BD as function of the velocity along line profile}

Furthermore, we investigate BDs for each line segment (as it was
described in \S 4), and plots of BD against velocity are present in
Fig. \ref{f16} (note here that a part of this Figure is available
electronically only). {Fig. \ref{f16} shows some common features
in the trends of the different line segments BD plotted vs. velocity
over an extended period of time ranging from 1995 to 2007. For
example, for almost all observations there is a minimum} in the far
blue and red wings (around $\pm$9000 km s$^{-1}$) as well as two
maxima in red and blue part (around  $\pm$6000 km s$^{-1}$). There
is also a central minimum (around 0 km s$^{-1}$) that in 2001 and
2002 is moving to +2000 km s$^{-1}$, while in 2005 and 2006 is
moving to -2000 km s$^{-1}$. It is interesting that from 2007 to the
end of the monitoring period only a peak in the blue part (between
-6000 to -4000  km s$^{-1}$) is dominant, while BD from -2000 to
+6000 km s$^{-1}$ tends to be constant having values between 3.5 and
4.5.

\onltab{3}{
\begin{table*}
\begin{center}
\caption[]{The errors of measurements ($e\pm\sigma$) for line
segments of H$\alpha$ and H$\beta$ given in percent. For each
segment the mean year flux is given in units
$10^{-15}$\,erg\,cm$^{-2}$\,s$^{-1}$\,\AA$^{-1}$
 for H$\beta$ and $10^{-14}$\,erg\,cm$^{-2}$\,s$^{-1}$\,\AA$^{-1}$ for H$\alpha$.}\label{t03}
\begin{tabular}{ccccccccccccc}

\hline
\hline

Year & \multicolumn{2}{c}{Ha(-4)}  &  \multicolumn{2}{c}{Ha(+4)} & \multicolumn{2}{c}{Ha(-3)} & \multicolumn{2}{c}{ Ha(+3)} \\
     &Flux&$(e\pm\sigma)$&Flux&($e\pm\sigma$)&Flux&($e\pm\sigma$)&Flux&($e\pm\sigma$)\\
\hline
1998    &   2.15&   11.17$\pm$1.34& 6.46&   13.26$\pm$5.49& 9.56&   7.48$\pm$2.51&  8.35&   8.61$\pm$5.87\\
2001    &   4.18&   5.04$\pm$3.26&  2.71&   2.89$\pm$1.54&  10.48&  2.00$\pm$2.30&  9.73&   1.62$\pm$0.03\\
2002    &   3.32&   3.29$\pm$2.90&  1.76&   27.65$\pm$10.92&7.24&   3.88$\pm$1.97&  6.25&   5.27$\pm$4.61\\
2004    &   4.03&   5.67$\pm$0.65&  2.71&   12.70$\pm$1.87& 12.47&  3.25$\pm$0.66&  9.16&   5.22$\pm$2.65\\
2007    &   7.21&   5.93$\pm$4.77&  5.56&   10.41$\pm$3.20& 20.95&  3.21$\pm$2.18&  12.00&  6.21$\pm$4.68\\
\hline
mean   &           &6.22$\pm$2.96&      &   13.38$\pm$8.99&      &  3.96$\pm$2.08&  &   5.39$\pm$2.51\\

&&&&&&\\
 \hline

 Year  & \multicolumn{2}{c}{ Hb(-4)}  & \multicolumn{2}{c}{Hb(+4)}  &  \multicolumn{2}{c}{Hb(-3)}  & \multicolumn{2}{c}{Hb(+3)} \\
       &Flux&($e\pm\sigma$)&Flux&($e\pm\sigma$)&Flux&($e\pm\sigma$)&Flux&($e\pm\sigma$)\\
\hline

1996&   9.24&   18.78$\pm$12.26&9.23&   8.56$\pm$3.46&  18.57&  9.51$\pm$1.05&  23.03&  9.47$\pm$0.53 \\
1998&   6.72&   35.15$\pm$21.79&20.29&  4.79$\pm$0.14&  18.21&  8.16$\pm$3.46&  19.20&  7.76$\pm$2.90 \\
2001&   10.33&  11.11$\pm$3.97& 7.57&   9.72$\pm$13.35& 16.38&  5.09$\pm$1.29&  17.68&  1.36$\pm$1.41 \\
2002&   9.94&   12.98$\pm$8.09& 4.94&   19.88$\pm$9.73& 13.38&  10.40$\pm$4.84& 12.09&  12.82$\pm$3.51 \\
2003&   8.66&   16.30$\pm$2.26& 3.97&   38.00$\pm$35.66&16.55&  5.24$\pm$0.58&  15.01&  5.88$\pm$1.27 \\
2004&   12.87&  11.33$\pm$7.25& 7.46&   29.39$\pm$31.57&24.06&  5.94$\pm$3.74&  21.25&  4.41$\pm$3.17 \\
2007&   17.70&  6.29$\pm$2.71&  18.76&  4.29$\pm$3.95&  39.19&  1.81$\pm$0.79&  28.94&  3.59$\pm$1.69 \\

\hline
mean&        &  15.99$\pm$9.34&      &  16.37$\pm$13.13&     &   6.59$\pm$2.97&      &  6.47$\pm$3.87&      \\

&&&&&&\\

\hline
Year  & \multicolumn{2}{c}{ Ha(-2)}  & \multicolumn{2}{c}{Ha(+2)} &\multicolumn{2}{c}{Ha(-1)} & \multicolumn{2}{c}{Ha(+1)} & \multicolumn{2}{c}{ Ha(0)}\\
       &Flux&($e\pm\sigma$)&Flux&($e\pm\sigma$)&Flux&($e\pm\sigma$)&Flux&($e\pm\sigma$)&Flux&($e\pm\sigma$)\\
\hline

1998 &  16.61 & 6.14$\pm$2.69&  10.09 & 6.32$\pm$4.90&  14.67& 5.61$\pm$2.20&   10.19&  8.03$\pm$4.36&  10.69&  11.76$\pm$2.78\\
2001 &  17.89 & 3.95$\pm$1.75&  11.07 & 1.79$\pm$2.51&  14.86& 1.90$\pm$1.24&   11.85&  2.19$\pm$1.79&  12.13&  8.61$\pm$2.63\\
2002 &  12.18 & 3.03$\pm$1.24&  8.07 &  2.36$\pm$2.37&  11.08& 3.51$\pm$2.66&   9.32&   2.12$\pm$1.16&  8.94&   6.21$\pm$4.95\\
2004 &  20.39 & 3.74$\pm$0.56&  12.21 & 2.59$\pm$1.01&  16.55& 5.79$\pm$2.84&   12.47&  4.71$\pm$3.17&  12.51&  5.39$\pm$2.43\\
2007 &  34.96 & 2.17$\pm$2.34&  18.78 & 5.48$\pm$4.22&  24.08& 1.78$\pm$1.85&   23.40&  4.66$\pm$3.02&  23.51&  4.11$\pm$3.25\\
\hline
mean &     &   3.80$\pm$1.48&    &  3.71$\pm$2.04&        &3.72$\pm$1.94&          &4.34$\pm$2.42& &    7.22$\pm$3.02\\

&&&&&&\\
\hline
 Year  & \multicolumn{2}{c}{ Hb(-2)}   & \multicolumn{2}{c}{ Hb(+2)} & \multicolumn{2}{c}{ Hb(-1)}  & \multicolumn{2}{c}{Hb(+1)}  &  \multicolumn{2}{c}{Hb(0)}\\
       &Flux&($e\pm\sigma$)&Flux&($e\pm\sigma$)&Flux&($e\pm\sigma$)&Flux&($e\pm\sigma$)&Flux&($e\pm\sigma$)\\
\hline

1996& 39.53&    7.63$\pm$2.38&   33.87&  3.87$\pm$2.51 &39.70&  5.42$\pm$2.29&  34.28&  4.27$\pm$2.04&  41.34&  6.48$\pm$0.50\\
1998& 35.39&    3.93$\pm$2.28&   23.30&  5.86$\pm$2.25 &32.53&  2.88$\pm$2.99&  24.02&  4.15$\pm$3.05&  26.53&  5.28$\pm$0.93\\
2001& 28.85&    3.75$\pm$4.86&  20.58&   2.86$\pm$1.89 &23.18&  3.68$\pm$3.51&  21.78&  3.88$\pm$0.005& 20.54&  4.33$\pm$1.56\\
2002& 21.70&    7.15$\pm$5.92&  17.42&  5.60$\pm$ 4.49 &20.89&  5.03$\pm$2.52&  20.94&  6.28$\pm$4.36&  18.96&  7.55$\pm$0.75\\
2003& 30.27&    2.63$\pm$1.54&  22.14&   6.57$\pm$1.64 &26.85&  4.50$\pm$0.31&  23.96&  2.07$\pm$0.44&  24.05&  4.96$\pm$2.26\\
2004& 39.90&    3.24$\pm$3.11&  29.87&   2.53$\pm$2.26 &38.13&  2.59$\pm$1.31&  30.37&  2.66$\pm$1.70&  30.41&  3.44$\pm$2.36\\
2007& 67.48&    1.16$\pm$0.69&   47.00& 1.86$\pm$1.16  &55.00&  1.35$\pm$1.24&  54.65&  1.96$\pm$1.48&  52.58&  3.03$\pm$1.83\\
\hline
mean&          &  4.21$\pm$2.36 &       &  4.16$\pm$1.85&   &   3.64$\pm$1.46& &    3.61$\pm$1.52&  &   5.01$\pm$1.61\\
&&&&&&\\
\hline
\hline
\end{tabular}
\end{center}
\end{table*}
}

\onltab{4}{
\begin{table*}
\begin{center}
\caption{The line-segment fluxes for H$\alpha$ line in units $10^{-14}$\,erg\,cm$^{-2}$\,s$^{-1}$\,\AA$^{-1}$. } \label{t04}
\begin{tabular}{rcccccccccc}

\hline
\hline
    &   MJD      &   Ha-4  &  Ha-3  &   Ha-2  &   Ha-1 & Ha0     & Ha1   &  Ha2  &Ha3   &  Ha4   \\
    & (2400000+) &  seg-4  &  seg-3 &   seg-2 &  seg-1 & seg0 0  & seg1  & seg2  & seg3 &  seg4  \\
 1  &    2       &     3   &     4  &    5    &  6    &    7  &    8  &    9     &  10  &   11   \\
\hline
 1&  +50162.6   & 3.98 & 11.03 & 21.11 & 19.77 & 23.32 & 18.50 & 17.36 & 11.93 & 7.63  \\
 2&  +50163.6   & 3.03 & 12.39 & 23.66 & 21.75 & 20.34 & 18.57 & 17.95 & 11.68 & 7.02  \\
 3&  +50276.6   & 2.05 & 10.60 & 21.35 & 20.26 & 19.82 & 15.68 & 15.92 & 10.99 & 6.11  \\
 4&  +50811.2   & 1.42 & 6.05  & 12.17 & 10.89 & 7.34  & 7.20  & 6.90  & 5.72  & 4.55  \\
 5&  +50867.6   & 0.64 & 5.99  & 12.00 & 11.20 & 6.91  & 7.42  & 7.06  & 5.67  & 4.06  \\
 6&  +51010.7   & 2.48 & 11.24 & 19.12 & 17.05 & 13.04 & 12.12 & 12.17 & 10.64 & 8.41  \\
 7&  +51019.7   & 2.15 & 9.86  & 17.06 & 15.41 & 10.74 & 10.36 & 10.59 & 8.88  & 6.59  \\
 8&  +51081.4   & 1.73 & 8.11  & 14.55 & 12.75 & 8.43  & 8.62  & 8.51  & 6.59  & 4.85  \\
 9&  +51082.4   & 2.21 & 8.54  & 15.02 & 12.84 & 10.13 & 9.31  & 8.93  & 7.05  & 5.58  \\
10&  +51083.4   & 2.00 & 9.09  & 15.82 & 13.71 & 9.92  & 9.47  & 8.96  & 7.18  & 5.83  \\
11&  +51149.6   & 2.00 & 9.14  & 15.88 & 13.82 & 10.00 & 9.50  & 8.97  & 7.19  & 5.59  \\
12&  +51427.3   & 3.58 & 10.81 & 18.27 & 16.27 & 11.70 & 11.36 & 9.92  & 9.53  & 6.62  \\
13&  +51756.3   & 4.07 & 10.62 & 18.06 & 15.31 & 12.87 & 11.94 & 10.59 & 9.57  & 4.09  \\
14&  +51925.6   & 4.47 & 10.82 & 17.18 & 15.30 & 11.84 & 11.99 & 11.29 & 10.00 & 2.88  \\
15&  +51929.6   & 4.30 & 10.77 & 18.49 & 15.08 & 13.03 & 12.15 & 11.28 & 10.23 & 2.96  \\
16&  +52034.7   & 4.09 & 10.44 & 18.09 & 14.61 & 12.89 & 12.07 & 11.30 & 9.46  & 2.58  \\
17&  +52043.9   & 3.74 & 9.76  & 17.33 & 13.99 & 11.56 & 11.22 & 10.50 & 9.18  & 2.42  \\
18&  +52074.9   & 4.35 & 10.51 & 18.49 & 14.54 & 12.66 & 11.85 & 11.08 & 9.49  & 2.59  \\
19&  +52075.9   & 3.75 & 9.95  & 17.85 & 14.96 & 10.18 & 11.36 & 10.57 & 9.26  & 2.40  \\
20&  +52192.6   & 3.95 & 9.05  & 15.34 & 12.89 & 10.17 & 10.22 & 9.05  & 7.45  & 2.29  \\
21&  +52327.2   & 3.32 & 6.42  & 10.50 & 9.42  & 8.68  & 8.48  & 6.89  & 5.58  & 2.19  \\
22&  +52339.0   & 3.62 & 6.85  & 11.17 & 10.29 & 8.94  & 8.30  & 6.92  & 5.61  & 1.68  \\
23&  +52430.0   & 2.96 & 6.29  & 10.80 & 9.85  & 8.40  & 8.34  & 7.15  & 5.25  & 0.80  \\
24&  +52450.4   & 3.06 & 6.20  & 10.40 & 10.06 & 6.83  & 8.14  & 7.37  & 5.86  & 1.91  \\
25&  +52469.4   & 2.85 & 6.41  & 10.76 & 10.52 & 6.98  & 8.32  & 7.40  & 5.76  & 1.56  \\
26&  +52562.3   & 3.52 & 8.46  & 14.76 & 13.34 & 10.97 & 11.59 & 10.34 & 8.03  & 2.24  \\
27&  +52591.7   & 3.54 & 9.12  & 15.31 & 13.17 & 10.25 & 11.04 & 9.65  & 7.01  & 1.57  \\
28&  +52770.2   & 2.84 & 7.70  & 12.55 & 10.85 & 7.62  & 8.55  & 7.97  & 6.13  & 1.41  \\
29&  +52902.8   & 4.19 & 10.08 & 16.43 & 13.27 & 9.41  & 9.86  & 9.09  & 7.04  & 1.81  \\
30&  +52933.7   & 2.63 & 10.13 & 17.53 & 13.58 & 10.44 & 10.61 & 10.36 & 7.96  & 1.77  \\
31&  +53066.4   & 3.34 & 10.89 & 19.68 & 14.94 & 10.90 & 11.29 & 11.29 & 8.83  & 2.47  \\
32&  +53170.0   & 3.63 & 11.63 & 19.05 & 14.67 & 10.27 & 10.38 & 10.32 & 7.47  & 1.99  \\
33&  +53237.8   & 3.90 & 12.09 & 20.20 & 16.38 & 11.35 & 11.45 & 10.81 & 8.26  & 2.43  \\
34&  +53359.1   & 4.10 & 12.74 & 20.65 & 17.10 & 13.83 & 13.78 & 13.68 & 10.20 & 2.95  \\
35&  +53360.2   & 4.47 & 13.43 & 21.64 & 18.04 & 14.57 & 14.27 & 14.04 & 10.70 & 3.46  \\
36&  +53477.0   & 4.42 & 15.04 & 24.99 & 19.71 & 14.23 & 15.06 & 15.50 & 11.38 & 2.85  \\
37&  +53479.0   & 4.81 & 15.57 & 25.39 & 19.36 & 16.95 & 16.11 & 16.20 & 11.98 & 2.68  \\
38&  +53711.4   & 6.21 & 20.41 & 35.49 & 23.70 & 23.58 & 22.77 & 20.41 & 14.37 & 4.34  \\
39&  +54036.6   & 6.02 & 18.83 & 32.61 & 22.55 & 21.10 & 21.08 & 17.27 & 10.86 & 3.31  \\
40&  +54112.6   & 8.69 & 22.24 & 36.02 & 25.60 & 26.00 & 25.16 & 20.57 & 13.71 & 6.22  \\
41&  +54113.7   & 7.89 & 21.13 & 35.09 & 25.44 & 25.03 & 24.06 & 19.71 & 12.83 & 5.30  \\
42&  +54143.6   & 7.90 & 21.40 & 35.26 & 26.13 & 26.51 & 25.52 & 20.31 & 13.30 & 5.43  \\
43&  +54242.9   & 7.39 & 20.17 & 33.68 & 22.36 & 23.26 & 22.97 & 20.09 & 13.09 & 5.48  \\
44&  +54321.8   & 6.48 & 20.23 & 34.70 & 23.62 & 21.91 & 22.12 & 17.50 & 11.27 & 5.63  \\
45&  +54323.8   & 7.49 & 21.76 & 37.07 & 24.96 & 24.48 & 24.82 & 20.26 & 13.25 & 6.78  \\
46&  +54346.7   & 7.01 & 21.30 & 35.34 & 23.07 & 23.01 & 23.01 & 18.10 & 10.99 & 4.53  \\
47&  +54389.7   & 6.04 & 20.02 & 33.25 & 22.08 & 22.12 & 22.42 & 17.98 & 10.51 & 4.52  \\
48&  +54408.6   & 6.30 & 20.03 & 33.43 & 22.59 & 22.11 & 22.54 & 17.70 & 10.67 & 4.50  \\
49&  +54412.7   & 6.37 & 20.27 & 33.43 & 22.28 & 21.54 & 21.70 & 16.94 & 10.30 & 4.96  \\
\hline
\end{tabular}
\end{center}
\end{table*}
}

\onllongtab{5}{
\begin{longtable}{rcccccccccc}
\caption[]{\label{t05} The line-segment fluxes for H$\beta$ line
in units $10^{-14}$\,erg\,cm$^{-2}$\,s$^{-1}$\,\AA$^{-1}$.}\\
\hline \hline
  &   MJD      &   Hb-4    &  Hb-3  &   Hb-2  &   Hb-1 & Hb0     & Hb1   &  Hb2  &Hb3   &  Hb4         \\
    & (2400000+) &  seg-4  &  seg-3 &   seg-2 &  seg-1 & seg0    & seg1  & seg2  & seg3 &  seg4   \\
\hline
 1  &    2       &     3   &     4  &     5   &    6   &    7    &    8  &    9  &  10  &   11         \\
\hline
\endfirsthead
\caption{Continued.}\\
\hline
  &    MJD      &   Hb-4    &  Hb-3  &   Hb-2  &   Hb-1 & Hb0     & Hb1   &  Hb2  &Hb3   &  Hb4         \\
    & (2400000+) &  seg-4  &  seg-3 &   seg-2 &  seg-1 & seg0    & seg1  & seg2  & seg3 &  seg4   \\
\hline
 1  &    2       &     3   &     4  &     5   &    6   &    7    &    8  &    9  &  10  &   11         \\
\hline
\endhead
\hline
\endfoot
\hline
\endlastfoot
  1& +49832.4  & 0.41   & 0.96  & 2.65  & 2.52  & 2.66  & 2.17  & 1.85  & 1.47  & 1.57    \\
  2& +49863.4 & 0.67    & 1.06  & 2.84  & 2.62  & 2.87  & 2.48  & 2.37  & 2.19  & 2.17    \\
  3& +50039.2 & 0.66    & 1.27  & 3.18  & 3.19  & 3.28  & 2.83  & 2.61  & 1.61  & 1.75    \\
  4& +50051.1 & 0.70    & 1.37  & 3.14  & 3.40  & 3.59  & 3.00  & 2.74  & 1.65  & 1.79    \\
  5& +50052.1 & 0.00    & 0.00  & 3.06  & 3.12  & 3.30  & 2.59  & 2.51  & 1.59  & 1.40    \\
  6& +50127.6 & 1.00    & 1.86  & 3.90  & 3.81  & 4.12  & 3.59  & 3.37  & 2.16  & 1.86    \\
  7& +50162.6 & 0.93    & 1.85  & 3.80  & 3.97  & 4.14  & 3.60  & 3.61  & 2.49  & 2.23    \\
  8& +50163.6 & 1.38    & 2.14  & 4.33  & 4.19  & 4.51  & 3.75  & 3.72  & 2.87  & 2.43    \\
  9& +50249.5 & 0.76    & 1.65  & 3.70  & 3.90  & 3.83  & 3.09  & 3.08  & 1.95  & 1.54    \\
 10& +50276.6 & 0.72    & 1.83  & 4.10  & 4.17  & 4.24  & 3.39  & 3.30  & 2.13  & 1.63    \\
 11& +50277.6 & 0.76    & 1.78  & 3.79  & 3.77  & 3.87  & 3.12  & 3.06  & 1.80  & 1.33    \\
 12& +50281.4 & 0.59    & 1.55  & 3.65  & 3.65  & 3.72  & 3.04  & 2.96  & 1.86  & 1.60    \\
 13& +50305.5 & 0.73    & 1.43  & 3.56  & 3.89  & 3.93  & 3.21  & 2.98  & 1.81  & 1.80    \\
 14& +50338.8 & 0.20    & 1.25  & 3.00  & 3.37  & 2.94  & 2.43  & 2.47  & 1.45  & 1.40    \\
 15& +50390.4 & 0.42    & 1.09  & 3.09  & 3.27  & 3.13  & 2.68  & 2.50  & 1.60  & 1.58    \\
 16& +50392.6 & 0.17    & 1.10  & 3.08  & 3.29  & 2.93  & 2.46  & 2.38  & 1.36  & 1.11    \\
 17& +50511.6 & 0.13    & 0.68  & 2.13  & 2.30  & 2.01  & 1.70  & 1.42  & 0.99  & 1.11    \\
 18& +50599.4 & 0.54    & 0.92  & 2.20  & 2.47  & 2.34  & 1.74  & 1.24  & 1.02  & 1.13    \\
 19& +50656.5 & 0.45    & 0.77  & 1.67  & 1.90  & 1.84  & 1.66  & 1.31  & 1.02  & 1.49    \\
 20& +50691.5 & 0.62    & 0.79  & 1.74  & 1.86  & 1.79  & 1.63  & 1.27  & 0.91  & 1.29    \\
 21& +50701.6 & 0.42    & 0.65  & 1.62  & 1.70  & 1.55  & 1.48  & 1.13  & 0.82  & 1.23    \\
 22& +50808.6 & 0.55    & 1.11  & 2.30  & 2.44  & 1.94  & 1.83  & 1.40  & 1.13  & 1.52    \\
 23& +50810.2 & 0.88    & 1.30  & 2.60  & 2.56  & 2.23  & 2.04  & 1.69  & 1.44  & 1.65    \\
 24& +50811.2 & 0.66    & 1.14  & 2.35  & 2.33  & 1.95  & 1.83  & 1.49  & 1.13  & 1.48    \\
 25& +50813.2 & 0.52    & 1.02  & 2.25  & 2.28  & 1.94  & 1.86  & 1.43  & 0.96  & 1.24    \\
 26& +50835.6 & 0.46    & 1.06  & 2.46  & 2.49  & 2.05  & 1.96  & 1.56  & 1.31  & 1.66    \\
 27& +50867.6 & 0.57    & 1.38  & 3.03  & 2.83  & 2.37  & 2.10  & 1.86  & 1.52  & 1.71    \\
 28& +50940.4 & 0.73    & 1.87  & 3.55  & 3.51  & 2.82  & 2.52  & 2.37  & 1.91  & 2.13    \\
 29& +50990.3 & 0.51    & 1.85  & 3.61  & 3.43  & 2.68  & 2.49  & 2.39  & 1.94  & 2.21    \\
 30& +51010.7 & 0.69    & 2.01  & 3.83  & 3.52  & 2.83  & 2.47  & 2.51  & 2.20  & 2.21    \\
 31& +51019.7 & 0.77    & 2.08  & 3.85  & 3.53  & 2.83  & 2.53  & 2.48  & 2.11  & 2.15    \\
 32& +51023.5 & 0.26    & 1.76  & 3.73  & 3.41  & 2.74  & 2.35  & 2.37  & 2.14  & 2.14    \\
 33& +51025.4 & 0.32    & 1.66  & 3.41  & 3.16  & 2.49  & 2.19  & 2.12  & 1.75  & 1.97    \\
 34& +51055.6 & 0.59    & 1.63  & 3.27  & 3.11  & 2.48  & 2.41  & 2.35  & 1.93  & 1.90    \\
 35& +51076.0 & 0.70    & 1.55  & 3.34  & 3.11  & 2.41  & 2.50  & 2.37  & 1.67  & 2.02    \\
 36& +51081.4 & 0.91    & 1.75  & 3.40  & 3.08  & 2.58  & 2.36  & 2.16  & 1.68  & 1.85    \\
 37& +51082.4 & 0.96    & 1.81  & 3.46  & 3.09  & 2.52  & 2.42  & 2.27  & 1.77  & 1.96    \\
 38& +51083.4 & 0.88    & 1.86  & 3.37  & 3.13  & 2.75  & 2.48  & 2.38  & 1.78  & 2.06    \\
 39& +51112.3 & 0.78    & 1.76  & 3.65  & 3.28  & 3.03  & 2.45  & 1.98  & 1.74  & 1.67    \\
 40& +51149.6 & 0.85    & 1.85  & 3.33  & 3.08  & 2.71  & 2.45  & 2.33  & 1.73  & 2.01    \\
 41& +51372.5 & 0.78    & 1.20  & 2.08  & 1.99  & 1.62  & 1.54  & 1.29  & 1.16  & 1.53    \\
 42& +51410.3 & 0.73    & 1.49  & 2.55  & 2.21  & 1.87  & 1.83  & 1.50  & 1.52  & 1.72    \\
 43& +51425.9 & 0.77    & 1.54  & 2.55  & 2.27  & 2.02  & 1.90  & 1.63  & 1.55  & 1.64    \\
 44& +51454.7 & 1.03    & 2.01  & 3.06  & 2.58  & 2.21  & 2.14  & 1.90  & 1.87  & 1.68    \\
 45& +51455.7 & 0.92    & 1.88  & 2.89  & 2.37  & 1.99  & 1.95  & 1.70  & 1.76  & 1.80    \\
 46& +51461.5 & 0.69    & 1.64  & 2.58  & 2.25  & 1.87  & 1.94  & 1.69  & 1.65  & 1.87    \\
 47& +51745.4 & 1.15    & 1.66  & 2.94  & 2.59  & 2.40  & 2.41  & 2.15  & 1.92  & 1.34    \\
 48& +51748.2 & 0.56    & 1.44  & 2.72  & 2.21  & 2.02  & 2.04  & 1.78  & 1.61  & 1.09    \\
 49& +51756.3 & 1.25    & 1.82  & 3.31  & 2.69  & 2.50  & 2.40  & 2.11  & 2.00  & 1.21    \\
 50& +51823.1 & 1.11    & 1.65  & 2.96  & 2.34  & 2.12  & 2.12  & 1.75  & 1.66  & 0.96    \\
 51& +51867.1 & 1.33    & 1.75  & 3.05  & 2.44  & 2.23  & 2.19  & 1.89  & 1.67  & 0.94    \\
 52& +51868.1 & 1.32    & 1.80  & 3.23  & 2.65  & 2.43  & 2.37  & 2.05  & 1.79  & 0.88    \\
 53& +51925.6 & 1.17    & 1.92  & 3.37  & 2.60  & 2.34  & 2.39  & 2.27  & 2.01  & 0.84    \\
 54& +51929.6 & 1.04    & 1.77  & 3.04  & 2.39  & 2.16  & 2.27  & 2.14  & 2.00  & 0.85    \\
 55& +51982.0 & 1.16    & 1.87  & 3.13  & 2.67  & 2.21  & 2.30  & 2.13  & 1.89  & 0.88    \\
 56& +52034.7 & 0.79    & 1.35  & 2.57  & 2.17  & 1.91  & 2.12  & 1.94  & 1.56  & 0.76    \\
 57& +52043.9 & 0.93    & 1.45  & 2.57  & 2.15  & 1.91  & 2.06  & 1.92  & 1.49  & 0.60    \\
 58& +52074.9 & 1.01    & 1.49  & 2.56  & 2.11  & 1.80  & 1.98  & 1.88  & 1.57  & 0.80    \\
 59& +52075.9 & 1.11    & 1.44  & 2.56  & 2.13  & 1.81  & 1.94  & 1.88  & 1.51  & 0.53    \\
 60& +52192.6 & 0.85    & 1.18  & 1.96  & 1.72  & 1.47  & 1.71  & 1.41  & 0.92  & 0.45    \\
 61& +52237.3 & 0.64    & 0.91  & 1.48  & 1.45  & 1.22  & 1.60  & 1.34  & 0.88  & 0.72    \\
 62& +52327.2 & 1.00    & 1.03  & 1.58  & 1.37  & 1.37  & 1.59  & 1.12  & 0.83  & 0.36    \\
 63& +52339.0 & 0.73    & 0.82  & 1.31  & 1.32  & 1.24  & 1.35  & 1.13  & 0.73  & 0.48    \\
 64& +52368.0 & 0.91    & 0.90  & 1.29  & 1.41  & 1.27  & 1.48  & 1.25  & 0.75  & 0.28    \\
 65& +52430.9 & 0.97    & 1.22  & 1.69  & 1.78  & 1.60  & 1.93  & 1.43  & 0.76  & 0.31    \\
 66& +52450.4 & 1.29    & 1.53  & 2.40  & 2.40  & 2.37  & 2.28  & 1.97  & 1.49  & 0.66    \\
 67& +52464.4 & 1.08    & 1.38  & 2.13  & 2.24  & 2.03  & 2.25  & 1.77  & 1.20  & 0.57    \\
 68& +52469.4 & 1.20    & 1.62  & 2.39  & 2.48  & 2.34  & 2.44  & 2.08  & 1.52  & 0.54    \\
 69& +52495.3 & 0.55    & 1.34  & 2.33  & 2.31  & 1.99  & 2.16  & 1.95  & 1.38  & 0.26    \\
 70& +52502.8 & 1.00    & 1.55  & 2.47  & 2.45  & 2.02  & 2.27  & 1.99  & 1.37  & 0.36    \\
 71& +52562.3 & 0.98    & 1.66  & 2.79  & 2.68  & 2.25  & 2.54  & 2.29  & 1.61  & 0.58    \\
 72& +52593.7 & 0.88    & 1.50  & 2.73  & 2.42  & 2.02  & 2.43  & 2.03  & 1.28  & 0.37    \\
 73& +52620.4 & 0.89    & 1.43  & 2.47  & 2.45  & 2.17  & 2.44  & 2.02  & 1.35  & 0.25    \\
 74& +52768.2 & 0.97    & 1.55  & 2.62  & 2.32  & 2.18  & 2.35  & 2.07  & 1.40  & 0.56    \\
 75& +52783.0 & 0.97    & 1.55  & 2.53  & 2.27  & 2.29  & 2.26  & 2.20  & 1.46  & 0.60    \\
 76& +52784.0 & 1.20    & 1.52  & 2.55  & 2.48  & 2.36  & 2.39  & 1.94  & 1.29  & 0.57    \\
 77& +52786.0 & 0.85    & 1.40  & 2.54  & 2.45  & 2.33  & 2.37  & 2.00  & 1.36  & 0.44    \\
 78& +52813.0 & 0.84    & 1.59  & 2.74  & 2.46  & 2.33  & 2.21  & 2.02  & 1.51  & 0.70    \\
 79& +52900.8 & 0.83    & 1.73  & 3.59  & 3.09  & 2.64  & 2.48  & 2.51  & 1.70  & 0.36    \\
 80& +52933.7 & 0.64    & 1.88  & 3.40  & 2.89  & 2.40  & 2.42  & 2.25  & 1.55  & 0.14    \\
 81& +52962.6 & 1.06    & 1.89  & 3.83  & 3.27  & 2.64  & 2.81  & 2.41  & 1.57  & 0.04    \\
 82& +52963.6 & 1.26    & 1.99  & 3.90  & 3.39  & 2.70  & 2.70  & 2.59  & 1.82  & 0.49    \\
 83& +53065.5 & 0.75    & 2.05  & 4.12  & 3.38  & 2.71  & 2.94  & 2.99  & 2.15  & 0.66    \\
 84& +53146.0 & 1.38    & 2.29  & 4.13  & 3.65  & 2.87  & 2.85  & 2.77  & 1.93  & 0.70    \\
 85& +53169.0 & 1.36    & 2.31  & 3.92  & 3.52  & 2.76  & 2.82  & 2.72  & 1.77  & 0.74    \\
 86& +53171.9 & 1.04    & 2.22  & 3.80  & 3.25  & 2.52  & 2.60  & 2.47  & 1.65  & 0.16    \\
 87& +53236.8 & 0.93    & 1.89  & 3.46  & 3.47  & 2.79  & 2.82  & 2.68  & 1.93  & 0.43    \\
 88& +53238.9 & 1.12    & 2.13  & 3.63  & 3.47  & 2.82  & 2.70  & 2.64  & 1.86  & 0.49    \\
 89& +53254.8 & 1.14    & 2.02  & 3.61  & 3.50  & 2.80  & 2.71  & 2.65  & 1.91  & 0.61    \\
 90& +53358.1 & 1.47    & 2.71  & 4.30  & 4.25  & 3.42  & 3.39  & 3.38  & 2.51  & 1.11    \\
 91& +53359.1 & 1.37    & 2.58  & 4.23  & 4.12  & 3.30  & 3.31  & 3.33  & 2.42  & 1.03    \\
 92& +53360.2 & 1.55    & 2.74  & 4.31  & 4.22  & 3.37  & 3.31  & 3.33  & 2.42  & 0.96    \\
 93& +53476.0 & 1.12    & 2.36  & 4.35  & 4.27  & 3.33  & 3.61  & 3.80  & 2.66  & 0.72    \\
 94& +53479.0 & 1.42    & 2.85  & 4.81  & 4.50  & 3.60  & 3.75  & 3.86  & 2.83  & 1.21    \\
 95& +53504.0 & 1.31    & 2.59  & 4.71  & 4.42  & 3.78  & 3.97  & 3.77  & 2.48  & 0.94    \\
 96& +53531.9 & 1.15    & 2.37  & 4.46  & 4.47  & 4.00  & 3.93  & 3.80  & 2.68  & 1.20    \\
 97& +53612.8 & 1.68    & 3.00  & 5.41  & 5.09  & 4.55  & 4.54  & 4.29  & 2.84  & 1.56    \\
 98& +53613.8 & 1.50    & 2.89  & 5.30  & 4.96  & 4.40  & 4.43  & 4.07  & 2.78  & 1.34    \\
 99& +53614.8 & 1.24    & 2.83  & 5.49  & 5.19  & 4.67  & 4.71  & 4.29  & 3.10  & 1.63   \\
100& +53711.4 & 1.96    & 4.03  & 7.07  & 6.01  & 5.25  & 5.09  & 4.98  & 3.49  & 1.29   \\
101& +53891.5 & 1.51    & 3.54  & 6.21  & 5.43  & 4.92  & 4.96  & 4.47  & 3.02  & 1.72   \\
102& +53979.8 & 1.47    & 3.58  & 6.03  & 4.94  & 4.45  & 4.65  & 4.19  & 2.92  & 1.48   \\
103& +53994.8 & 1.04    & 3.27  & 5.56  & 4.69  & 4.20  & 4.38  & 3.92  & 2.71  & 1.03   \\
104& +53995.8 & 1.31    & 3.38  & 5.82  & 5.00  & 4.49  & 4.59  & 4.03  & 2.80  & 1.57   \\
105& +53996.7 & 1.55    & 3.46  & 5.91  & 4.91  & 4.33  & 4.60  & 4.01  & 2.70  & 1.37   \\
106& +53997.7 & 1.42    & 3.58  & 6.17  & 4.92  & 4.44  & 4.62  & 4.02  & 2.89  & 1.22   \\
107& +53998.8 & 1.27    & 3.46  & 5.86  & 4.73  & 4.42  & 4.49  & 3.89  & 2.66  & 0.99   \\
108& +54036.6 & 1.24    & 3.27  & 5.79  & 4.83  & 4.29  & 4.37  & 3.66  & 2.68  & 1.08   \\
109& +54037.6 & 1.38    & 3.49  & 5.91  & 4.95  & 4.47  & 4.47  & 3.96  & 2.70  & 1.44   \\
110& +54040.7 & 1.35    & 3.36  & 5.93  & 4.84  & 4.35  & 4.40  & 3.87  & 2.58  & 1.29   \\
111& +54088.2 & 1.71    & 3.23  & 5.46  & 4.63  & 4.31  & 4.41  & 3.77  & 2.44  & 1.70    \\
112& +54112.6 & 1.43    & 2.94  & 5.17  & 4.53  & 4.22  & 4.31  & 3.75  & 2.46  & 1.65    \\
113& +54113.7 & 1.34    & 2.89  & 5.14  & 4.50  & 4.19  & 4.26  & 3.74  & 2.42  & 1.59    \\
114& +54143.6 & 1.73    & 3.01  & 5.05  & 4.59  & 4.52  & 4.53  & 3.83  & 2.52  & 1.88    \\
115& +54242.9 & 2.02    & 3.53  & 6.19  & 5.15  & 4.73  & 4.83  & 4.28  & 2.70  & 1.68    \\
116& +54243.9 & 2.42    & 3.88  & 6.52  & 5.15  & 4.68  & 5.10  & 4.88  & 3.26  & 1.95    \\
117& +54271.9 & 1.89    & 3.73  & 6.47  & 5.31  & 4.74  & 4.91  & 4.81  & 3.17  & 1.79    \\
118& +54273.0 & 2.49    & 3.93  & 6.84  & 5.53  & 5.41  & 5.41  & 5.07  & 3.22  & 2.00    \\
119& +54321.8 & 1.85    & 4.01  & 6.90  & 5.38  & 5.02  & 5.52  & 4.90  & 3.06  & 1.70    \\
120& +54323.8 & 1.65    & 3.93  & 6.81  & 5.40  & 5.26  & 5.62  & 5.07  & 3.33  & 1.97    \\
121& +54324.8 & 1.75    & 4.04  & 6.90  & 5.37  & 5.05  & 5.44  & 4.75  & 2.99  & 1.70    \\
122& +54346.7 & 1.96    & 4.24  & 7.46  & 5.81  & 5.72  & 5.82  & 5.04  & 3.11  & 1.86    \\
123& +54347.7 & 2.03    & 4.42  & 7.34  & 5.68  & 5.53  & 5.79  & 5.00  & 3.02  & 2.06    \\
124& +54348.7 & 1.71    & 4.17  & 7.20  & 5.55  & 5.35  & 5.60  & 4.79  & 2.86  & 1.86    \\
125& +54389.7 & 2.22    & 4.43  & 7.44  & 6.00  & 5.48  & 5.75  & 4.95  & 3.04  & 2.00    \\
126& +54391.6 & 2.05    & 4.36  & 7.38  & 6.07  & 5.92  & 6.13  & 5.16  & 3.17  & 1.99    \\
127& +54406.7 & 1.84    & 3.98  & 6.92  & 5.78  & 5.61  & 5.82  & 4.77  & 2.78  & 2.00    \\
128& +54408.6 & 1.98    & 4.14  & 7.15  & 6.09  & 5.88  & 5.84  & 4.84  & 2.90  & 2.04    \\
129& +54412.7 & 1.70    & 3.98  & 6.89  & 5.78  & 5.66  & 5.71  & 4.71  & 2.95  & 2.06    \\
\hline
\end{longtable}
}

\begin{figure}
\centering
\includegraphics[width=9cm]{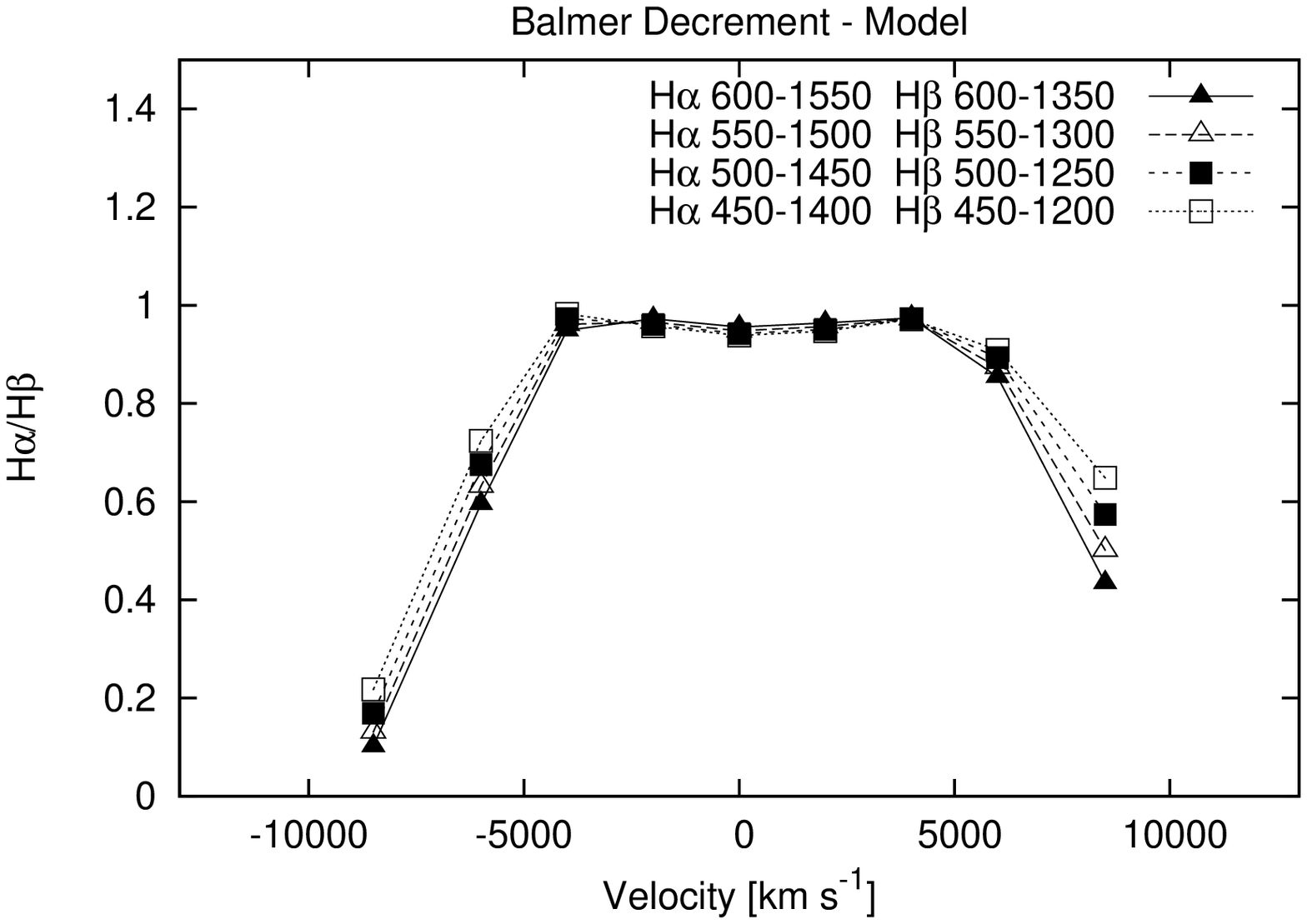}
\includegraphics[width=9cm]{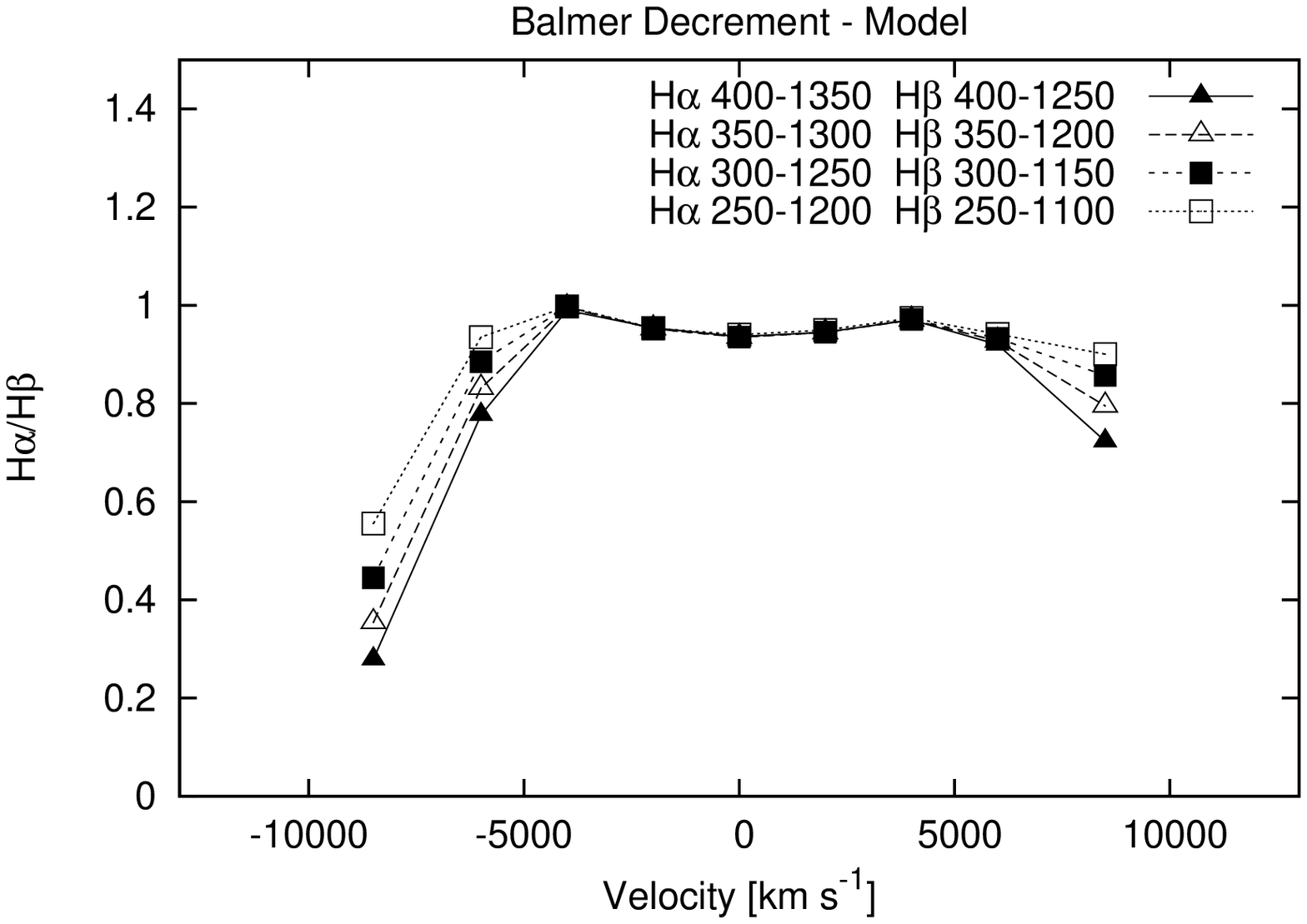}
\includegraphics[width=9cm]{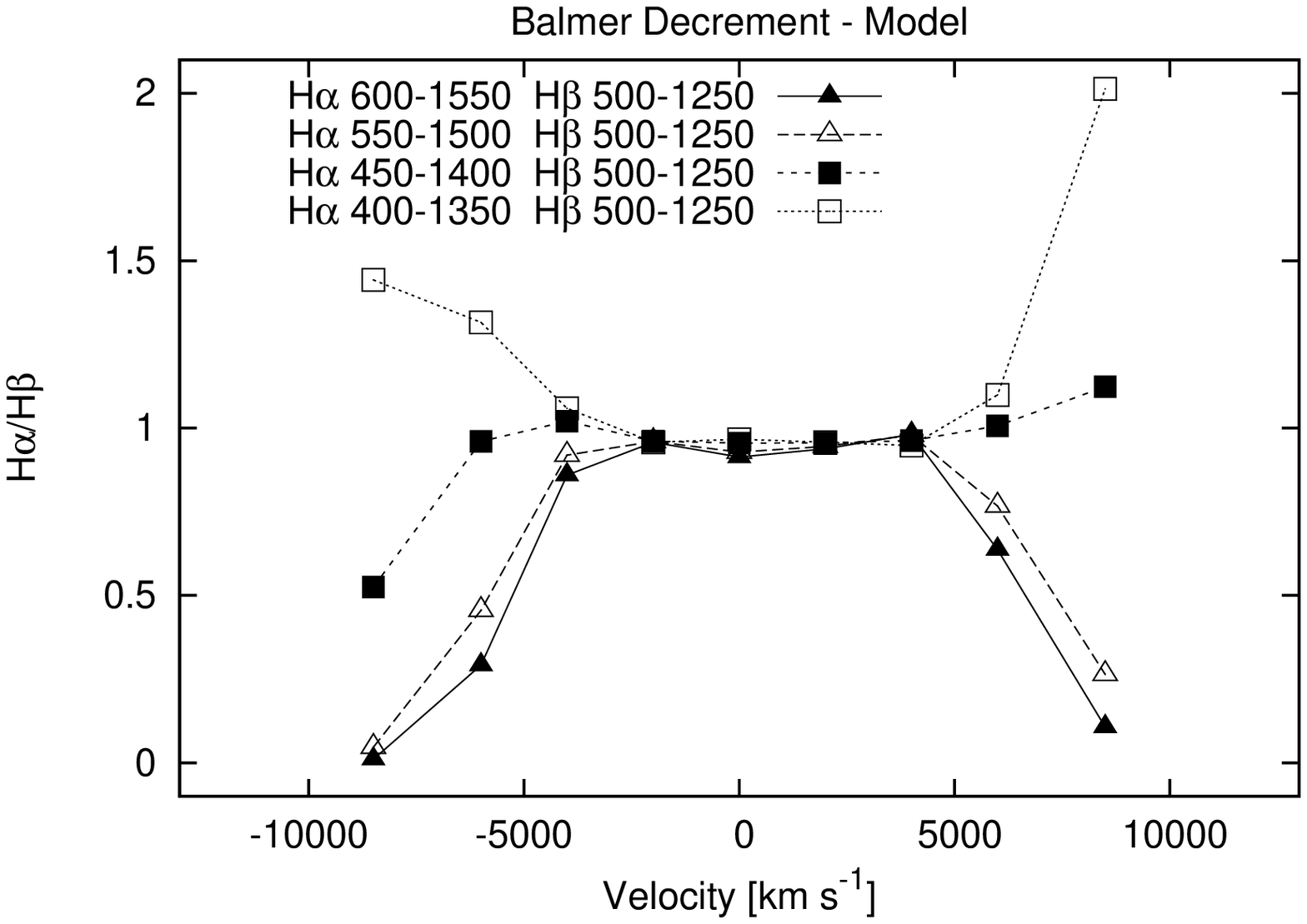}
\caption{The  Balmer decrement versus velocity. First two
panels --  the inner radius is the same, but different positions of
the H$\alpha$ and H$\beta$  disk with respect to the black hole are
assumed. The bottom panel shows the cases where the one of the two
emitting regions is closer to the central black hole. The inner and
outer radii assumed in  models are labeled on each panel. The BD
points are normalized to the one at -4000 km s$^{-1}$.}\label{f191}
\end{figure}

\subsection{Modeling of the Balmer decrement}

It is known that the BD depends on physical processes. { But BD
as function of the velocities along the line profile may also depend
on the geometry of the H$\alpha$ and H$\beta$ emitting region, as
e.g. if the line emission region of  H$\alpha$ and H$\beta$ are not
the same one may expect that the H$\alpha$ and H$\beta$ flux ratio
will be a function of the gas velocity.}  Therefore, we modeled the
shape of the BD vs. velocity.

In order to test the influence of the geometry to the BD vs.
velocity along the line, we used modeled disk profiles (see \S 3.2),
but now we consider two cases: {a) the disk regions emitting
H$\alpha$ and H$\beta$ have different sizes but the same inner radii
that vary accordingly (as discussed in \S 3.2); b) the inner radius
of the H$\alpha$ emitting region differ from the fixed radius of the
H$\beta$ region by $\pm 50 R_{\rm g}, \pm 100 R_{\rm g}$.}

Our simulations of the BD vs. velocity are given in Fig. \ref{f191},
where it can be seen that the disk model (assuming different
dimensions of the emission disk that emits H$\alpha$ and H$\beta$)
can reproduce very similar BD profiles along velocity field,
especially as they are observed  in Period I. There is a difference
in shape that is probably caused by the physical conditions in the
disk as well as due to the central component emission. We found
bell-like profiles of BD, and they also have two peaks (in red and
blue part), as well as a small minimum in the center. The shape of
the BD vs. velocity observed in Period II, where a peak in the blue
part is prominent (see Fig. \ref{f16}), and in some cases has a
deeper red minimum than blue one, cannot be obtained in modeled BD
vs. velocity. This may be due to the influence of the central
component, but also, due to the outburst in Period II, or due to
different physical processes in the disk.

{As it can be seen in Fig. \ref{f191} (bottom panel, open
squares), a big difference between modeled BD vs. velocity and
observed one is in the case where the H$\alpha$ emitting region is
closer to the black hole for 100 R$_g$. This indicates that the disk
part emitting H$\alpha$ cannot to be  significantly closer to the
central black hole than the one that emits the H$\beta$ line.}

\section{Discussion}

In this paper we performed detailed study of the line parameter
variations (peak separation, variation of the line segments, Balmer
decrement)  during a 13-year long period. As we noted above, the
line profiles of the 3C 390.3 in the monitoring period have a
disk-like shape, i.e. there are mainly two peaks, where the blue one
is more enhanced. In some periods, there is a central peak, that
probably is coming from an additional emission region or extra
emission caused by some perturbation in the disk \citep[see
e.g.][]{jov10}. In order to confirm the disk emission hypothesis we
modeled the peak position variation using a semi-relativistic model,
and found that peak position variations can be explained with the
disk model.

We also investigated line-segment flux variation and found that
the ratios of the line segments are well fitted by a disk model. It
is interesting that the response of the line-segment flux to
line-segment flux is better in Period II than in Period I. The same
is in the case of the response  of the segment fluxes to the
continuum flux. It seems that in Period I, the variability in the
line profile was not caused only by an outburst in the ionizing
continuum,  but also by some perturbations in the accretion disk
\citep[see e.g.][]{jov10}. While in Period II, it could be that the
disk structure is not changing much and that the variability in the
line parameters is primarily caused by the outburst in the
continuum.

The Balmer decrement as a function of the continuum has a
decreasing trend  in Period I (with the increase of the continuum)
and is constant in Period II. Also, there is a different shape of
the BD vs. velocity in different periods. The characteristic shape
of the BD vs. velocity has two peaks (blue and red) which do not
correspond to the line peaks (they are in the far wings, see Fig.
\ref{f16}), there are minima in the far blue and red wing, as well
as a minimum in the center. The shape of the BD vs. velocity is
similar to what we expect from an accretion disk, where the
dimensions of the disk for H$\alpha$ and H$\beta$ are different. In
the period (Period II) of the highest brighteners  in the continuum
and in lines, the  BD vs. velocity shapes show a blue peak and
more-less flat red part, that is probably caused by some other
emission, additional to the disk one, and also by different physical
conditions in these two sub-regions. Our modeled shapes of the BD
vs. velocity show that the case of the disk emission where the
dimensions of the H$\alpha$ and H$\beta$ emitting disks are
different, can explain bell-like shapes of the BD vs. velocity. Note
here that we did not consider physical processes in the emitting
disk which can affect the shape of the BD vs. velocity. Comparing
the dimensions of the disk, we came to the same conclusion as from
the CCF (see Paper I), that the H$\alpha$ emitting disk is larger
than the H$\beta$ one, but also that the inner radius of both disks
seems to be similar.

\subsection{Variability and the BLR structure of 3C 390.3}

Before we discuss the BLR structure, let us recall some results that
we obtained in Paper I and in this paper: i) the CCF analysis shows
that the dimension of the BLR that emits  H$\beta$  is smaller than
one that emits H$\alpha$; ii) there is a possibility that there are
quasi-periodic oscillations, which may be connected with the
instability in the disk or disk-like emitting region; iii)
concerning the line profile variations, the variability observed in
13-year period can be divided into two periods, before and after the
outburst in 2002; iv) there are three peaks, two which are expected
in the disk emission, and one - central, that is probably coming
from disk perturbations or from an additional emitting region; v)
the shapes of H$\beta$ and H$\alpha$ observed in the same period are
similar, especially in Period I (note here that in Period II, there
are differences in the center and red wings); vi) our simulation of
the variability, taking into account only different disk position
with respect to the central black hole, can qualitatively explain
the observed broad line parameter variations; vii) comparing the
shifts of the modeled and observed H$\beta$ line, we found that
there is a blue-shift in the observed H$\beta$ comparing to the
modeled one, that is also different in two periods: around 300 km
s$^{-1}$ in Period I and around 850 km s$^{-1}$ in Period II; viii)
the inspection of the delays of the individual line segments
 (see \S 4.3) indicates presence of a disk and also
a wind, i.e. the disk-wind model may be present in the BLR of 3C 390.3.

Additionally, note here that the disk-like structure of the 3C 390.3
BLR is favored in several recent papers \citep[see
e.g.][]{fl08,jov10} assuming an elliptical disk \citep[][]{fl08} or
perturbations in the disk \citep[][]{jov10}.

\begin{figure}
\centering
\includegraphics[width=9cm]{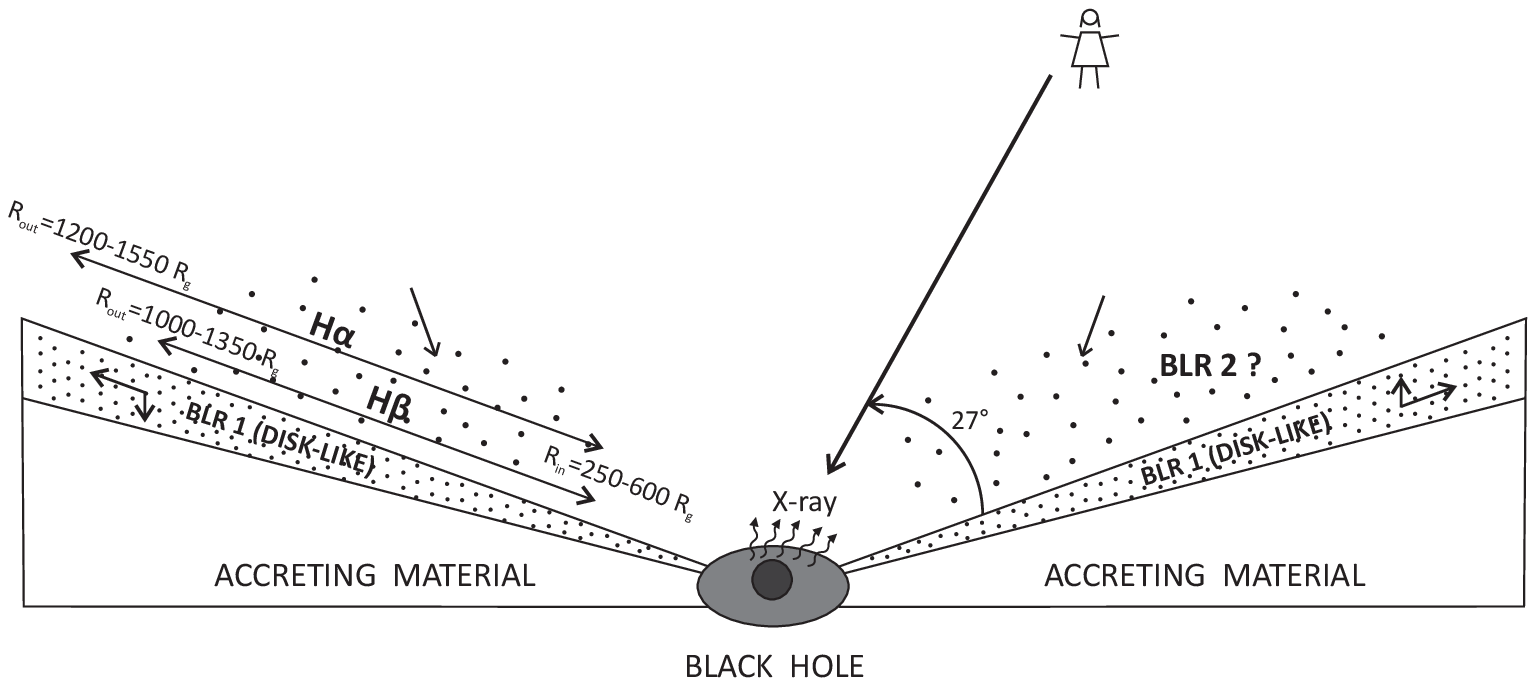}
\caption{The model of the BLR that is composed of the disk-like BLR1, and an outflowing/inflowing BLR2.}\label{f20}
\end{figure}

Taking into account all mentioned above, one can speculate about the
BLR structure and the nature of these variations. Let us consider
that, in principle, there is an accreting material that is optically
thick, and the light from a central source is able to photoionize
only a small thin region above (below) the accreting material (or
thick disk, see Fig. \ref{f20}), let us call this region a
``disk-like broad line region'' (disk-like BLR1). Emitting gas in
this region has disk-like motion, since it is located in the disk
periphery and follows the disk kinematics. The brightness of broad
lines which are coming from this region will be caused by the
central ionizing continuum source, but the line  parameters will
depend on the dimensions of the region as well as the location of
the BLR1 with respect to the central black hole. Of course, other
parameters as e.g. emissivity or local velocity dispersion can
slightly affect the line profiles, but it seems that the change in
the inner/outer radius is the most important.

The shape of the broad lines confirm the disk-like geometry, and
correlation between the broad-line and continuum fluxes confirms the
influence of the central source to the line intensity (see Paper I).
The optical continuum and the H$\beta$ flux variations are probably
related to changes in the X-ray emission modulated by variable
accretion rate, i.e. the change of the surface temperature of the
disk as a result of the variable X-ray irradiation \citet{U00}. But
question is, what can cause the changes in the dimension (as well in
position) of the broad line emitting region? It is obvious that it
is not the continuum flux, since the variation in the continuum does
not correlate with the line parameter variation.

The variation in the line parameters may be related to the
perturbations in the accreting material that is feeding the
disk-like BLR with scattered gas. The density and optical depth of
such gas from the accreting material may cause that in different
epochs the disk-like BLR has different dimensions and positions with
respect to the central black holes. As, e.g. if scattered material
(gas) from accreting material is absent in some parts, it will
affect the line parameters similar as change in the dimensions
(position) of the disk-like BLR.

On the other side, the observed broad double-peaked lines have a
blue-shift with respect to the modeled one. This may indicate that
there is some kind of wind in the disk, probably caused by the
radiation of the central source. Recently \citet{to10} performed  an
uniform and systematic search for the blueshifted Fe K$\alpha$
absorption lines in the X-ray spectra of 3C 390.3 observed with
Suzaku and detected absorption lines at energies greater than 7 keV.
That implies the origin of the Fe K$\alpha$ is in the highly ionized
gas outflowing with mildly relativistic velocities (in the velocity
range from 0.04 to 0.15 light speed). Taking into account that the
optical line emission is probably originating farther away than the
X-ray, one can expect significantly smaller outflow velocities in
the optical lines.

One can expect that  perturbations in the thick disk (or accreting
material) can affect the emitting disk-like region. It may cause
some perturbation or spiral shocks in the disk-like BLR \citep[see
e.g.][]{ch94,jov10}. \citet{ch94} found that the observed
variability of the double-peaked broad emission lines (seen in some
active galactic nuclei) can be due to the existence of two-armed
spiral shocks in the accretion disk. Using this model they
successfully fitted the observations of  3C 390.3 made at different
epochs with self-similar spiral shock models which incorporate
relativistic corrections.

Alternatively, \citet{jov10} fitted the observations of  3C
390.3 from different epochs with a model that includes moving
perturbation across the disk.  The perturbations/shocks in the disk
model can explain different line profiles, and also the central peak
\citep[see e.g.][]{jov10}. But the central peak might be emitted
from a broad line region that does not follow the disk geometry, it
may be very similar to the two-component model, i.e. that we have
composite emission from the disk and an additional region \citep[see
e.g.][]{pop04,i06,b09}. Also, \citet{tig10}  considered that the BLR
of 3C390.3 has a complex structure composed from two components: 1)
BLR1 -- the traditional BLR (Accretion Disk), 2) BLR2 -- a
sub-region (outflow) that surrounds the compact radio jet.  In
 this paper the existence of a jet-excited outflowing BLR is suggested,
which may  question the general assumption of a virialized BLR.

It is interesting that the central component is present during the
whole monitoring period, and that in 1995--1996 it was located in
the center, and after that shifted to the the red part of line. Such
redshift is not expected if there is an outflow in the BLR. But let
us recall the results obtained in the study of dynamics of the
line-driven disk-wind, where kinematics of the gas shows that both
infall and outflow can occur in different regions of the wind at the
same time \citep[see][]{pr98,pr00}, which depends on radiation
force. Also, recently \cite{ga09} reported about the possibility
that an inflow is present in the BLR of AGN.

Finally, one can consider a model as given in Fig. \ref{f20}: the
disk-like BLR1 follows the kinematics of the accreting material, and
from time to time perturbations can appear in this region (similar
as in the fully accretion disk). The emitting gas in the BLR1 is
coming (ejected) from the accreting material and it is ionized by
the central source. The radiation pressure may contribute in such
way that we have a wind (an outflow) in this region. There may be
also a region above the disk-like BLR1 (a BLR2 in Fig. \ref{f20}),
where the ionized gas may have motion with respect to the central
black hole, and depending on the radiation pressure the velocities
are different in different periods.

\subsection{Different nature of variation in broad lines of 3C 390.3}

It is interesting to discuss the difference between the variation in
Period I and Period II. It is obvious that the disk-parameters
variation can explain the observed line-parameter variations (as
e.g. in the peak separation and line-segment fluxes variations). In
Period I, there were several quasi-periodical low-intensity
outbursts (see Paper I), and it seems that the line, and
consequently the disk-like geometrical parameters are changed. It
may be that in this period the outbursts are caused by some
perturbations in the disk (in the inner part that emits the X-ray
radiation as well as in the outer part emitting the broad lines).
These perturbations affected the line profiles and probably partly
line fluxes (especially in the far wings). As it can be seen from
modeled variation the inner/outer radius is significantly changed.

In the Period II, there is a strong outburst, that caused the
increase of the brightness in the continuum and lines.  The lines
stay brighter, but the structure (inner/outer radius) of the
disk-like region is not changing much. There is a different response
of the continuum to the line segments in Period I and Period II,
i.e. in Period II there is a higher correlations between the
continuum and line-segment fluxes.

On the other hand, the BD in Period I has a decreasing trend with
the increase of the continuum flux, while in Period II, it stays
almost constant. This may indicate different nature of the
variability. It seems that in Period I, beside the ionizing
continuum influence on the line intensities (especially in the far
wings), there was the additional effect that influence the line
profiles and intensities. Probably it was caused by some
perturbations or shocks in the disk-like BLR  \citep[][]{ch94,jov10}
that produced changes in the structure of the disk-like BLR.

\section{Conclusions}

In this paper, we present line profile variations of 3C 390.3 in a
long period. Due to the change of line profiles, we divided the
observations into two periods (before and after the minimum in 2002:
Period I and II, respectively) and found difference in the line
segments and Balmer decrement variations in these two periods. From
our investigation, the main conclusions are the following:

i) the line profiles during the monitoring period are changing,
always showing the disk-like profile, with the higher blue peak.
There is also the central peak that may come from the emission
region additional to the disk, but as it was mentioned in
\citet{jov10}, it may also be caused by the perturbation in the
disk.

ii) the far-wings flux variation in the first period, where the
far-red wing flux does not respond well to the continuum flux, 
is probably caused by some physical processes in the innermost part
of the disk. The observed changes in H$\alpha$ and H$\beta$ may be
interpreted in the framework of a disk model with changes in the
location and size of the disk line emitting regions. In Period II,
the good correlation between the continuum and line-segment flux
suggests that the brightness of the disk is connected with the
ionizing continuum, and that the structure of the disk does not
significantly change.

iii) Balmer decrement is also different for these two periods. In
the first period the BD decreases with continuum flux, while in the
second period the BD stays more-less constant (around 4.5). The
segment BD shows two maxima (around $\pm$6000 km s$^{-1}$) which do
not correspond to the red and blue peak, but instead they are
farther in the blue and red wing (than peaks velocities). Also one
minimum around zero velocity is present. This minimum changed
position between $\pm$2000 km s$^{-1}$ around zero velocity. This
central minimum (as well as shifted maxima) may be caused by the
additional (to the accretion disk) emission that perhaps is present
in the 3C 390.3 broad lines. These results suggest that, in
addition to the physical conditions across disk, the size of the
H$\alpha$ and H$\beta$ emitting regions of the disk plays an
important role. We modeled bell-like BD vs. velocity profiles in
the case when the H$\alpha$ disk is larger than H$\beta$ one.

iv) the variation observed in the line parameters can be well
modeled if one assumes changes in position of the emitting disk with
respect to the central black hole. The emission of the disk-like
region is dominant, but there is the indication of the additional
emission. Therefore, to explain the complex line-profile variability
one should consider a complex model that may have a disk geometry
together with outflows/inflows (see Fig. \ref{f20}).

An important conclusion of this work is that, even if the
disk-like geometry plays a dominant role, the variability of the
H$\alpha$ and H$\beta$ line profiles and intensities (and probably
partly in the continuum flux) has different nature for different
periods. It seems that in Period I, the perturbation(s) in the disk
caused (at least partly) the line and continuum amplification, while
in Period II the ionizing continuum caused the line amplification
without big changes in the disk-like structure.

\section*{Acknowledgments}

This work was supported by INTAS
(grant N96-0328), RFBR (grants N97-02-17625 N00-02-16272,
N03-02-17123, 06-02-16843, and N09-02-01136), State program
'Astronomy' (Russia), CONACYT research grant 39560-F and 54480
(M\'exico) and the Ministry of Science and Technological Development
of Republic of Serbia through the project Astrophysical Spectroscopy
of Extragalactic Objects. L. \v C. P.,W. K. and D. I. are grateful
to the Alexander von Humboldt fondation for support in the frame of
program "Research Group Linkage". {We thank the anonymous referee for useful suggestions that improved
the clarity of this manuscript.}

\Online


\begin{thebibliography}{}


\bibitem [\protect\citeauthoryear{Arshakian et al.} {2010}]{tig10}
 Arshakian, T. G., León-Tavares, J., Lobanov, A. P., Chavushyan, V. H., Shapovalova, A. I., Burenkov, A. N.,
Zensus, J. A. 2010, MNRAS, 401, 1231

\bibitem [\protect\citeauthoryear{Bon et al.} {2009}]{b09} Bon, E., Popovi\'c, L. \v C., Gavrilovi\'c, N., La Mura, G., Mediavilla, E. 2009,
MNRAS, 400, 924

\bibitem [\protect\citeauthoryear{Chakrabarti \& Wiita} {1994}]{ch94} Chakrabarti, S. K., Wiita, P. J. 1994, ApJ, 434, 518 ch94

\bibitem[\protect\citeauthoryear{Chen \& Halpern} {1989}]{ch89} Chen, K., \& Halpern, J.~P.\ 1989, \apj,  344, 115

\bibitem[\protect\citeauthoryear{Chen et al.} {1989}]{chp89}  Chen, K., Halpern, J. P. \& Filippenko, A. V. 1989, ApJ, 339, 742


\bibitem[\protect\citeauthoryear{Dietrich et al.} {1988}]{di88}  Dietrich, M., Peterson, B. M., Albrecht, P. et al. 1988, ApJS, 115, 185

\bibitem[\protect\citeauthoryear{Eracleous et al.} {1997}]{er97} Eracleous, M., et al. 1997, \apj, 490, 216

\bibitem[\protect\citeauthoryear{Eracleous \& Halpern} {1994}] {eh94} Eracleous, M. \& Halpern, J. P. 1994, ApJS, 90, 1

\bibitem[\protect\citeauthoryear{Eracleous \& Halpern} {2004}] {eh04} Eracleous, M. \& Halpern, J. P. 2004, ApJS, 150, 181

\bibitem[\protect\citeauthoryear{Eracleous et al.} {2009}] {er09} Eracleous, M., Lewis, K. T., Flohic, H. M. L. G. 2009, NewAR, 53, 133

\bibitem[\protect\citeauthoryear{Flohic \& Eracleous} {2008}]{fl08} Flohic, H.~M.~L.~G., \& Eracleous, M.\ 2008, \apj, 686, 138

\bibitem[\protect\citeauthoryear{Horne et al.} {2004}]{h04} Horne, K., Peterson, B. M., Collier, S. J., Netzer, H. 2004, PASP, 116, 465

\bibitem[\protect\citeauthoryear{Gaskell} {1996}]{gas96} Gaskell, C.M. 1996, \apj, 464, L107

\bibitem[\protect\citeauthoryear{Gaskell} {2009}]{ga09} Gaskell, C. M. 2009, NewAR, 53, 140

\bibitem[\protect\citeauthoryear{Ili\'c et al.} {2006}]{i06} Ili\'c, D., Popovi\'c, L. \v C., Bon, E., Mediavilla, E. G., Chavushyan, V. H. 2006, MNRAS, 371, 1610

\bibitem[\protect\citeauthoryear{Jovanovi\'c et al.} {2010}]{jov10} Jovanovi\'c, P., Popovi\'c, L. \v C., Stalevski, M., Shapovalova, A. I. 2010, \apj, 718, 168

\bibitem[\protect\citeauthoryear{Kollatschny} {2003}]{ko03} Kollatschny, W. 2003, \aap, 407, 461

\bibitem[\protect\citeauthoryear{Kollatschny \& Bischoff} {2002}]{ko02}  Kollatschny, W., \& Bischoff, K. 2002, \aap, 386, L19

\bibitem[\protect\citeauthoryear{K\"onigl \& Kartje} {1994}] {kk94} K\"onigl, A., Kartje, J. F. 1994, ApJ, 434, 446

\bibitem[\protect\citeauthoryear{Lewis et al.} {2010}]{le10} Lewis, K. T., Eracleous, M., Storchi-Bergmann, T. 2010, ApJS, 187, 416

\bibitem[\protect\citeauthoryear{Perez et al.} {1988}]{pe88} Perez, E., Mediavilla, E., Penston, M. V., Tadhunter, C., Moles, M.
1988, MNRAS, 230, 353

\bibitem[\protect\citeauthoryear{Popovi\'c et al.} {2004}]{pop04} Popovi\'c, L.\v C., Mediavilla, E.; Bon, E.; Ili\'c, D. 2004, \aap, 423, 909

\bibitem[\protect\citeauthoryear{Proga et al.} {1998}]{pr98} Proga, D., Stone, J. M., Drew, J. E. 1998 MNRAS, 295, 595

\bibitem[\protect\citeauthoryear{Proga et al.} {2000}]{pr00} Proga, Daniel; Stone, James M.; Kallman, Timothy R. 2000, ApJ, 543, 686

\bibitem[\protect\citeauthoryear{Sergeev et al.} {2002}]{se02} Sergeev, S. G., Pronik, V. I., Peterson, B. M., Sergeeva, E. A., Zheng, W.  2002 \apj, 576, 660

\bibitem[\protect\citeauthoryear{Sergeev et al.} {2010}]{se10} Sergeev, S. G., Klimanov, S. A., Doroshenko, V. T., Efimov, Yu. S., Nazarov, S. V., Pronik, V. I.
  2010 MNRAS, tmp.1496S

\bibitem[\protect\citeauthoryear{Shapovalova et al.} {2001}]{sh01} Shapovalova, A.I., Burenkov, A.N. , et al., 2001, A\&A, 376, 775

\bibitem[\protect\citeauthoryear{Shapovalova et al.} {2004}]{sh04} Shapovalova, A. I., Doroshenko, V. T., Bochkarev, N. G., et al. 2004, \aap, 422, 925

\bibitem [\protect\citeauthoryear{Shapovalova et al.} {2010}]{sh10} Shapovalova, A. I., Popovi\'c, L. \v C.,  Burenkov, A. N. et al. 2010, \aap, 517A, 42

\bibitem [\protect\citeauthoryear{Sulentic et al.} {2000}]{sul00} Sulentic, J. W., Marziani, P., Dultzin-Hacyan, D. 2000, ARA\&A, 38, 521

\bibitem [\protect\citeauthoryear{Tombesi et al.} {2010}]{to10} Tombesi, F., Sambruna, R. M., Reeves, J. N., Braito, V., Ballo, L., Gofford, J., Cappi, M.,
 Mushotzky, R. F. 2010, ApJ, 719, 700

\bibitem [\protect\citeauthoryear{Ulrich} {2000}] {U00} Ulrich, M.-H.\ 2000, \aapr, 10, 135

\bibitem [\protect\citeauthoryear{Van Groningen \& Wanders} {1992}] {vw92} Van Groningen, E., \& Wanders, I.1992, PASP, 104, 700

\bibitem [\protect\citeauthoryear{Veilleux \& Zheng} {1991}] {vz91} Veilleux, S. \& Zheng, W. 1991, ApJ, 377, 89.

\bibitem [\protect\citeauthoryear{Welsh \& Horne} {1991}]  {w01} Welsh, W. F., Horne, K.  1991, ApJ, 379, 586

\bibitem [\protect\citeauthoryear{Zamfir et al.} {2010}]{za10} Zamfir, S., Sulentic, J. W., Marziani, P., Dultzin, D. 2010, MNRAS, 403, 1759

\bibitem[\protect\citeauthoryear{Zheng} {1996}]{zh96} Zheng, W. 1996, \aj, 111, 1498



\end{thebibliography}
\end{document}